\documentclass[usenatbib,usegraphicx]{mn2e}

\usepackage{pslatex,color,tabularx,float,multicol,caption}
\usepackage{amsmath,amsfonts,amssymb,amsxtra,eufrak,url}

\def\idm#1{{\mbox{\scriptsize #1}}}

\def\vec#1{{\pmb #1}}

\newcommand{\yr}{\mbox{yr}}

\newcommand{\au}{\mbox{au}}

\newcommand{\msun}{\mbox{m}_{\odot}}

\newcommand{\mE}{\mbox{m}_{\oplus}}

\newcommand{\Mmean}{\mathcal{M}}

\title[Migration and 9:7 MMR]
{On the {migration-induced} formation of the 9:7 mean motion resonance}
\author[Cezary Migaszewski]{Cezary Migaszewski$^{1,2}$\thanks{E-mail:migaszewski@umk.pl}\\
$^{1}$Institute of Physics and CASA$^*$, Faculty of Mathematics and Physics, University of Szczecin, Wielkopolska 15, 70-451 Szczecin, Poland\\
$^{2}$Centre for Astronomy, Faculty of Physics, Astronomy and Informatics, Nicolaus Copernicus University, Grudziadzka 5, 87-100 Toru\'n, Poland}
\begin{document}
%
\date{Accepted 2017 April 10. Received 2017 April 10; in original form 2017 January 23}
\pagerange{\pageref{firstpage}--\pageref{lastpage}} \pubyear{2017}
\maketitle
\label{firstpage}

\captionsetup[figure]{labelfont=bf,font=small}

\begin{abstract}

We study formation of 9:7 mean motion resonance (MMR) as a result of convergent migration of two planets embedded in a disc. Depending on the migration parameters, initial orbits and planets' masses the system may pass through the resonance or enter it (permanently or temporarily). We illustrate that a stable equilibrium of the averaged system (a periodic orbit of the N-body model) is surrounded by the region of permanent resonance capture{, whose} size depends on the migration parameters {and the} planets mass ratio. A system located inside this region tends towards the equilibrium (the capture is permanent){, while} a system located outside the region evolves away from the equilibrium and leaves the resonance. We verify recent results of Delisle et al. and Xu~\&~Lai where they show that for $m_1 \lesssim m_2$ {($m_1, m_2$ are the inner and outer planets' masses, respectively)} the equilibrium is unstable when the migration is added, so the capture cannot be permanent. We show that for particular migration parameters the situation may be reversed (the equilibria are unstable for $m_1 \gtrsim m_2$). We illustrate that 9:7~MMR consists of two modes separated with a separatrix. The inner one is centred at the equilibrium and the outer one has no equilibrium in its centre. A system located outside the region of stable capture evolves from the inner into the outer mode. The evolution occurs along families of periodic orbits of the averaged system, that play a crucial role in the dynamics after the resonance capture.

\end{abstract}

\begin{keywords}
Planetary systems -- planets and satellites: dynamical evolution and stability -- planet-disc interactions
\end{keywords}

\section{Introduction}

It is generally accepted that planets migrate in protoplanetary discs and that the migration plays an important role in formation of mean motion resonances \citep[e.g.][]{Snellgrove2001,Lee2002,Murray2002}. It is also known that for slow and smooth migration a system of two planets evolves along families of periodic orbits \citep[e.g.,][]{Beauge2003,Beauge2006,Hadjidemetriou2006,Hadjidemetriou2010,Migaszewski2015,Voyatzis2016}. The cited papers, though, are focused mainly on first order MMRs. Recently, \cite{Migaszewski2017} discussed the importance of the periodic orbits of the second order MMR, 9:7, in the possible formation and dynamics of a system of two super-Earths around Kepler-29. 

\cite{Migaszewski2015} showed that the system which undergoes the convergent migration not only follows the branch of periodic orbits, but also that the final system resulting from the smooth migration with efficient circularisation ends up as an exactly periodic configuration. As discussed in \citep{Migaszewski2017}, for the second order MMR, like 9:7, the final configuration is never exactly periodic, although the system may be located relatively close to the periodic system. In this work we drill the topic and try to understand why in certain situations the system passes through and in some other cases it enters the resonance. Moreover, we want to find conditions under which the system stays in the resonance either permanently or temporarily.

Our interest in the 9:7 MMR is not only motivated by the discovery of the Kepler-29 system and its resonant nature \citep{Fabrycky2012,JontofHutter2016,Migaszewski2017}. It is also motivated by the fact that resonances of orders higher than one have not been studied extensively in terms of the migration. The papers devoted to the topic use various approaches and focus on different aspects of the resonance formation. For instance \cite{Xiang-Gruess2015} studied the migration of two-planet systems in wide ranges of the migration parameters and obtained several systems that ended up in higher order MMRs. The work was focused on finding constrains on the migration parameters (like the migration and circularisation rates) that lead to formation of various MMRs. \cite{Quillen2006} and \cite{Mustill2011} studied the resonance capture probability for second order MMRs as a function of masses and migration parameters. There has been also pointed out that the possibility of the capture can be determined by analysing the dynamics of the averaged system in vicinity of the stable equilibrium when the migration is added to the model \citep[e.g.,][]{Delisle2015,Xu2016}. It was found in the cited papers that for a few Earth mass planets and the circularisation rates comparable for both planets the capture cannot be permanent if the inner planet is less massive than the outer one. 

Unlike in the cited papers, in this work we focus on particular resonance (9:7~MMR) and study its structure in details. Apart from the equilibrium of the averaged system (a periodic orbit of the unaveraged system), which was shown to play an important role in the {migration-induced} resonance formation, we analyse a role of periodic configurations of the averaged system in the process.

The paper is organised as follows. In section~2 we introduce a model of the system of two migrating planets. We present an example which illustrates probabilistic nature of the resonance capture. We chose three initial  configurations differing only slightly one from another that evolve in three qualitatively different ways. In Section~3 we present the averaged equations of motion that will be used in most of the analysis. The study of the dynamics of the conservative model (without the migration) is presented in Section~4. We describe the structure of the phase space, i.e., equilibria, periodic orbits, chaotic regions corresponding to the separatrices. In Section~5 we study the evolution of the system of migrating planets. We start from the dynamics of the system which is already in the resonance. We explain the role of the equilibrium (in particular, we describe the region of stability around it) and the periodic orbits. Next, we focus on the process of entering into the resonance. In the last part of this section we study the importance of the planets' mass ratio as well as a particular model of migration for the possibility of the permanent capture in the resonance. Conclusions are gathered in the last section.

\section{A model of the system}

In order to introduce the migration and circularisation of the planets into the N-body equations of motion we use a simple parametric model. For the $i$-th planet ($i=1,2$) the additional acceleration (expressed in the astrocentric reference frame) is a linear combination of the astrocentric velocity $\vec{v}_i$ and a velocity in a circular orbit at the distance of planet $i$, $\vec{v}_{c,i}$ \citep[e.g.,][]{Papaloizou2000,Beauge2006,Moore2013,Voyatzis2016}. It may be parametrised as follows:
\begin{equation}
\vec{f}_i = -\frac{\vec{v}_i}{2\,\tau_{a,i}} - \frac{\vec{v}_i - \vec{v}_{c,i}}{\tau_{e,i}},
\label{eq:migration_force}
\end{equation}
where the time-scales of migration and circularisation of planet~$i$ are denoted by
$\tau_{a,i}$ and $\tau_{e,i}$, respectively. We assumed \citep[after][]{Migaszewski2017} that 
\begin{equation}
{\tau_{a,i} = \tau_0 \, \left( \frac{r_i}{1\,\au} \right)^{-\alpha}, \quad \tau_{e,i} = \tau_{a,i}/\kappa,}
\label{eq:tau_a}
\end{equation}
where $\tau_0$, $\alpha$ {and $\kappa$} are constant. 

The model of the force given above is widely used in parametric studies of the migration that are focused on the mechanical part of the problem rather than its astrophysical aspect, although other models are used in the literature as well \citep[e.g.,][]{Hadjidemetriou2010}. The parameters can be adopted from wide ranges of values. We will choose them to be realistic, so it would be possible, in principle, to find such a physical model of the disc that would lead to the migration governed by the parametric model of those values of the parameters.

An important limitation of the parametric model is that it assumes that if planet~$i$ perturbs the mass distribution in the disc, the perturbation results in a force acting only on planet~$i$ and does not affect planet~$j$, $j \neq i$. For low mass planets, in a range of a few Earth masses, the assumption is relatively safe. For more massive planets it may be more risky \citep{Podlewska-Gaca2012,Baruteau2013}. In this work we study the migration of two super-Earths, therefore the single-planet approximation may be safely applied.

\subsection{Probabilistic capture}

\begin{figure*}
\centerline{
\vbox{
\hbox{
\includegraphics[width=0.33\textwidth]{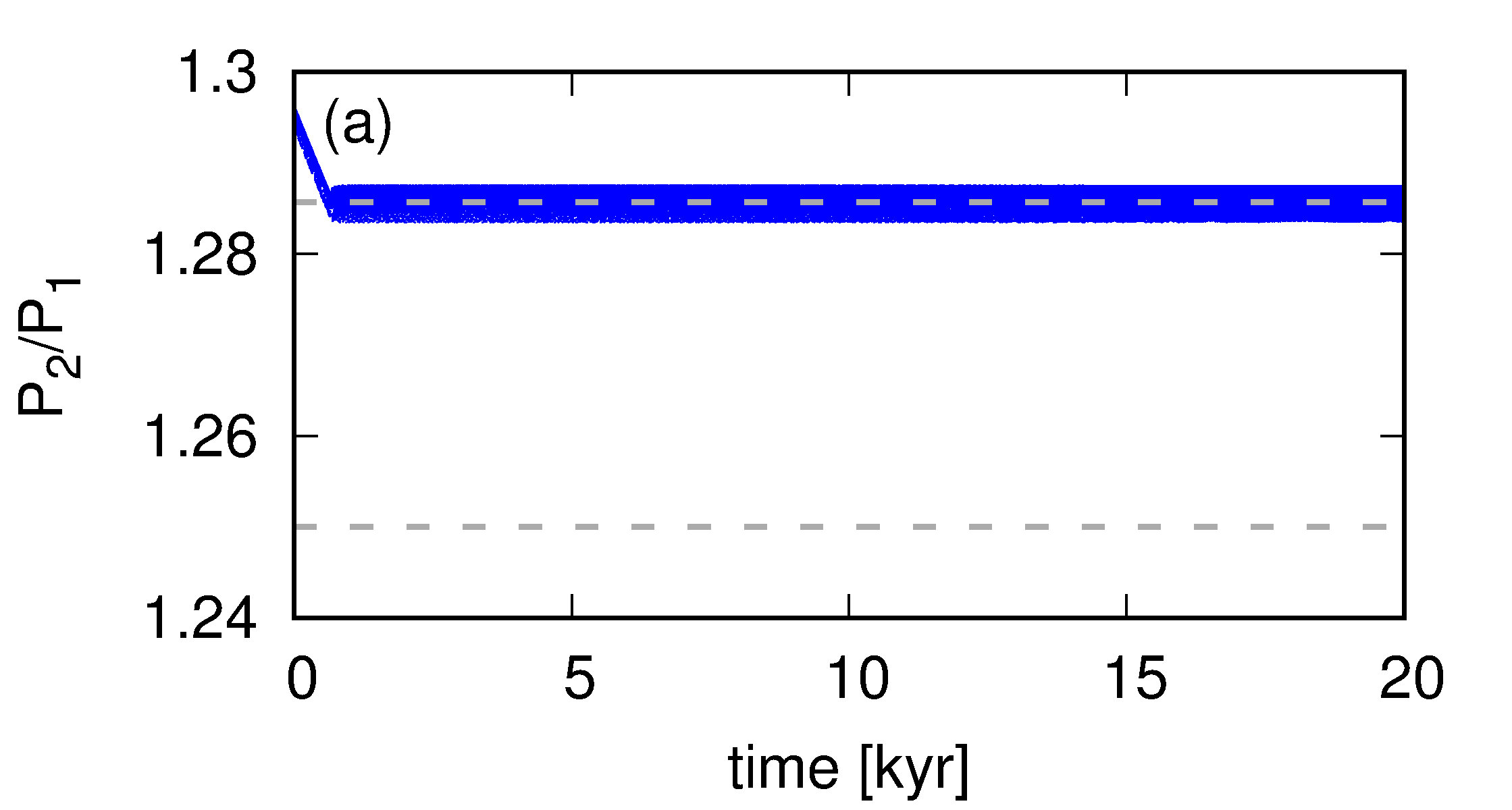}
\includegraphics[width=0.33\textwidth]{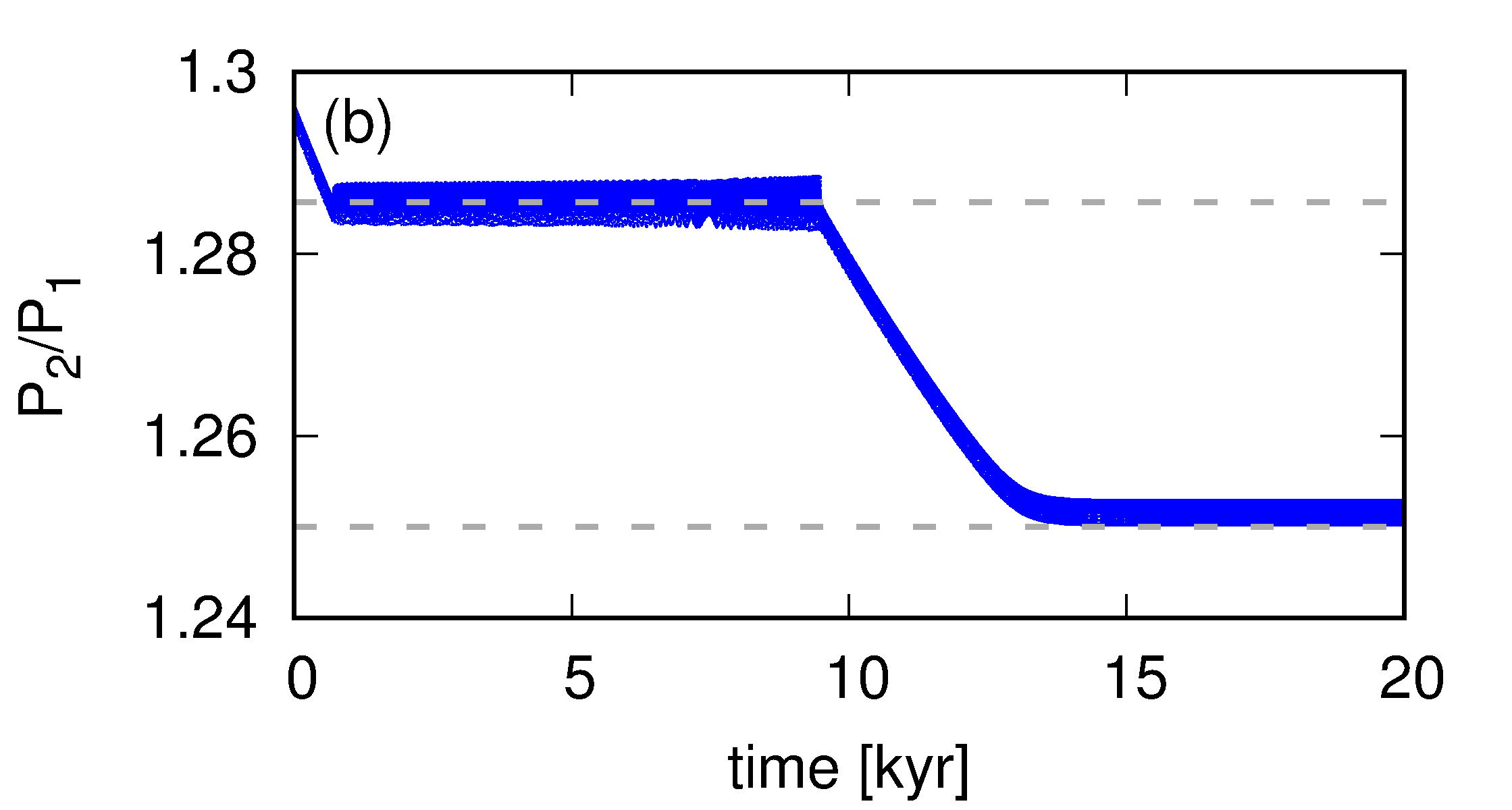}
\includegraphics[width=0.33\textwidth]{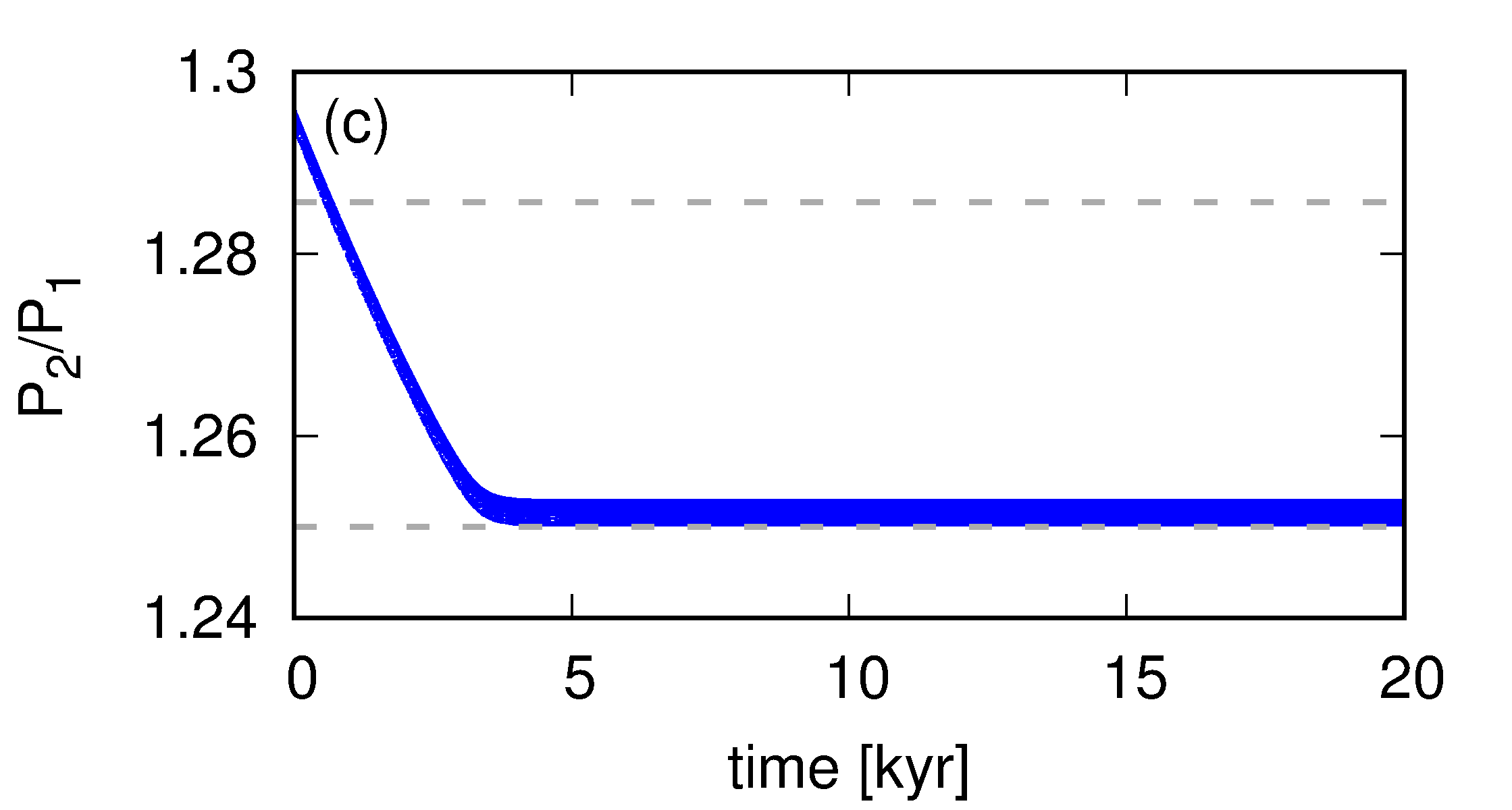}
}
\hbox{
\includegraphics[width=0.33\textwidth]{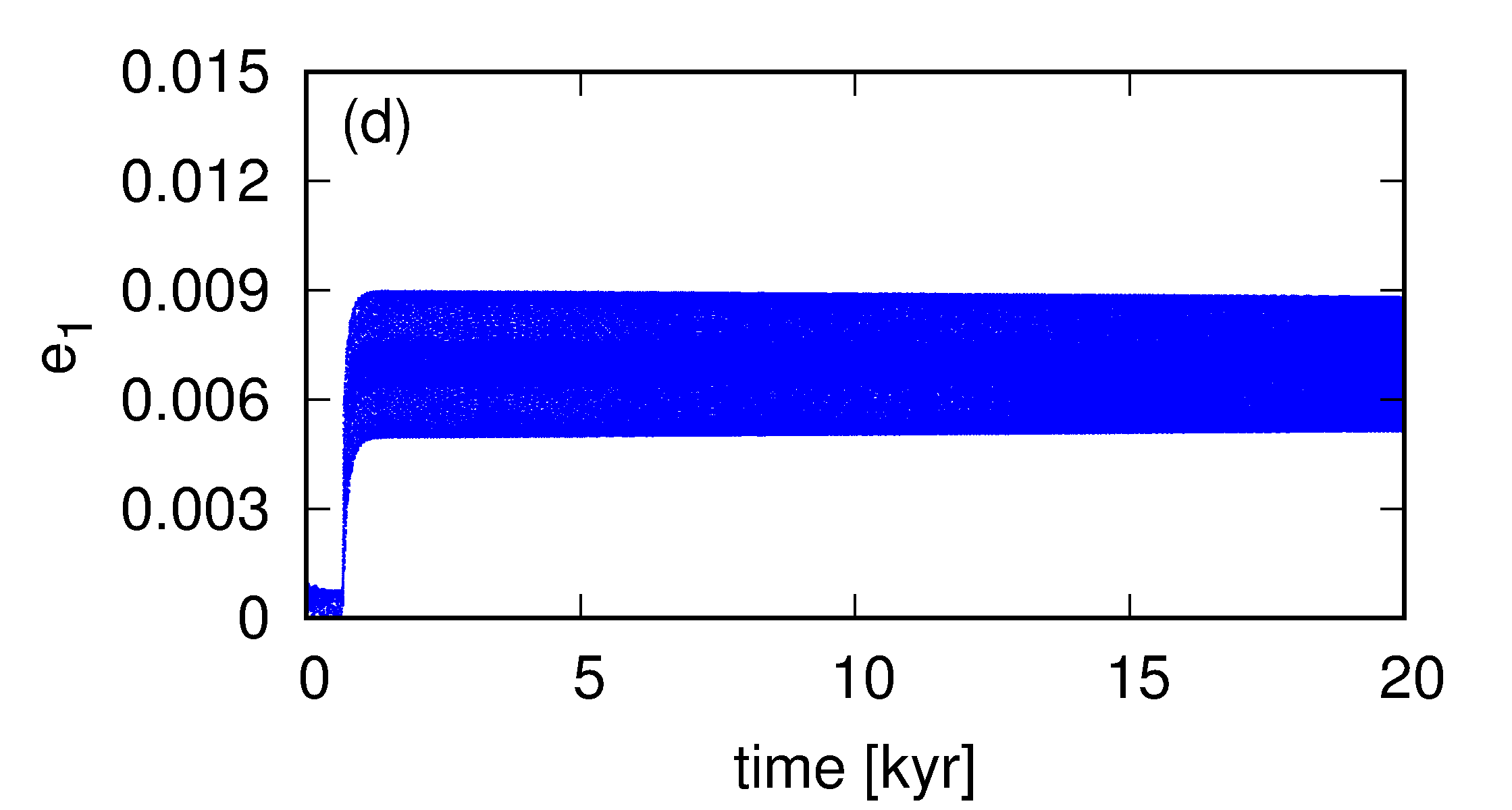}
\includegraphics[width=0.33\textwidth]{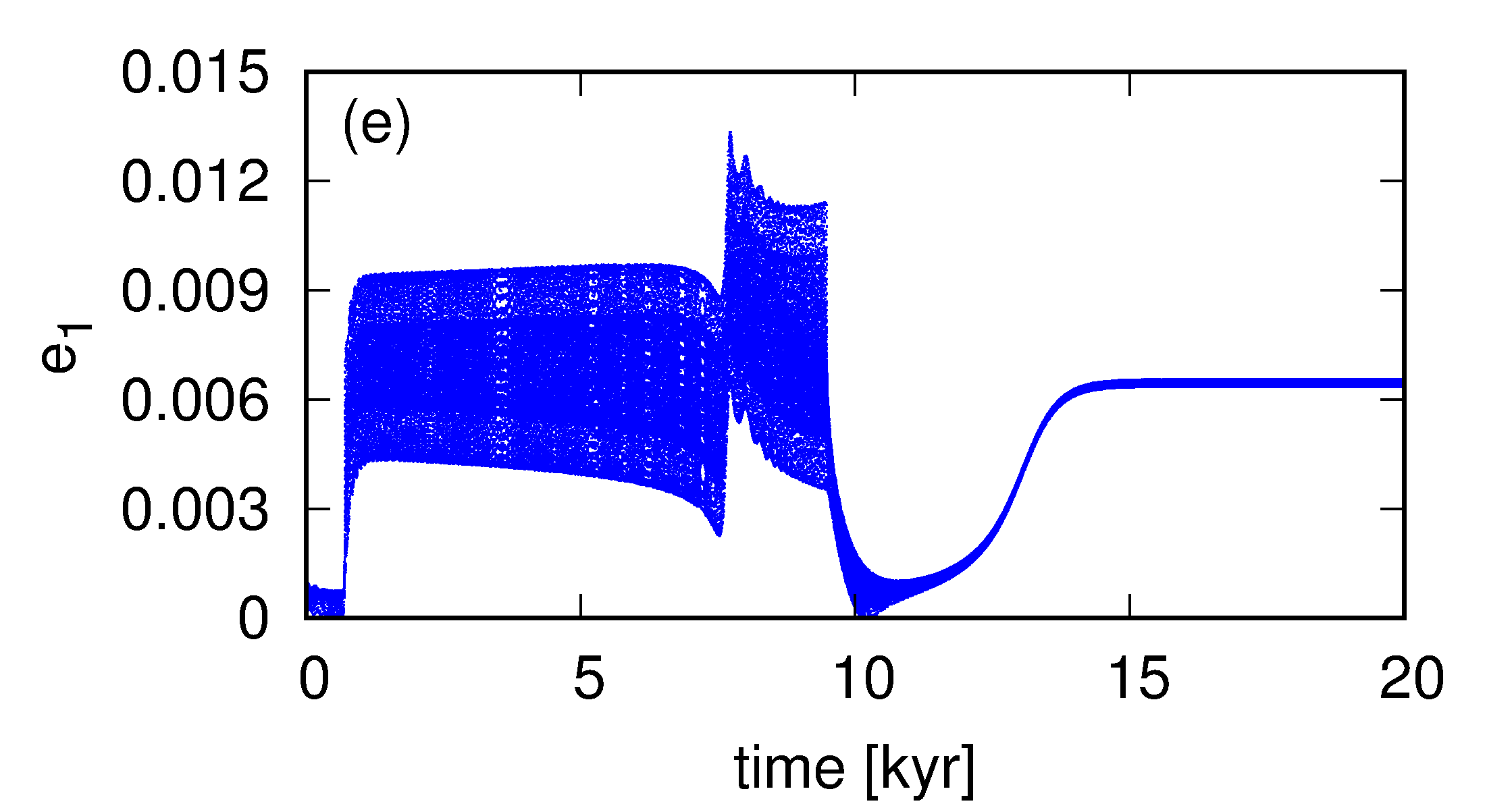}
\includegraphics[width=0.33\textwidth]{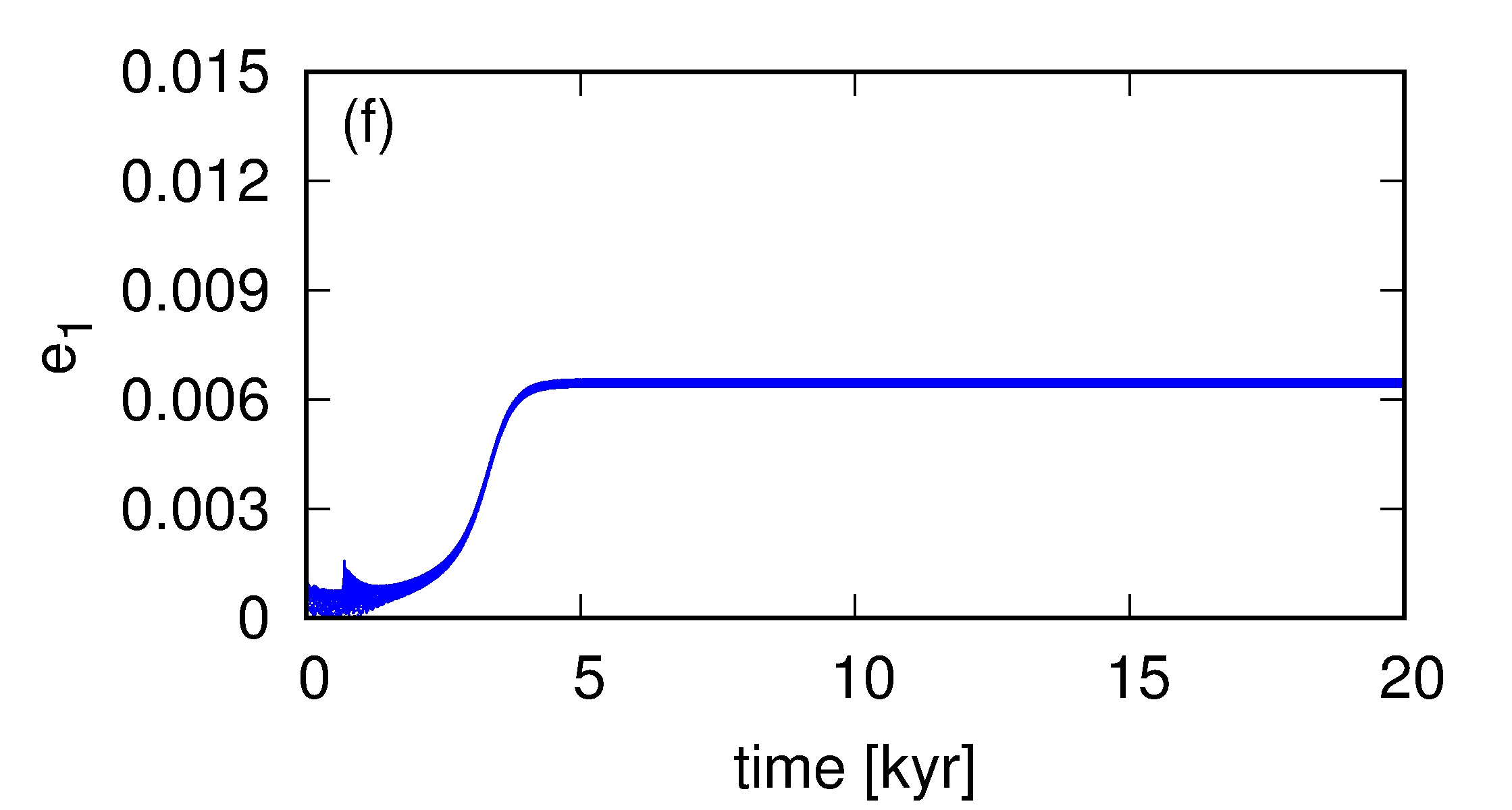}
}
}
}
\caption{Evolution of the period ratio (the top row) and the eccentricity $e_1$ (the bottom row) for three example migration simulations. Parameters of the migration for all three cases are: $\tau_0 = 1900\,\yr$, $\alpha = 1.2$, $\kappa = 120$. Initial orbital elements for the three cases differ only by $a_2 = 0.11874\,\au$ (the left-hand columns), $a_2 = 0.11875\,\au$ (the middle column), $a_2 = 0.11873\,\au$ (the right-hand column). Remaining initial Keplerian elements are $a_1 = 0.1\,\au$, $e_1 = e_2 = 0.0001$, $\varpi_1 = \varpi_2 = \Mmean_1 = \Mmean_2 = 0$. Masses of the planets $m_1 = m_2 = 6\,\mE$, the mass of the star $m_0 = 1\,\msun$. Grey dashed lines in the top row show the positions of 9:7 and 5:4~MMRs.}
\label{fig:migr_ex1}
\end{figure*}

Entering into a second-order resonance is much more complex than for the first-order MMRs. Depending on initial orbits as well as the migration and circularisation rates a migrating system can pass the resonance or stay in it, permanently or temporarily \citep{Quillen2006,Mustill2011}. We chose an example which illustrates the complexity of the migration into 9:7~MMR, see Fig.~\ref{fig:migr_ex1}. The parameters of the migration are given in the caption as are the planets' masses and initial orbits. Each column is for one of three simulations. The top row shows the evolutions of the period ratio, $P_2/P_1$ (the outer planet's period over the inner planet's period), while the bottom row illustrates the behaviour of the inner planet's eccentricity, $e_1$. The three simulations presented in Fig.~\ref{fig:migr_ex1} differ one from another only by the initial semi-major axes of the second planet, i.e., in the simulations shown from the left-hand to the right-hand panels, $a_2 = 0.11874, 0.11875, 0.11873\,\au$. We can see that initial differences are $\sim 10^{-5}\,\au$ for $a \sim 0.1\,\au$. Such a small difference, though, leads to very different evolution and final configurations.

Shortly after the beginning of the simulation the system shown in the left-hand column of Fig.~\ref{fig:migr_ex1} enters into 9:7~MMR and stays there, as we show farther, permanently. The second system (middle column) enters into 9:7~MMR, stays there for a few thousands of years, leaves 9:7~MMR and enters into 5:4~MMR, in which the system stays permanently. Moreover, we observe that between $t \sim 7500\,\yr$ and $t \sim 9500\,\yr$, the period ratio remains around 9/7, while the eccentricity behaviour changes. The third simulation (the right-hand column of Fig.~\ref{fig:migr_ex1}) shows that the system passes through 9:7~MMR, revealing only small excitation of the eccentricity and then goes into 5:4~MMR.

\begin{figure}
\centerline{
\vbox{
\hbox{\includegraphics[width=0.48\textwidth]{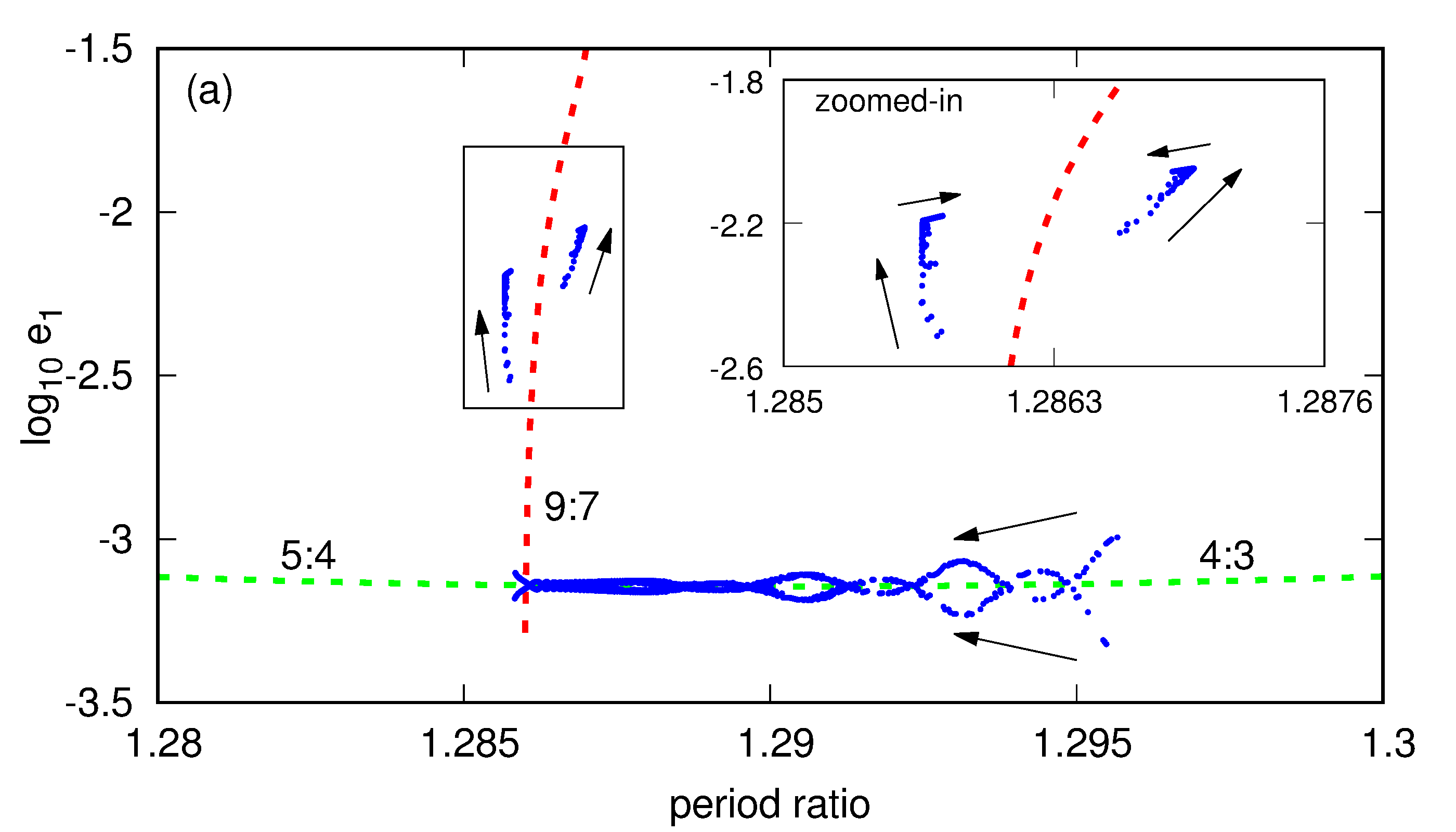}}
\hbox{\includegraphics[width=0.48\textwidth]{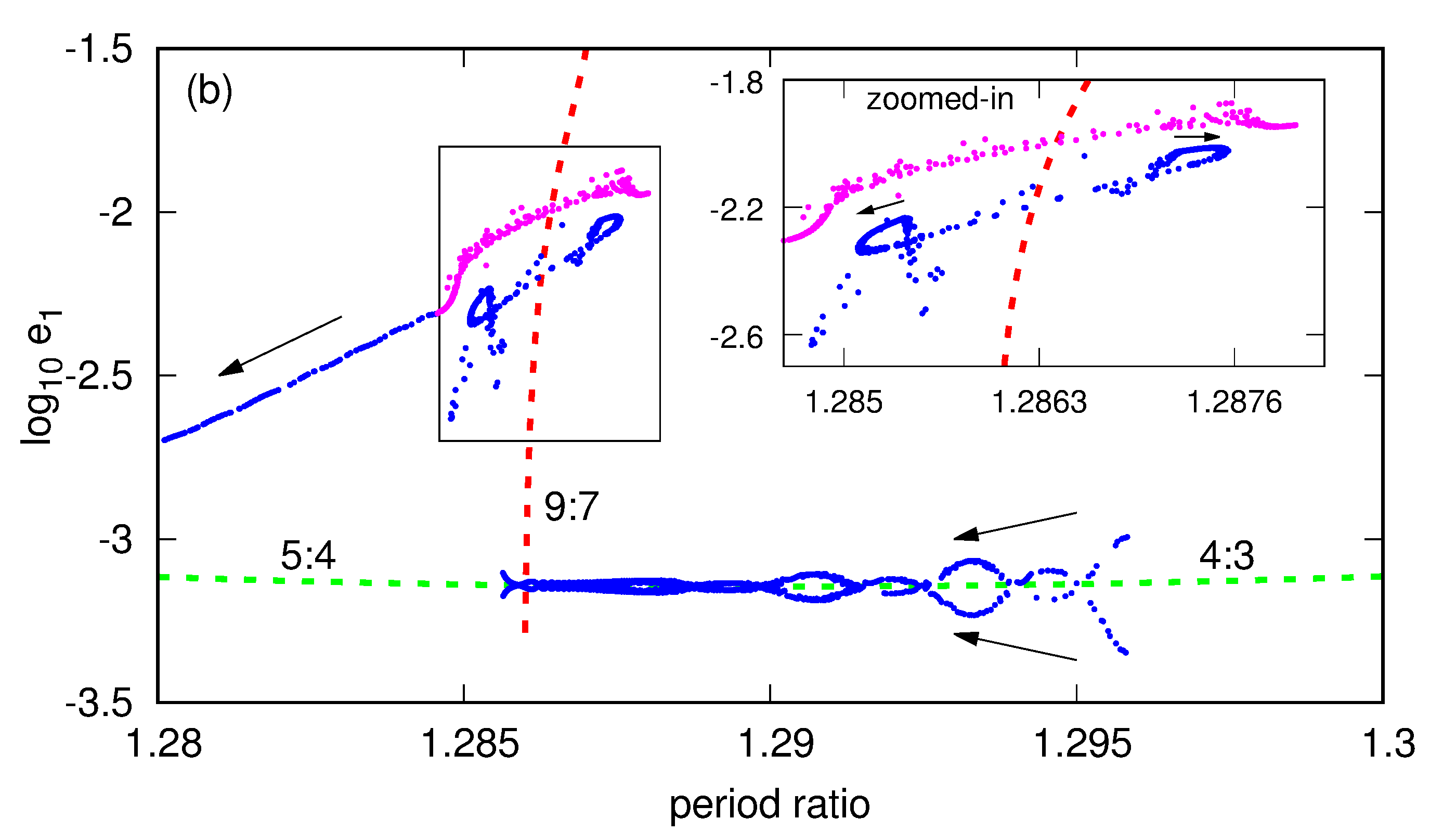}}
\hbox{\includegraphics[width=0.48\textwidth]{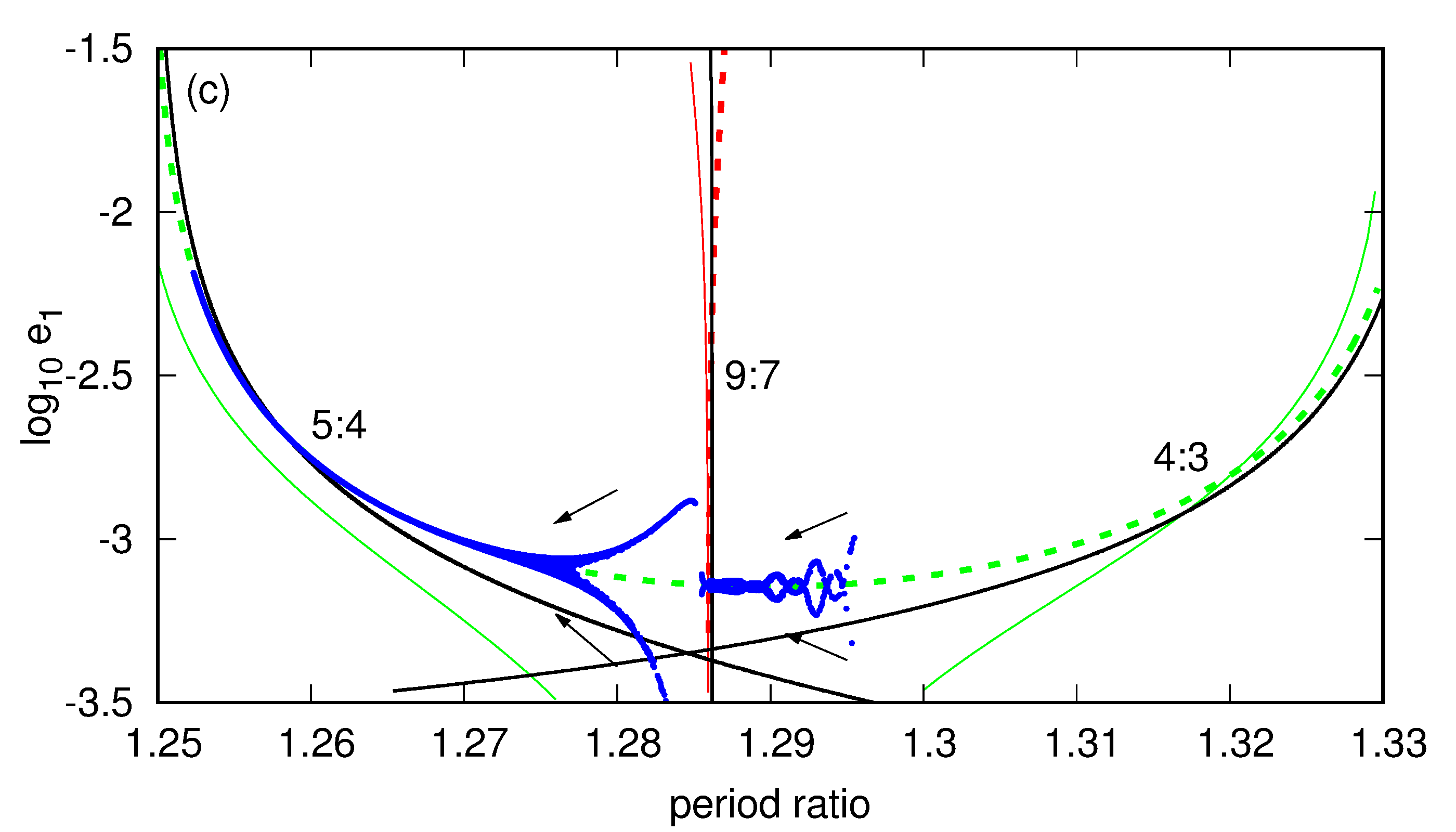}}
}
}
\caption{The same three simulations illustrated in Fig.~\ref{fig:migr_ex1}, where the positions of the systems are plotted at the diagrams only when $|\Mmean_1 - \pi|<\epsilon$ and $|\Mmean_2 - \pi|<\epsilon$ ($\epsilon = 0.5~$degree). The evolution of each system is over plotted at the periodic orbits {branches, shown with dashed curves} ($\Mmean_1 = \Mmean_2 = \pi$, see the text for details). Magenta colour of points in the zoomed fragment in panel~(b) corresponds to $t \in (7650, 9500)\,\yr$, see the text for discussion. In panel~(c) there are additional curves shown. The red solid curve is for the periodic configurations branch of 9:7~MMR (with the phases of $\Mmean_1 = \Mmean_2 = 0$), while the green solid curves are for 5:4~MMR (phases are $\Mmean_1 = 0, \Mmean_2 = \pi$) and for 4:3~MMR (with phases $\Mmean_1 = \pi, \Mmean_2 = 0$). Black solid curves are for the positions of equilibria of the averaged model.}
\label{fig:migr_ex2}
\end{figure}

As already mentioned in the Introduction, the periodic orbits play a crucial role in the migration induced formation of MMRs. This fact is demonstrated in Fig.~\ref{fig:migr_ex2} which shows branches of periodic orbits in the $(P_2/P_1, \log_{10}e_1)-$plane (green dashed curves are for the branches of 5:4 and 4:3~MMRs, while the red dashed curve is for 9:7~MMR; the branches are best visible in the bottom panel). Subsequent panels from~(a) to~(c) show the results for the first, second and third simulations illustrated in Fig.~\ref{fig:migr_ex1}. The periodic configurations are found with a help of the direct approach described in \citep{Migaszewski2017}. They were computed for initial values of the difference between the inner and the outer planets' longitudes of pericentres $\Delta\varpi \equiv \varpi_1 - \varpi_2 = \pi$ and the mean anomalies for both planets $\Mmean_1 = \Mmean_2 = \pi$.

The choice of the initial values of the angles is such that the resonant angles of 9:7~MMR as well as the two first order MMRs 5:4 and 4:3 equal appropriate values. The resonant angles of 5:4 for $P_2/P_1 \geq 5/4$ for which there exist periodic configurations are the following:
\begin{eqnarray}
\phi_{5:4}^{(1)} & = & 4 \lambda_1 - 5 \lambda_2 + \varpi_1 = 0,\nonumber \\
\phi_{5:4}^{(2)} & = & 4 \lambda_1 - 5 \lambda_2 + \varpi_2 = \pi,\nonumber
\end{eqnarray}
{where $\lambda_1, \lambda_2$ are the mean longitudes of the inner and the outer planets, respectively.} The values are reversed when the period ratio is smaller than the nominal value of 5:4~MMR.
This is the case for all first order MMRs, therefore the periodic configurations of 4:3 for $P_2/P_1 \leq 4/3$ occurs when
\begin{eqnarray}
\phi_{4:3}^{(1)} & = & 3 \lambda_1 - 4 \lambda_2 + \varpi_1 = \pi,\nonumber \\
\phi_{4:3}^{(2)} & = & 3 \lambda_1 - 4 \lambda_2 + \varpi_2 = 0.\nonumber
\end{eqnarray}
It is easy to verify that for $\Mmean_1 = \Mmean_2 = \pi$ and the anti-aligned orbits the resonant angles of 5:4 and 4:3~MMR equal the values given above. The critical values of the resonant angles of 9:7~MMR read:
\begin{eqnarray}
\phi_{9:7}^{(1)} & \equiv \phi_1 = & 7 \lambda_1 - 9 \lambda_2 + 2 \varpi_1 = \pi,\nonumber \\
\phi_{9:7}^{(2)} & \equiv \phi_2 = & 7 \lambda_1 - 9 \lambda_2 + 2 \varpi_2 = \pi,\nonumber \\
\phi_{9:7}^{(3)} & \equiv \phi_3 = & 7 \lambda_1 - 9 \lambda_2 + \varpi_1 + \varpi = 0,\nonumber \\
\end{eqnarray}
which occurs for $\Mmean_1 = \Mmean_2 = \pi$, but also for $\Mmean_1 = \Mmean_2 = 0$. In the presented illustration we chose the first case, as it covers all the resonances considered.

The evolution of the first example is shown in blue colour in the top panel of Fig.~\ref{fig:migr_ex2}. The points are plotted only if both mean anomalies differ from the critical values of $\pi$ by less than half a degree. We do not check the value of $\Delta\varpi$ as shortly after the simulation starts it is already close to $\pi$. The direction of the evolution at the $(P_2/P_1, \log_{10} e_1)$-diagram is shown with black arrows. Initially the system evolves along the 4:3~MMR periodic branch. The amplitude of the oscillation around the periodic configuration decreases with time. When the system reaches the branch of 9:7~MMR, the amplitude is already very small. The system disappears from the diagram for some period of time, as there is no configurations with both $\Mmean_1, 
\Mmean_2$ close to $\pi$. Next, the system appears above the 4:3/5:4 branch (the green curve), close to the 9:7 branch (the red curve) and it moves up. The zoomed plot of this fragment shows that the system tends towards the red curve, however very slowly. This system, as illustrated in the left-hand column of Fig.~\ref{fig:migr_ex1}, will stay in 9:7~MMR permanently.

The second example is illustrated in Fig.~\ref{fig:migr_ex2}b. Initially the system evolves in the same way as example~1 (which is expected, as the relative difference in $a_2$ is $\sim 10^{-4}$). Similarly to example~1, after reaching $P_2/P_1 \approx 9/7$, the system goes above the green curve and evolves around the red curve. The zoomed fragment shows, though, that unlike example~1, the system tends away from the branch of periodic orbits and leaves the resonance after certain period of time. The magenta colour marks the location of the system for $t \in [7650, 9500]\,\yr$. That corresponds to the change of the behaviour of $e_1$ shown in Fig.~\ref{fig:migr_ex1}e. For $t \gtrsim 9500\,\yr$ the system tends towards 5:4~MMR. 

The third example, which is illustrated in Fig.~\ref{fig:migr_ex2}c, evolves initially similarly to the first two systems. The difference is that the third system passes through 9:7~MMR and goes directly to 5:4~MMR. The only visible result of the passage is that the amplitude of the oscillations of the system around the 4:3/5:4~MMR branch of periodic orbits is increased. The amplitude, though, is damped quickly after the 9:7~MMR encounter and next the system evolves along the 5:4~branch with almost zero amplitude. Its final position at the branch is dictated by the values of $\alpha$ and $\kappa$. For a given $\alpha$, the final eccentricity is higher when $\kappa$ is smaller (i.e., when the eccentricity damping is less efficient). Moreover, as we will show further in this work, $\kappa$ also governs the damping of the oscillations around the periodic orbits. High $\kappa$ means fast damping, so the final configuration in a first order MMR is exactly or almost exactly periodic. The damping of the oscillations around the second order MMR is more complex, which will be the subject to study in the next sections.

\begin{figure}
\centerline{
\vbox{
\hbox{\includegraphics[width=0.4\textwidth]{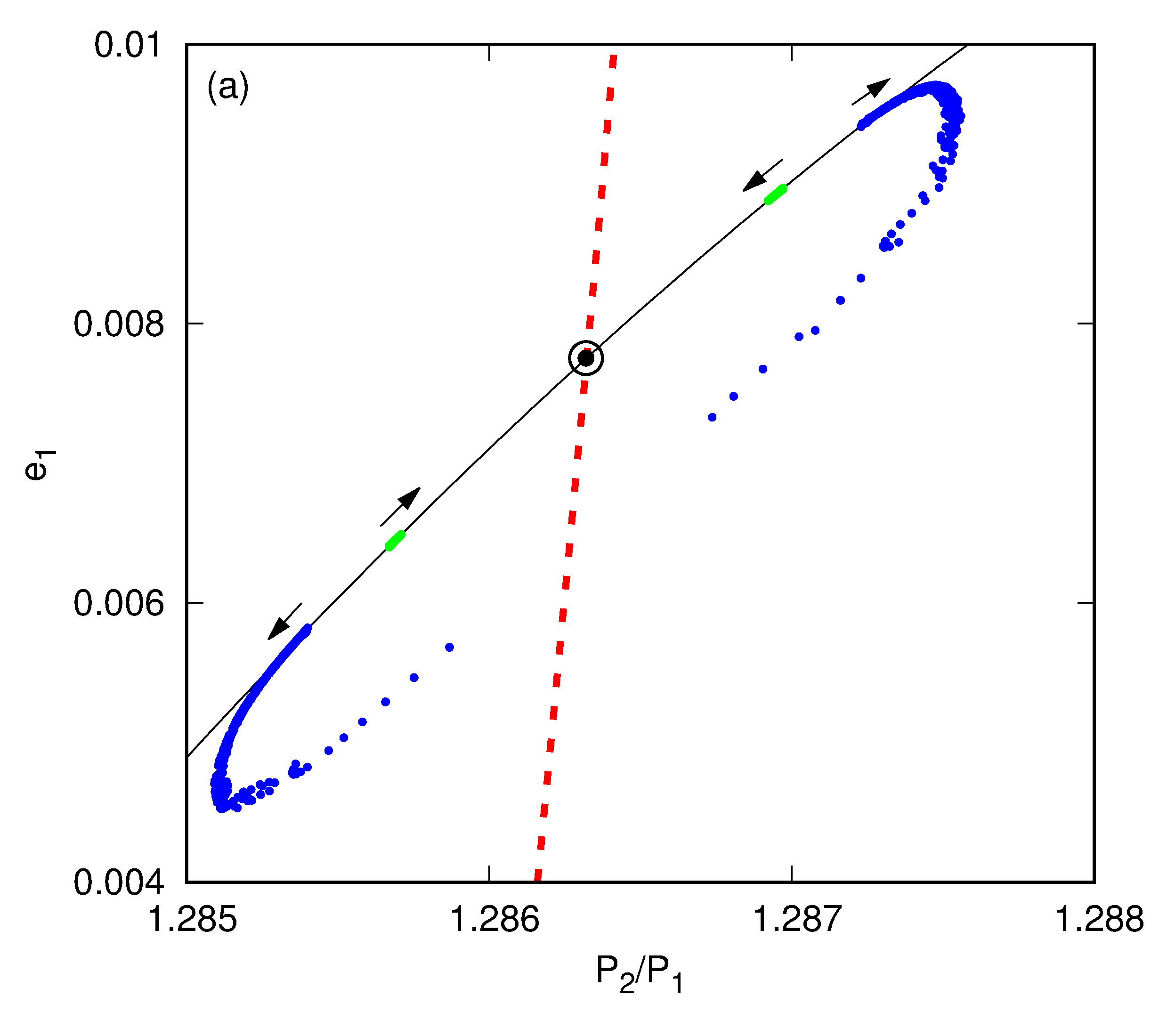}}
\hbox{\includegraphics[width=0.4\textwidth]{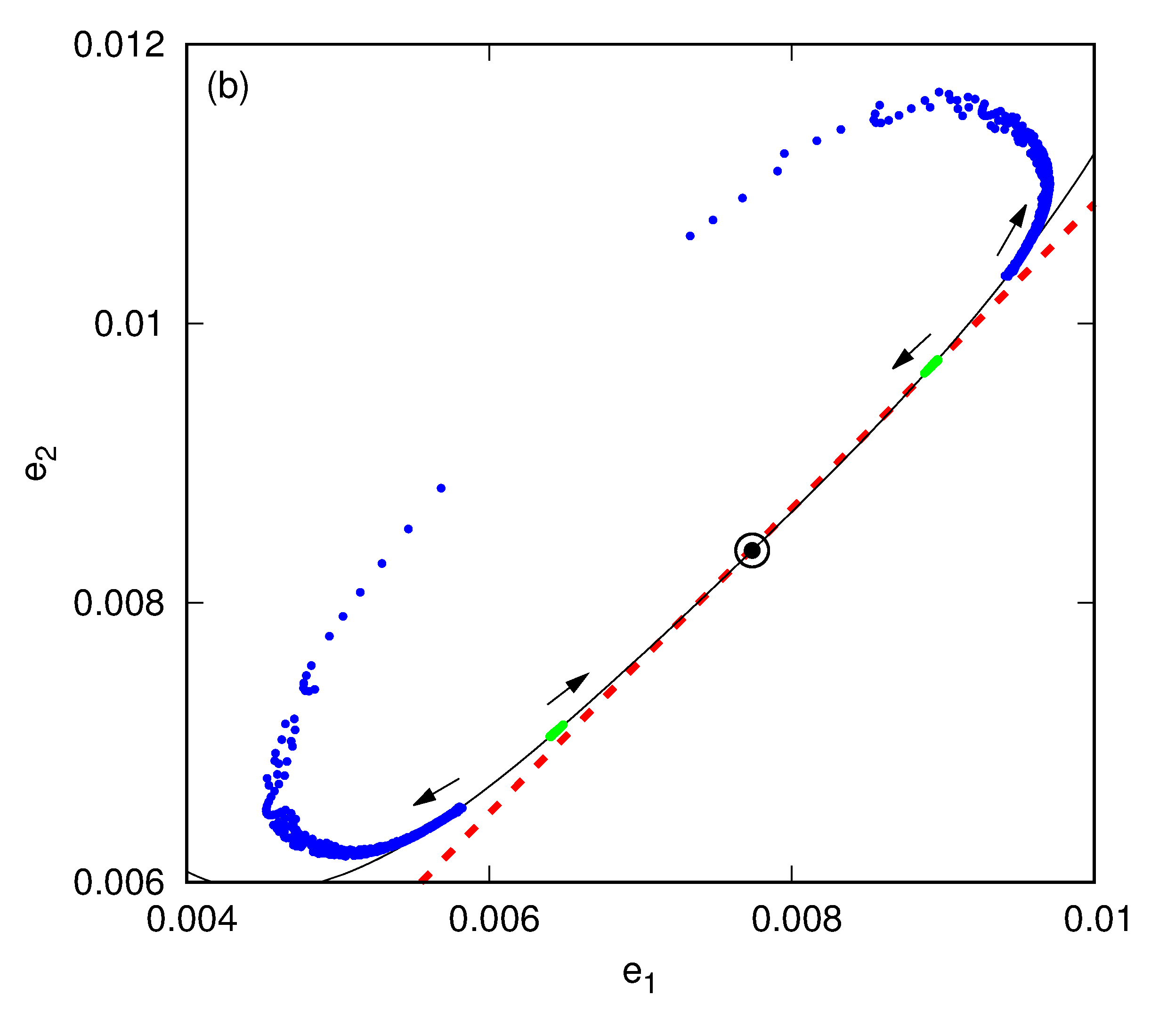}}
}
}
\caption{Two simulations illustrated in Fig.~\ref{fig:migr_ex2}a and~b presented at $(P_2/P_1, e_1)-$ and $(e_1, e_2)-$planes (panels~a and~b, respectively) for $t \in (1500, 7500)\,\yr$. See the text for discussion.}
\label{fig:migr_ex3}
\end{figure}

Figure~\ref{fig:migr_ex3} illustrates the difference between the evolution of the first and the second example systems. It shows the location of the systems at two diagrams, $(P_2/P_1, e_1)$ and $(e_1, e_2)$ for $t \in (1500, 7500)\,\yr$, which is the period of time counted from just after the entrance of the systems into 9:7~MMR up to the time corresponding to the transition between two regimes of motion in the evolution of example~2. The evolution of example~1 is shown in green colour, while the second example is plotted in blue. A red dashed curve shows the branch of 9:7~MMR periodic orbits. The observation we have already made is that the first system evolves towards the periodic configuration, while the second one moves away from it (see the arrows for the evolution direction). Locally both tracks can be approximated by polynomial curves (shown with black solid curves). For panel~(a) we used a quadratic curve, while for panel~(b) the curve is of the forth order. A black circle in each panel indicates the point towards which the system of example~1 tends. At the $(P_2/P_1, e_1)-$diagram that point corresponds to the intersection of the black and the red curves, while at the $(e_1, e_2)-$diagram the curves are in this point tangent one to another.

Apparently, both systems evolve along a single curve (in a vicinity of the periodic orbit). One can also notice that the initial location of the system at the black curve determines its motion along it, i.e., whether it occurs towards to or away from the black circle. It is thus straightforward that there exists such a point at the evolution track for which the system does not move {either way}. Such a point would indicate the critical distance from the periodic orbit below which the system stays in the resonance permanently. We will show further that the black curve corresponds to the branch of periodic orbits of the averaged system. In order to do so, we introduce in the next section the averaged model of motion.

\section{The averaged system}

The resonant dynamics of a two-planet system can be approximated by the averaged Hamiltonian $\overline{H}$ of two degrees of freedom with two first integrals of motion, i.e., the total angular momentum $C \equiv G_1 + G_2$ and the spacing parameter $K \equiv (p+q) L_1 + p L_2$ \citep{Michtchenko2001,Beauge2003}. The indices $p$ and $q$ indicates the (p+q):p resonance, and for 9:7~MMR $p=7, q=2$. $G_i$ and $L_i$ are the actions from the Delanuay angle-action canonical variables, i.e., $L_i \equiv \beta_i \sqrt{\mu_i a_i}$ and $G_i \equiv L_i \sqrt{1 - e_i^2}$, where $\beta_i \equiv (1/m_0 + 1/m_i)^{-1}$ are the reduced masses, $\mu_i \equiv k^2 (m_0 + m_i)$
and $k$ is the Gauss' gravitational constant. After the canonical transformation of the Delanuay variables, one can find the actions $I_i = L_i - G_i$ and the angles $\sigma_i = (1+s) \lambda_2 - s \lambda_1 - \varpi_i$, where $s \equiv p/q$ and $\lambda_i$ is the mean longitude of planet~$i$. Therefore, the averaged Hamiltonian $\overline{H} = \overline{H} (I_1, I_2, \sigma_1, \sigma_2; C, K)$, see \cite{Michtchenko2001,Beauge2003} for details.

The averaging of the Hamiltonian can be done numerically or analytically. For sake of self-consistency of the material, we repeat here a few informations about the averaging from \citep{Migaszewski2017}. The averaged Hamiltonian is given by a generic formula of the form of
\begin{equation}
\label{eq:Hamiltonian}
\overline{H} = -\frac{\mu_1 \beta_1}{2 a_1} -\frac{\mu_2 \beta_2}{2 a_2} - \frac{k^2 m_1 m_2}{a_2} \overline{R},
\end{equation}
where the averaged disturbing function is given by the integral
\begin{equation}
\label{eq:disturbing_function}
\overline{R} = \frac{1}{2\pi} \int_0^{2\pi} \frac{a_2}{\| \vec{r}_1 - \vec{r}_2 \|} dQ, \quad Q \equiv \frac{\lambda_1 - \lambda_2}{q}.
\end{equation}
In this work we use the analytical approximation of the integral above, which is given by the sum of the secular and resonant terms, i.e., $\overline{R} = \overline{R}_{\idm{sec}} + \overline{R}_{\idm{res}}$, where $\overline{R}_{\idm{sec}}$ and $\overline{R}_{\idm{res}}$ are given in terms of power series in eccentricities and inclinations by a recipe from \citep{Murray1999}. Explicit form of the expansion for 9:7~MMR up to the forth order in the eccentricities and with the inclinations assumed to be $0$ is given in the appendix of \citep{Migaszewski2017}.

\subsection{The equations of motion}

The evolution of the Hamiltonian system (without the migration terms that will be given below) is governed by the following set of canonical equations of motion
\begin{equation}
\dot{\sigma}_i = \frac{\partial \overline{H}}{\partial I_i}, \quad
\dot{I}_i = -\frac{\partial \overline{H}}{\partial \sigma_i}, \quad i = 1,2.
\end{equation}
One can also introduce the non-singular equations \citep{Michtchenko2001}, that are more suitable for small eccentricities when compared to the $(I_i, \sigma_i)$-set. The new variables are defined as follows:
\begin{equation}
x_i \equiv \sqrt{2 I_i} \cos{\sigma_i}, \quad y_i \equiv \sqrt{2 I_i} \sin{\sigma_i}
\end{equation}
and the canonical equations of motion read
\begin{equation}
\dot{y}_i = \frac{\partial \overline{H}}{\partial x_i}, \quad
\dot{x}_i = -\frac{\partial \overline{H}}{\partial y_i}, \quad i=1,2.
\end{equation}
The partial derivatives in the canonical equations are computed for fixed values of the integrals $C$ and $K$.

The migration can be incorporated into the equations of motion by adding extra terms to the $\dot{x}_i$ and $\dot{y}_i$ formulae as well as by adding two equations governing the time variation of $C$ and $K$. First we average out the evolution of the semi-major axes and eccentricities for each planet if the motion is governed by the N-body equations with addition of the force which mimics the migration and circularisation, Eq.~\ref{eq:migration_force}.

If $\tau_a$ is given by Eq.~\ref{eq:tau_a} and $\tau_e = \tau_a/\kappa$, one can find approximate formulae for the averaged $\dot{a}_i$ and $\dot{e}_i$:
\begin{eqnarray}
\label{eq:migration_terms}
\dot{a}_i^{\idm{(migr)}} &\approx& -\frac{a_i}{\tau_a(a_i)} \left( 1 + \frac{5}{8} \left[ 1 - \frac{4}{5} \alpha \right] \kappa_i e_i^2 - \frac{3}{4} \left[ 1 - \frac{1}{3} \alpha \right] \alpha e_i^2 \right),\nonumber\\
\label{eq:adot_edot}
\dot{e}_i^{\idm{(migr)}} &\approx& -\frac{e_i}{\tau_e(a_i)} \left( 1 - \frac{\alpha}{2 \kappa_i} - \frac{13}{32} \left[ 1 - \frac{3}{13} \alpha - \frac{8}{13} \alpha^2 \right] e_i^2 \right.\\
&&+ \left. \frac{3}{8} \left[ 1 + \frac{1}{2} \alpha - \frac{1}{6} \alpha^2 \right] \frac{\alpha e_i^2}{\kappa_i} \right).\nonumber
\end{eqnarray}
The $\alpha$-dependence of the above equations stems from the fact that $\tau_a$ depends on $r$. If $\tau_a$ depended on $a$ instead of $r$, one would have to put $\alpha=0$ in Eq.~\ref{eq:adot_edot}. We will discuss the differences between the models with $\tau_a = \tau_a(r)$ and $\tau_a = \tau_a(a)$ in the last part of this work. In most of the study we use the $r$-dependent model of $\tau_a$.

The canonical variables $x_1, y_1, x_2, y_2$ as well as the integrals of the Hamiltonian system $C$ and $K$ (that are not constant when formulae Eq.~\ref{eq:adot_edot} are taken into account) depend functionally on $a_1, e_1, \sigma_1, a_2, e_2, \sigma_2$. It is then possible to find the following equations of motion for the system of two migrating planets in a vicinity of 9:7~MMR:
\begin{eqnarray}
\label{eq:equations_with_migration}
\dot{y}_i &=& \frac{\partial \overline{H}}{\partial x_i} + \sum_{j=1}^2 \left( \frac{\partial y_i}{\partial a_j} \dot{a}_j^{\idm{(migr)}} + \frac{\partial y_i}{\partial e_j} \dot{e}_j^{\idm{(migr)}}\right), \quad i=1,2, \nonumber\\
\dot{x}_i &=& -\frac{\partial \overline{H}}{\partial y_i} + \sum_{j=1}^2 \left( \frac{\partial x_i}{\partial a_j} \dot{a}_j^{\idm{(migr)}} + \frac{\partial x_i}{\partial e_j} \dot{e}_j^{\idm{(migr)}}\right), \quad i=1,2, \nonumber\\
\dot{C} &=& \sum_{j=1}^2 \left( \frac{\partial C}{\partial a_j} \dot{a}_j^{\idm{(migr)}} + \frac{\partial C}{\partial e_j} \dot{e}_j^{\idm{(migr)}}\right), \nonumber\\
\dot{K} &=& \sum_{j=1}^2 \left( \frac{\partial K}{\partial a_j} \dot{a}_j^{\idm{(migr)}} + \frac{\partial K}{\partial e_j} \dot{e}_j^{\idm{(migr)}}\right).
\end{eqnarray}
We note that the force which mimics the migration and circularisation, Eq.~\ref{eq:migration_force}, does not lead to any change in $\sigma_1$ and $\sigma_2$, thus $\dot{\sigma}_i^{\idm{(migr)}} = 0$.

\begin{figure*}
\centerline{
\vbox{
\hbox{
\includegraphics[width=0.45\textwidth]{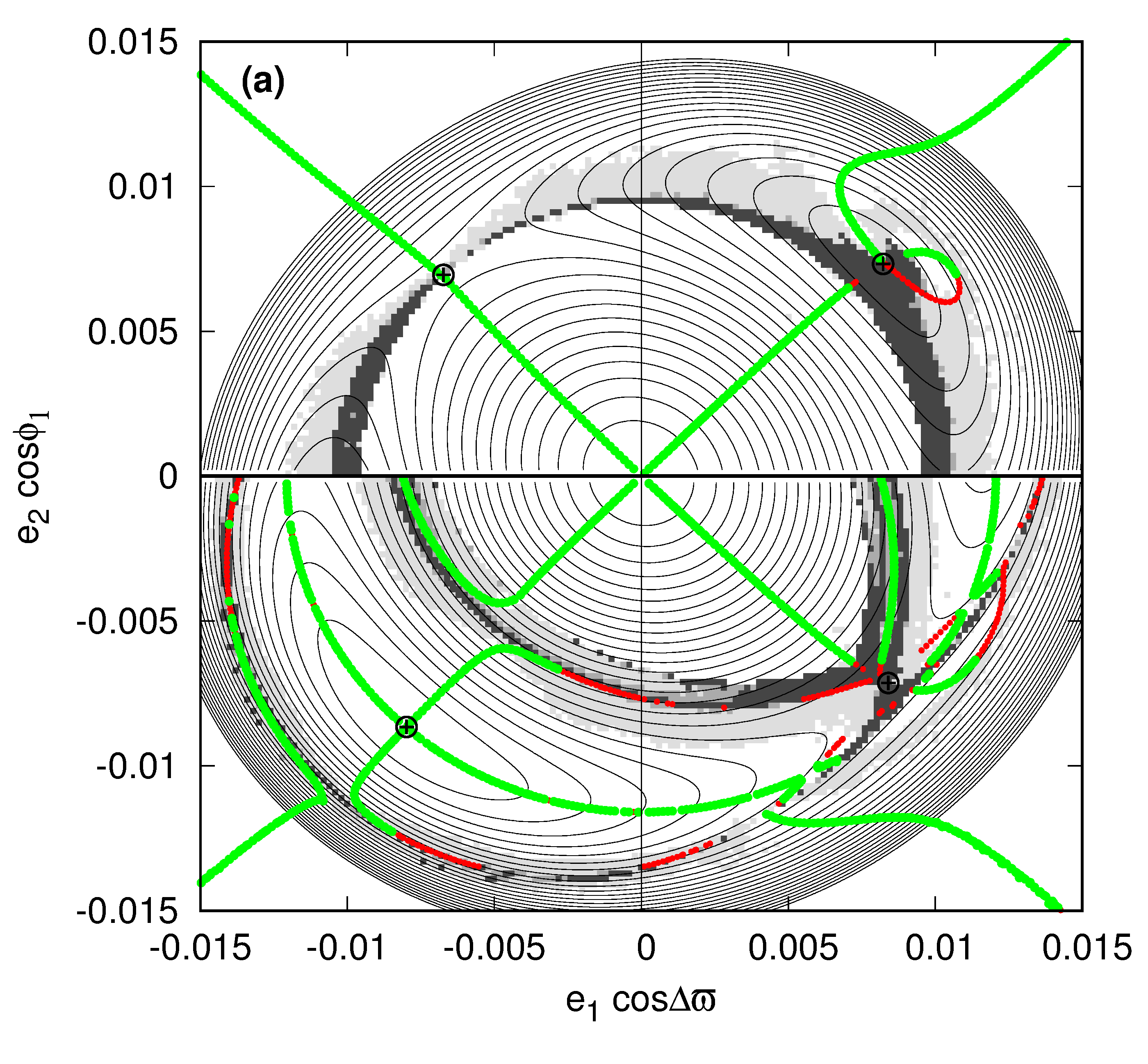}
\includegraphics[width=0.45\textwidth]{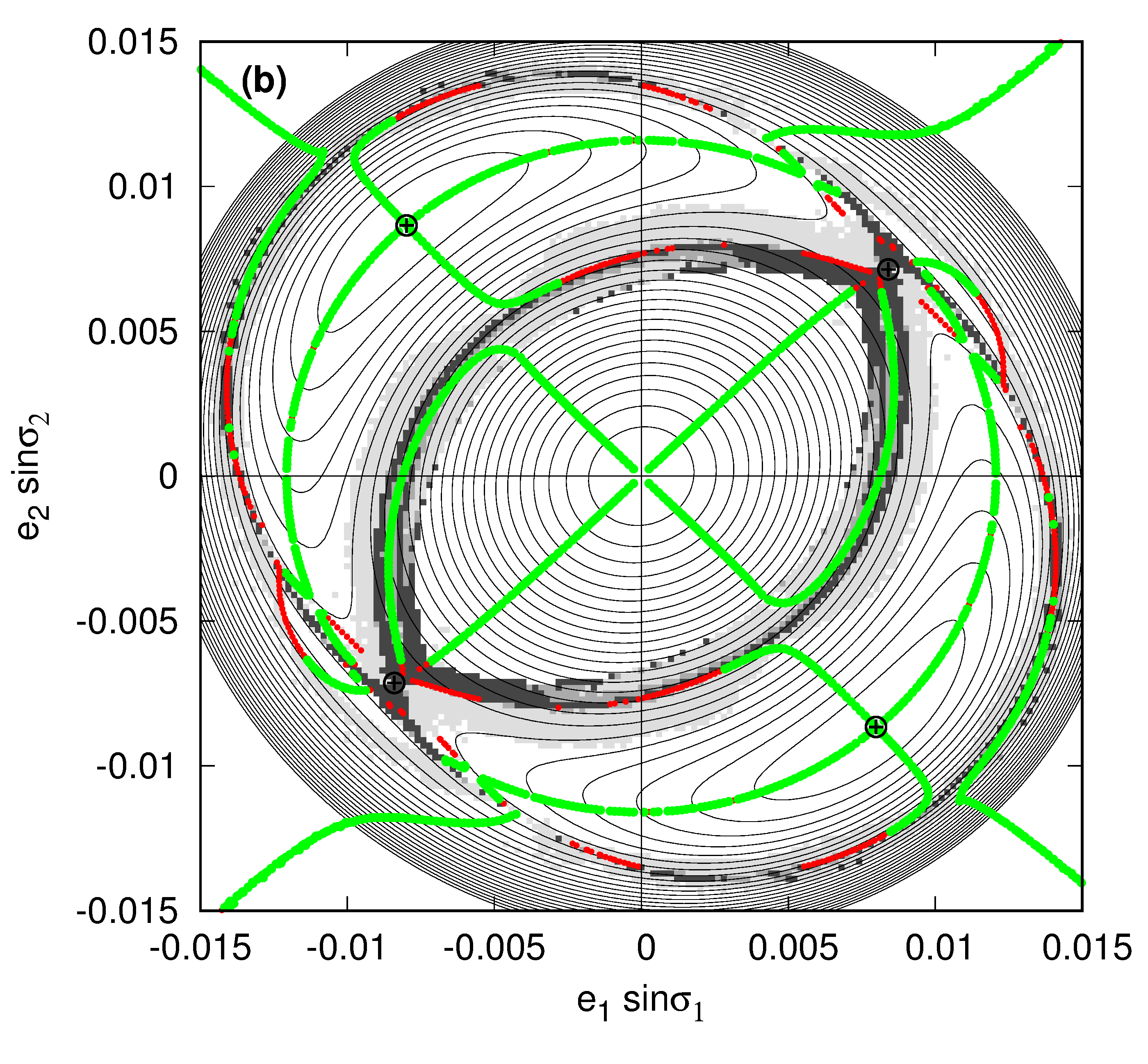}
}
\hbox{
\includegraphics[width=0.45\textwidth]{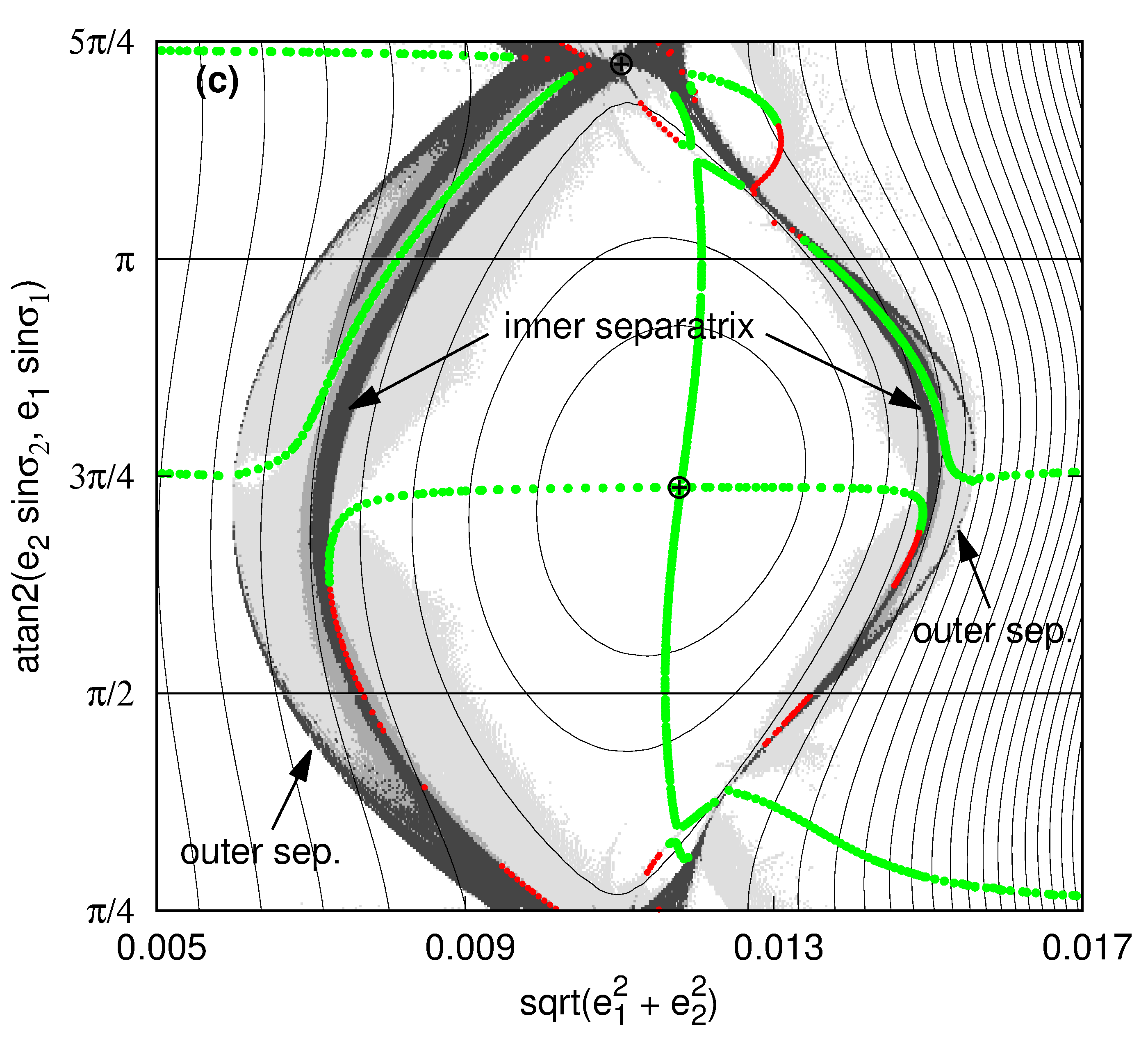}
\includegraphics[width=0.45\textwidth]{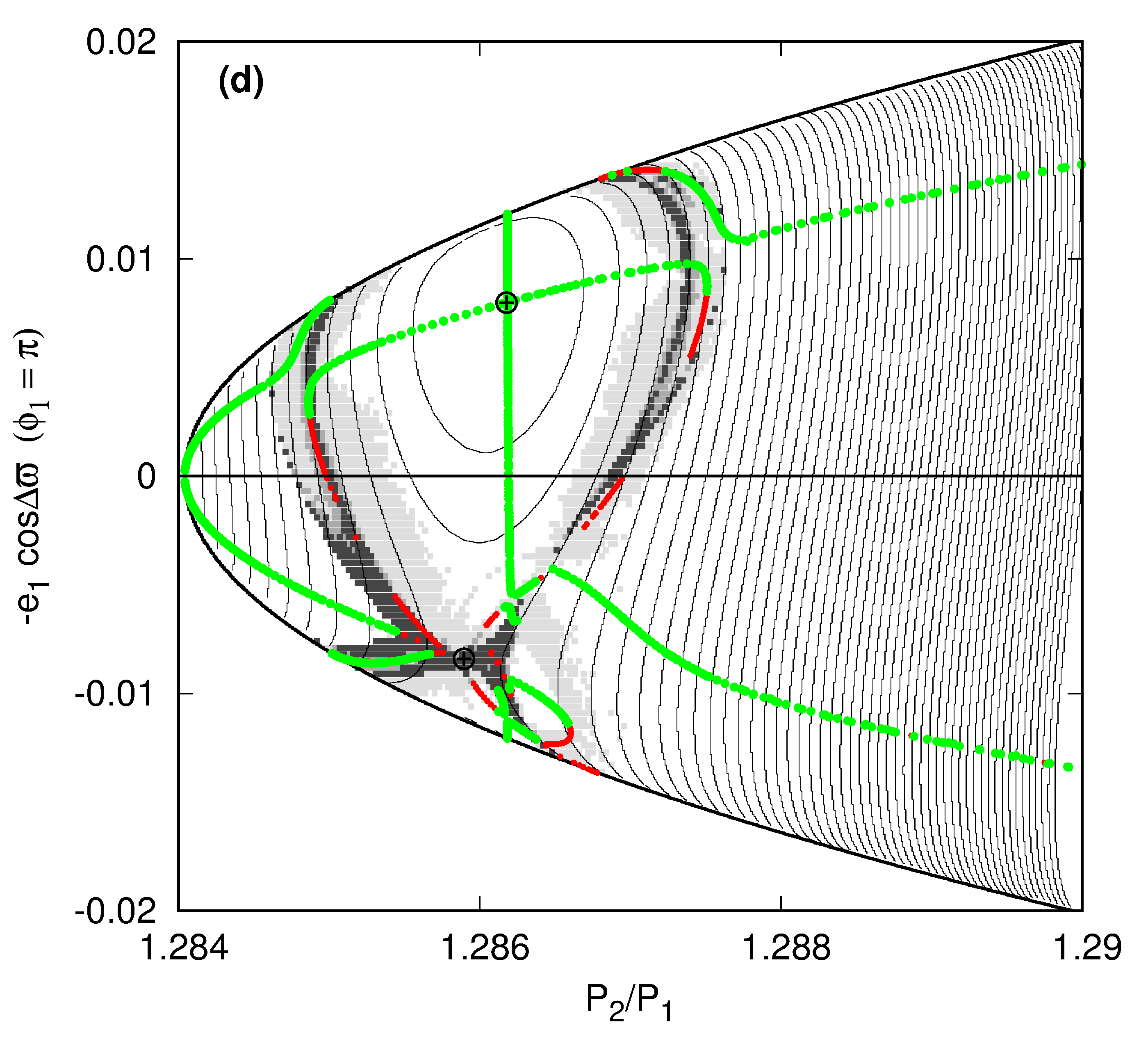}
}
}
}
\caption{Energy plots presented at four diagrams that reveal dynamical features discussed in this paper. At each panel black solid curves denote curves of constant values of the averaged energy. Cross/circle black points denote the equilibria of the averaged system. Green and red curves illustrate branches of periodic orbits of the averaged system, where green colour means stable orbits and red -- unstable. Light/medium/dark grey points indicate chaotic evolution in terms of the spectral number. The darker the shade is, the more chaotic is the evolution of the system. Subsequent panels from~(a) to~(d) show the energy plots at the $\Sigma$-plane, $\mathcal{S}$-plane, polar representation of the $\mathcal{S}$-plane and $(P_2/P_1, -e_1 \cos\Delta\varpi)$-plane. See the text for details.}
\label{fig:energy_plots}
\end{figure*}

\section{Dynamics of the conservative system}

We start the study from an overview of the dynamics of the two-planet system without the migration. In this section we study the dynamics at energy plots. We also introduce the concept of periodic orbits of the averaged system as well as its connection with chaos.

\subsection{The representative plane of initial conditions}

In the further part of this work we will find it very useful a concept of the so called representative plane of initial conditions, $\Sigma-$plane, \citep[e.g.,][]{Beauge2003,Migaszewski2017}, which can be used to construct the energy plots, that reveal the equilibria. The representative plane is a plane of the eccentricities, each point of which, $(e_1, e_2)$, corresponds to values of $a_1, a_2$ that can be computed for given values of $C$ and $K$, or, what will be more useful when the migration is considered, for a given value of the scale-free angular momentum $c \equiv C/K$. Two remaining variables $\sigma_1$ and $\sigma_2$ are chosen such that both derivatives $\partial \overline{H}/\partial \sigma_i$ ($i=1,2$) vanishes for every point $(e_1, e_2)$ of the plane. It can be shown that it occurs for one of four combinations of the angles, i.e., $(\sigma_1, \sigma_2) \in \{ (0, 0), (0, \pi), (\pm \pi/2, \pm \pi/2), (\pm \pi/2, \mp \pi/2) \} $. Note that a simultaneous change of the signs of both angles in the two last combinations does not lead to any change in the Hamiltonian. As $\varpi = \sigma_2 - \sigma_1$ and $\phi_1 = -2\sigma_1$, the representative values of $(\Delta\varpi, \phi_1)$ are $\{(0,0), (0,\pi), (\pi,0), (\pi,\pi)\}$.

A consequence of the definition given above is that a point in the plane for which $\partial \overline{H}/\partial I_i = 0$ ($i=1,2$) is an equilibrium of the averaged Hamiltonian system (which corresponds to a periodic configuration within the N-body model of motion). Those values correspond to $\Delta\varpi = 0$ or $\pi$, therefore the $\Sigma-$plane
is called the symmetric representative plane. Equilibria of the averaged system, though, can also exist for $\sigma_i$ different from the representative values given above \citep[e.g.,][for 2:1~MMR in moderate-to-high eccentricities regime]{Beauge2003}. Asymmetric configurations with $\Delta\varpi$ different from $0$ or $\pi$ will not be considered in this work. Such equilibria have not been found for any resonance in small eccentricities regime.

Figure~\ref{fig:energy_plots}a presents the energy levels (black solid curves) at the $\Sigma$-plane $(e_1 \cos\Delta\varpi, e_2 \cos\phi_1)$. Elliptic and hyperbolic points in the plane indicate equilibria (marked with black cross/circle symbols). The energy plot has been obtained for $C$ and $K$ values corresponding to the stable equilibrium of 9:7~MMR taken from the branch of equilibria illustrated with black solid curve in Fig.~\ref{fig:migr_ex2}c (almost vertical line in the centre of the plot). We chose an equilibrium with $e_1 = 0.008$ (for which $e_2 \approx 0.0086$), as the example whose migration was discussed in Section~2 has the equilibrium eccentricities close to these values (see Fig.~\ref{fig:migr_ex3}). The stable equilibrium is marked with black cross-circle symbol in the $(\pi, \pi)$-quarter of the $\Sigma$-plane. The equilibria located in the remaining three quarters are unstable. 

Remaining panels of Fig.~\ref{fig:energy_plots} presents the energy levels at different planes. They are used for better visualisation of the energy plot.  Panel~(b) shows the plot at the $\mathcal{S}$-plane, whose coordinates are $(e_1 \sin\sigma_1, e_2 \sin\sigma_2)$, therefore $\sigma_i$ can be $+\pi/2$ or $-\pi/2$. The {$\mathcal{C}$-plane} with $\sigma_1,\sigma_2$ equal to $0$ or $\pi$, $(e_1 \cos\sigma_1, e_2 \cos\sigma_2)$, is not shown here as it is irrelevant for the purpose of this work. The $\mathcal{S}$-plane possesses the same information as the two bottom quarters of the $\Sigma$-plane, while the $\mathcal{C}$-plane represents the two upper quarters. Panel~(c) is a polar representation of the $\mathcal{S}$-plane, while panel~(d) shows the diagram of $(P_2/P_1, -e_1 \cos\Delta\varpi)$. Thick solid curves in panel~(d) show the borders of the permitted motion and is defined with $e_2=0$. The diagram with $e_2$ instead of $e_1$ is similar to panel~(d) and was not shown.

\begin{figure*}
\centerline{
\vbox{
\hbox{\includegraphics[width=0.48\textwidth]{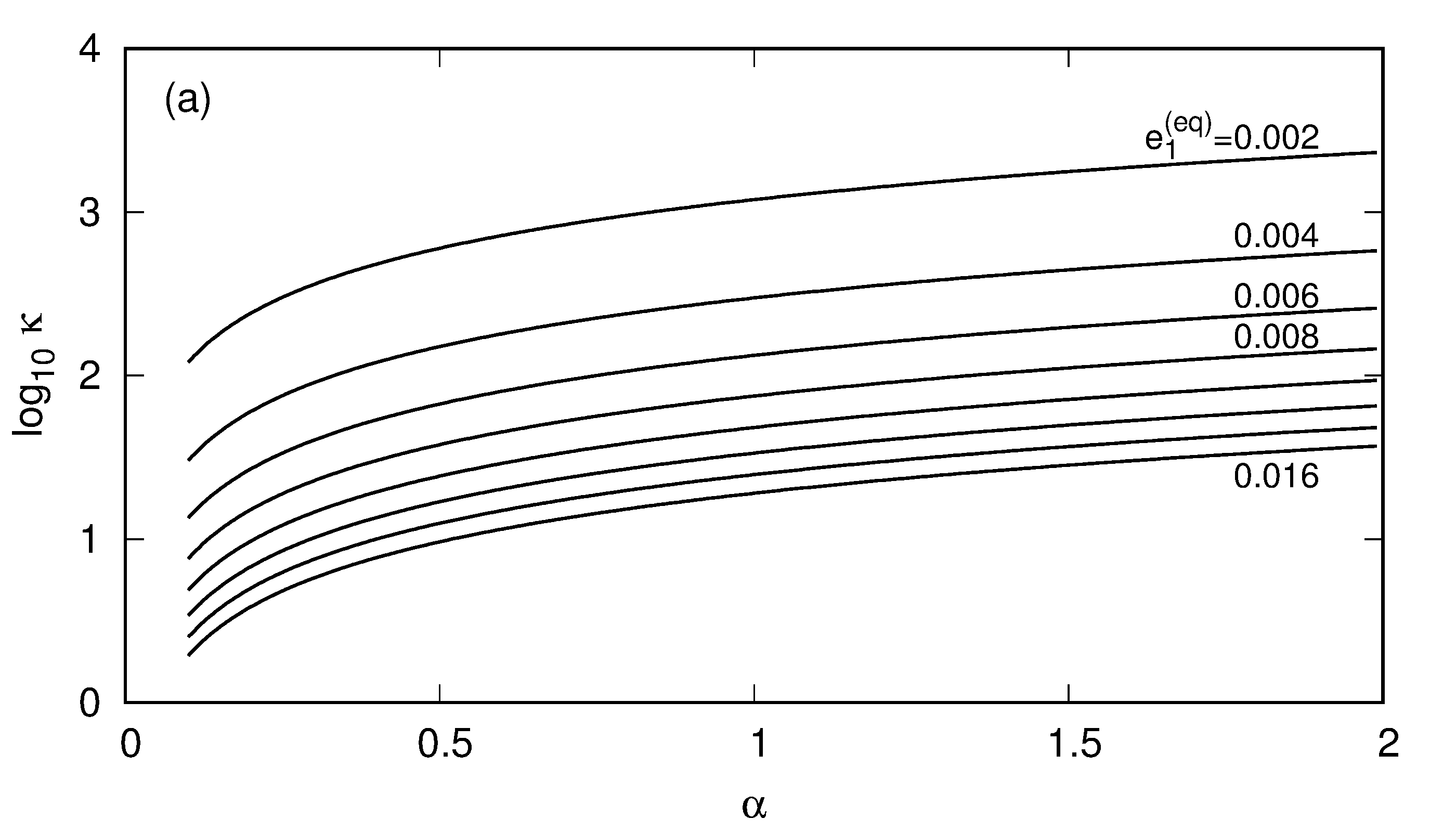}
\includegraphics[width=0.48\textwidth]{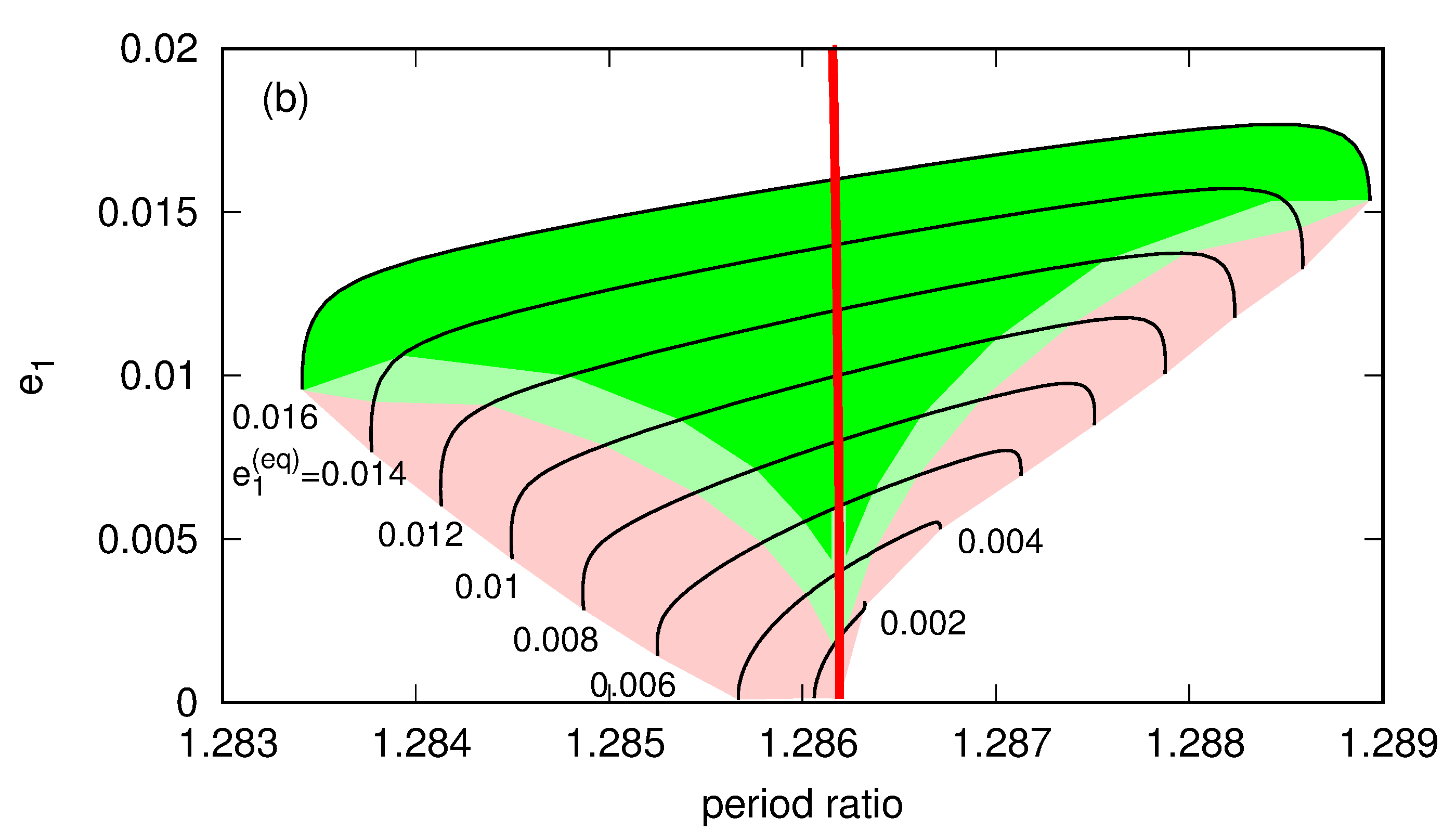}}
}
}
\caption{Panel~(a): Critical values of $e_1$ as a function of $\alpha$ and $\kappa$ parameters of the migration model. Levels of constant values of $e_1 = 0.002, 0.004, 0.006, 0.008, 0.01, 0.012, 0.014, 0.016$ are given with black solid curves. Masses of the planets $m_1 = m_2 = 6\,\mE$. Panel~(b): The red curve denotes the branch of equilibria of the averaged system. The black curves denote the branches of periodic configurations of the averaged system obtained for different equilibrium values of $e_1=0.002, 0.004, \dots, 0.016$ (see the labels). Green and light green areas are the {stability} zones computed for $\alpha=0.2$ and $\alpha=1.8$, respectively. The pink area denotes the region of divergence. See the text for details.}
\label{fig:eq_ecc}
\end{figure*}

\subsection{Chaos and periodic orbits}

Grey points in the panels of Fig.~\ref{fig:energy_plots} indicate chaotic evolution of the system in terms of the spectral number \citep{Michtchenko2001a}. We integrate a given system for $2^{18}$ time-steps (which corresponds to $\sim 2500$ periods related to the resonant dynamics) and compute the Fast Fourier Transform (FFT). The spectral number is defined as a number of peaks in the power spectrum of amplitudes greater than a given limit, here $10^{-3}$ of the highest peak value. Darkest grey colour means that $SN \geq 100$, lighter grey is for systems with $SN \in [50, 100)$, while the lightest denotes $SN \in [10, 50)$.

Green and red curves in Fig.~\ref{fig:energy_plots} denote branches of stable and unstable periodic orbits of the averaged system. A configuration is called periodic if there exists a period $T$ after which the system returns to its initial state, i.e., $\vec{x}(T) = \vec{x}(0)$, where $\vec{x} = (a_1, e_1, \sigma_1, a_2, e_2, \sigma_2)$. Green and red points in the panels of Fig.~\ref{fig:energy_plots} corresponds to the systems with $\delta \equiv \| \vec{x}(T) - \vec{x}(0) \| \lesssim 10^{-8}$. In order to check the stability of a given periodic configuration we change the initial eccentricities by a small value of $\pm 10^{-5}$ ($C$ and $K$ are kept unchanged) and compute $\delta$. The second step is to compute the spectral number, $SN$, for the periodic configuration. Then, for a stable periodic system both $\delta$ and $SN$ are small, while for an unstable configuration both the quantities have high values. We chose $10^{-4}$ as the limit value for $\delta$ and $100$ for $SN$.

\subsection{Structure of the resonance}

The structure of the resonance is best seen in Fig.~\ref{fig:energy_plots}c, i.e., at the polar version of the $\mathcal{S}$-plane. The centre of the resonance is given by the intersection of two branches of stable periodic orbits. The region of the resonance is encompassed by the separatrix (narrow regions of chaotic motions). The horizontal branch of the periodic orbits bends down at its left- and right-hand sides, passing into the branches of the unstable periodic configurations. The bifurcation takes place at the inner separatrix, which encompasses the inner part of the resonance (the one centred at the equilibrium). At both sides of the inner resonance there exists the outer part of the resonance, which is encompassed by the outer separatrix (the outer resonance is not centred at any equilibrium). The evolution of the systems in the two separated regions of the resonance is characterised by librations of both $\Delta\varpi$ and $\phi_1$ angles around $\pi$.

The resonance can be also investigated at the period ratio--eccentricity plane (Fig.~\ref{fig:energy_plots}d). The branch of periodic orbits which is horizontal in Fig.~\ref{fig:energy_plots}c corresponds to the track of the evolution of the system due to migration illustrated in Fig.~\ref{fig:migr_ex3} (compare Fig.~\ref{fig:migr_ex3}a with Fig.~\ref{fig:energy_plots}d). The sign of the $y$-axis of Fig.~\ref{fig:energy_plots}d was changed in order to better show the correspondence. We will show in the next section that the branch of periodic orbits discussed here plays an important role in the migration induced formation of 9:7~MMR.

\section{Migration}

In the previous section we discussed the structure of the resonance, i.e., the equilibria, periodic orbits, division of the resonance into two parts. We will show now that after adding the migration terms to the equations of motion (Eqs.~\ref{eq:migration_terms} and~\ref{eq:equations_with_migration}) the system migrates along the branches of periodic orbits (of the averaged system). We will also demonstrate that a system which enters the resonance far enough from the equilibrium evolve away from it and can leave the resonance in case of long enough migration. 

\subsection{Periodic orbits and equilibrium eccentricities}

As we already mentioned, after entering the resonance a system evolves along certain path in the phase space (see Fig.~\ref{fig:migr_ex3}a). This path is common for both {types of evolution, i.e., towards and away from the resonance centre}. We showed above that this path corresponds to the branch of periodic orbits (see Fig.~\ref{fig:energy_plots}d). The example from Section~2 {illustrates that the system which evolves away from the equilibrium} passes from one mode of the resonance to another (see Fig.~\ref{fig:migr_ex1}e for the evolution of the eccentricity). We observe a change in the libration centres and amplitudes of the eccentricities. The system remains in the second mode of the resonance for limited time and leaves the resonance. Figures~\ref{fig:energy_plots}c demonstrates that the resonance indeed consists of two modes, which we called the inner (centred at the equilibrium) and the outer resonance (with no equilibrium). We will show in this section that the transition between the two modes of 9:7~MMR can be observed within the averaged model of the system as a passage from the inner to the outer MMR.

\begin{figure}
\centerline{
\vbox{
\hbox{\includegraphics[width=0.48\textwidth]{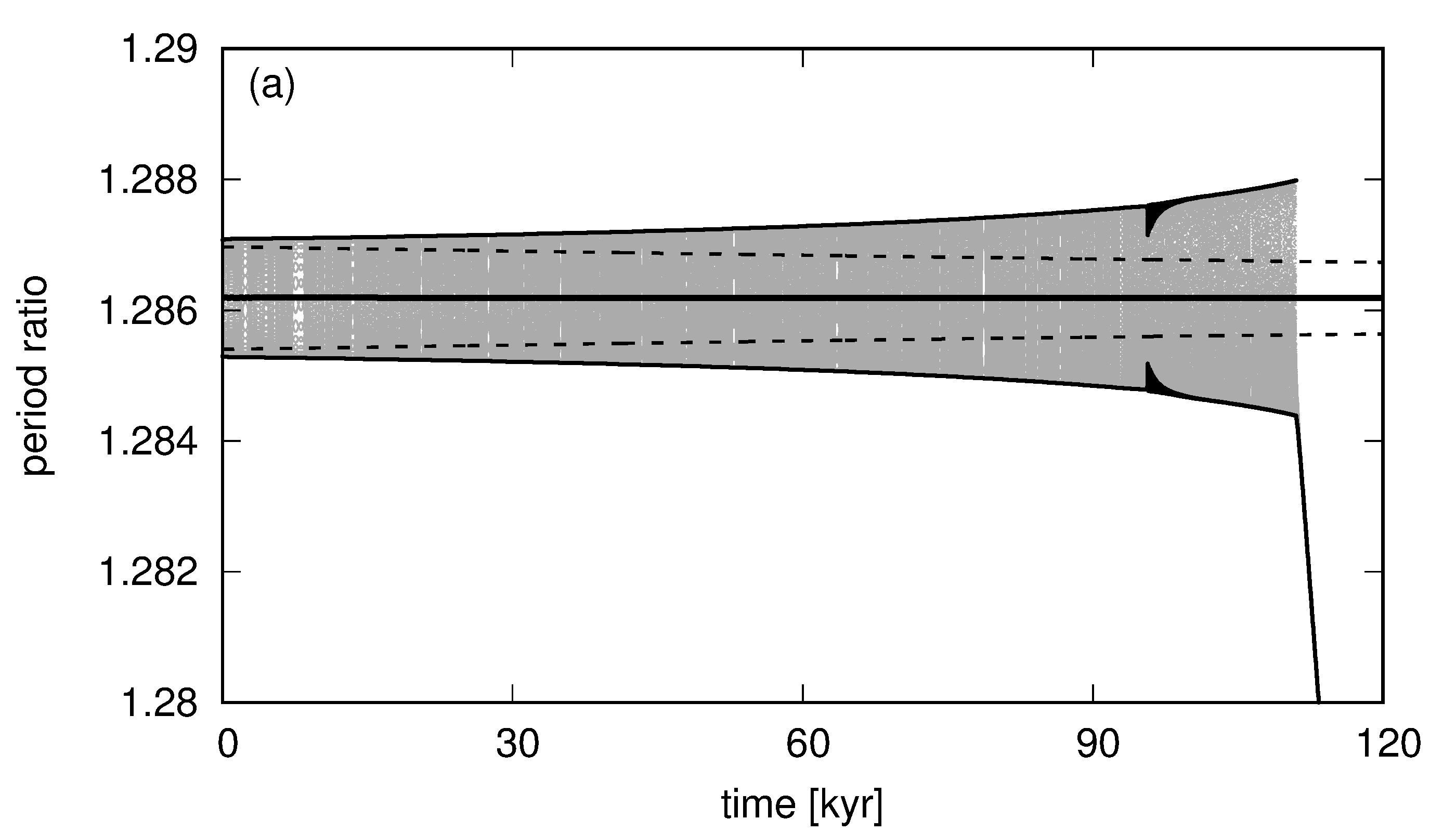}}
\hbox{\includegraphics[width=0.48\textwidth]{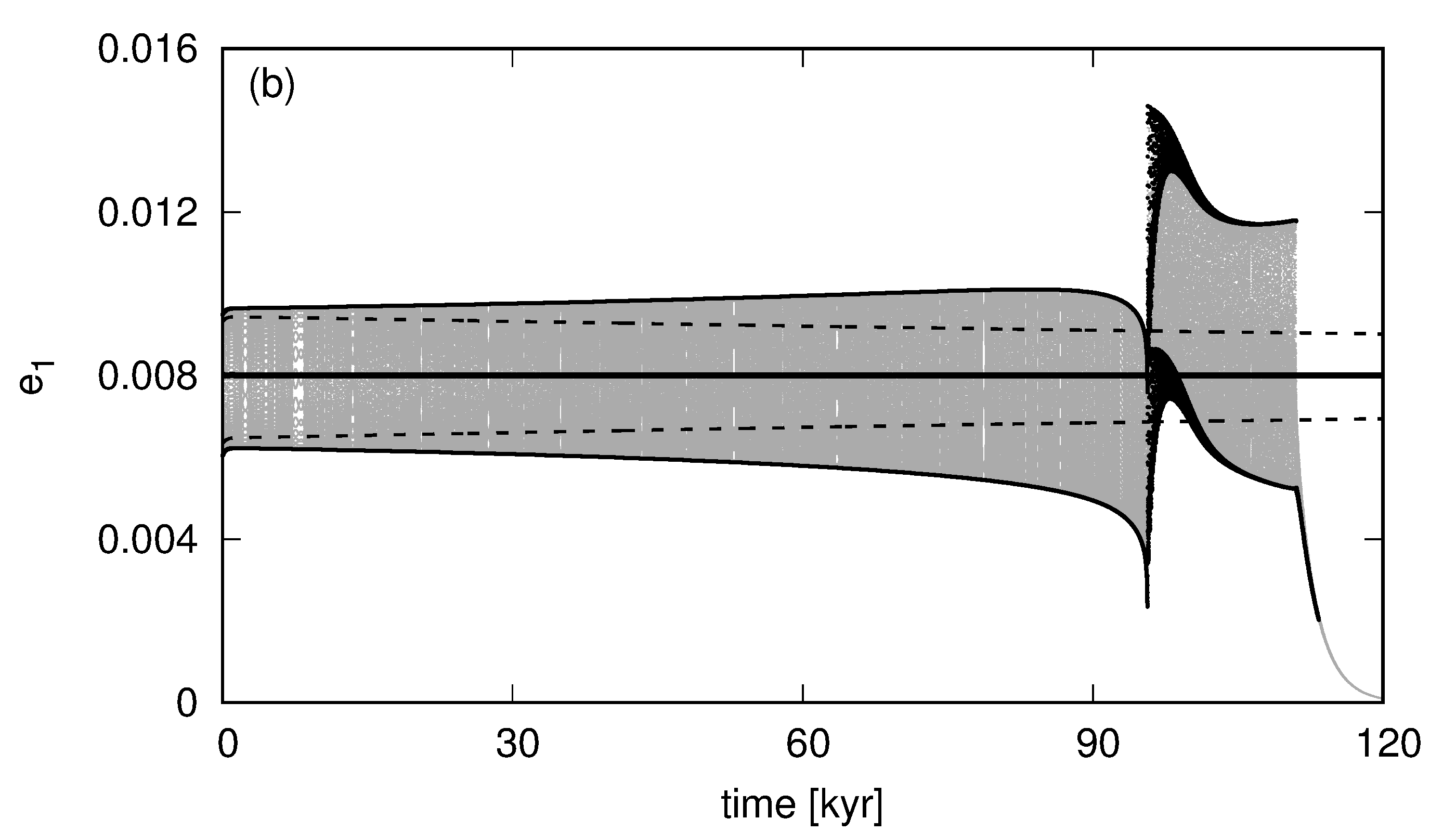}}
\hbox{\includegraphics[width=0.48\textwidth]{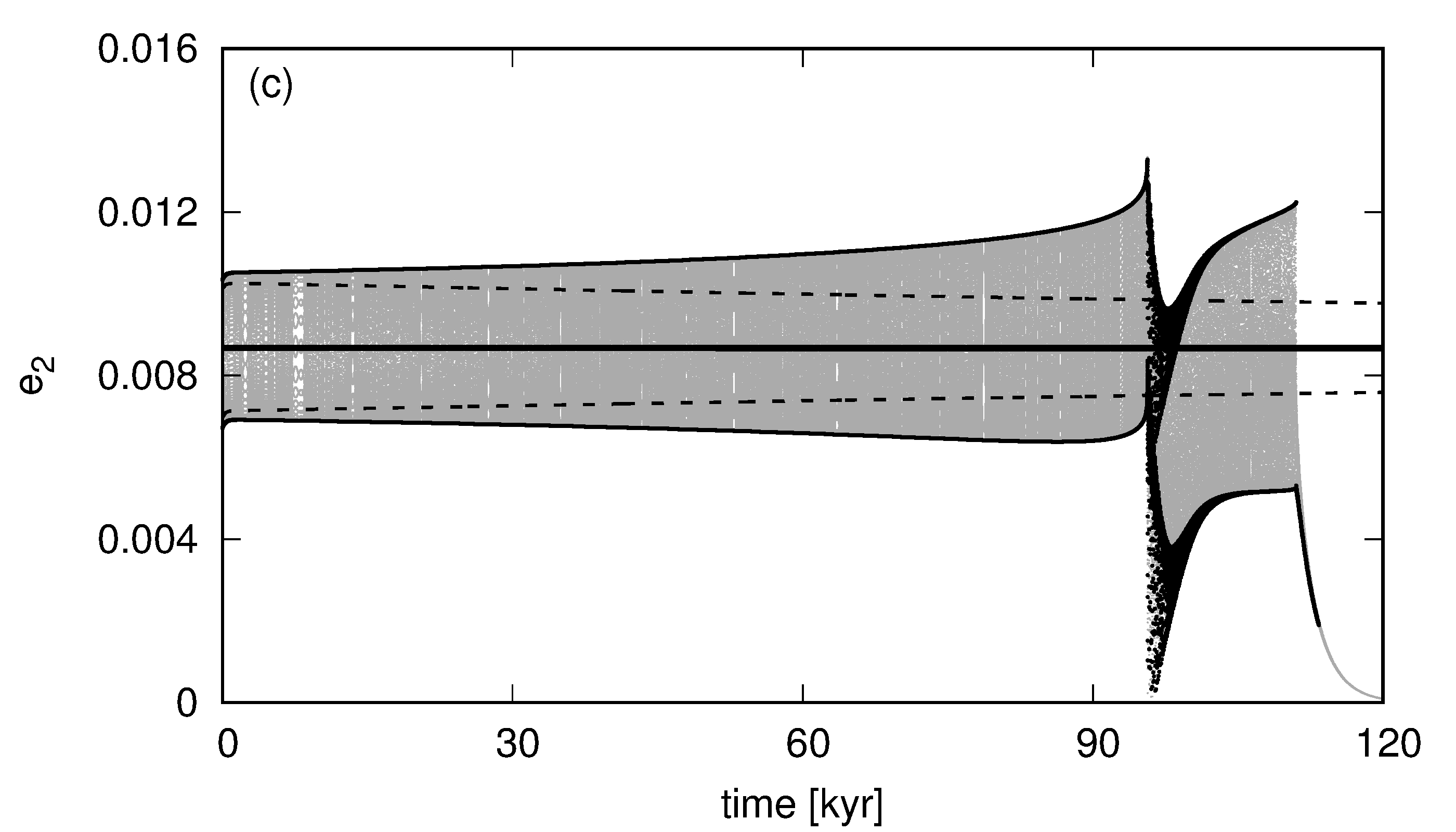}}
}
}
\caption{The evolution of three example initial systems chosen from the branch of periodic configurations of the averaged system (see Fig.~\ref{fig:eq_ecc}b, the branch of periodic orbits corresponds to the equilibrium value of $e_1=0.008$). Subsequent panels illustrate the evolution of the period ratio, $e_1$ and $e_2$. Black thick horizontal lines are for the equilibrium configuration. Dashed curves show the evolution of the system initially inside the {stability} region shown with the green colour in Fig.~\ref{fig:eq_ecc}b. Only epochs with $\phi_1=\pi$ are plotted. Black points denote the evolution of the system initially {outside the stability zone} (red colour in Fig.~\ref{fig:eq_ecc}b) for $\phi_1=\pi$, while grey dots show the whole evolution of the orbital elements. The parameters of the migration are $\tau_0=5~$kyr, $\alpha=1.0$, $\kappa=75.05$. Initial $a_1 = 0.1\,\au$.}
\label{fig:conv_div_example}
\end{figure}

Similarly to the periodic orbits branch which corresponds to the path of the system which undergoes the migration, the stable equilibrium (the centre of the resonance) corresponds to the point towards which the system {evolves during the migration in case of permanent capture} (black circle in Fig.~\ref{fig:migr_ex3}a).
A system initially located in the equilibrium stays in this point during the migration only for certain values of the migration parameters {(the radial dependence of the migration time-scale, $\alpha$, and the ratio between the migration and circularisation time-scales, $\kappa$)}. The migration {time-scale} is not important {so long} as the migration is slow compared to the {time-scale for resonant libration}. On the other hand, for those particular values of $\alpha$ and $\kappa$, a system which entered the resonance {close enough to its centre (inside the stability zone)}, will move towards this point.

Figure~\ref{fig:eq_ecc}a illustrates the equilibrium values of $e_1$ (the plot for $e_2$ is similar) as a function of $(\alpha, \kappa)$. Clearly, for higher values of $\kappa$ the eccentricities are lower. Additionally, for given values of the equilibrium eccentricities, when $\alpha$ is smaller, $\kappa$ also has to be smaller. {The dependence of the equilibrium eccentricities on both $\alpha$ and $\kappa$ may seem unexpected, as the $\kappa$-dependence alone is usually discussed in the literature. In fact, the equilibrium eccentricities depend on the ratio between the time-scale of the period ratio variation (which corresponds to the relative migration time-scale) and the circularisation time-scale, i.e., $\kappa_X \equiv \tau_X/\tau_e$, where $\tau_e = (\tau_{e,1} + \tau_{e,2})/2$, $\tau_X \equiv -X/\dot{X}$ and $X \equiv P_2/P_1$. When $\alpha$ is lower, the relative migration (for the same $\tau_1$) is slower and $\kappa_X$ is higher, which leads to lower equilibrium eccentricities. When only one planet is migrating, as it is commonly studied in the literature, the distinction between $\kappa$ and $\kappa_X$ is not necessary, and the equilibrium eccentricities depend on $\kappa$ and not on $\alpha$.} 

Figure~\ref{fig:eq_ecc}b presents the branch of equilibria (the red solid, almost vertical curve) together with branches of periodic orbits (only the stable part of the horizontal branch from Fig.~\ref{fig:energy_plots}c is shown) obtained for given values of the equilibrium eccentricities (or a given value of the scale-free angular momentum $c$).

\begin{figure}
\centerline{
\vbox{
\hbox{\includegraphics[width=0.48\textwidth]{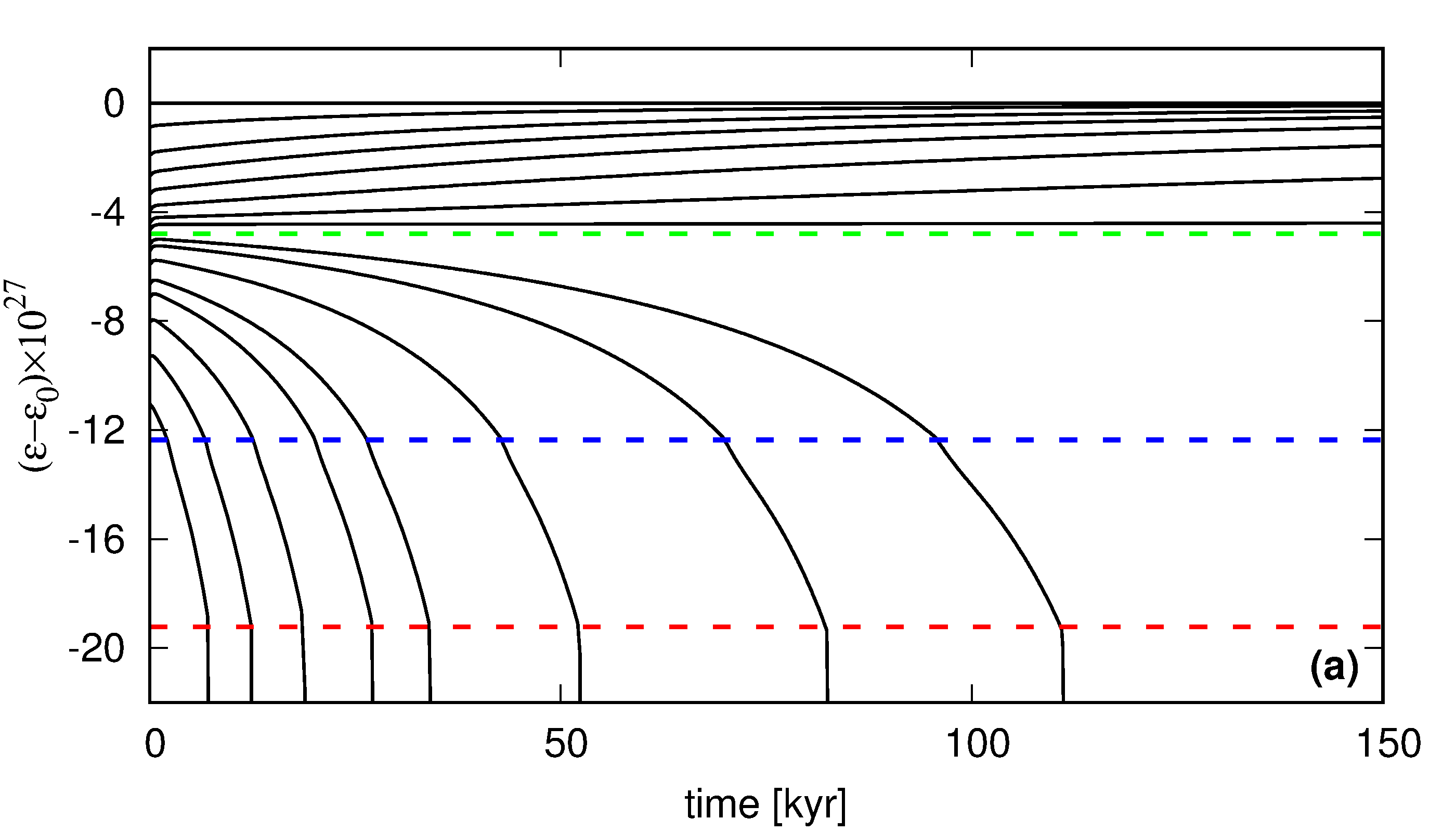}}
\hbox{\includegraphics[width=0.48\textwidth]{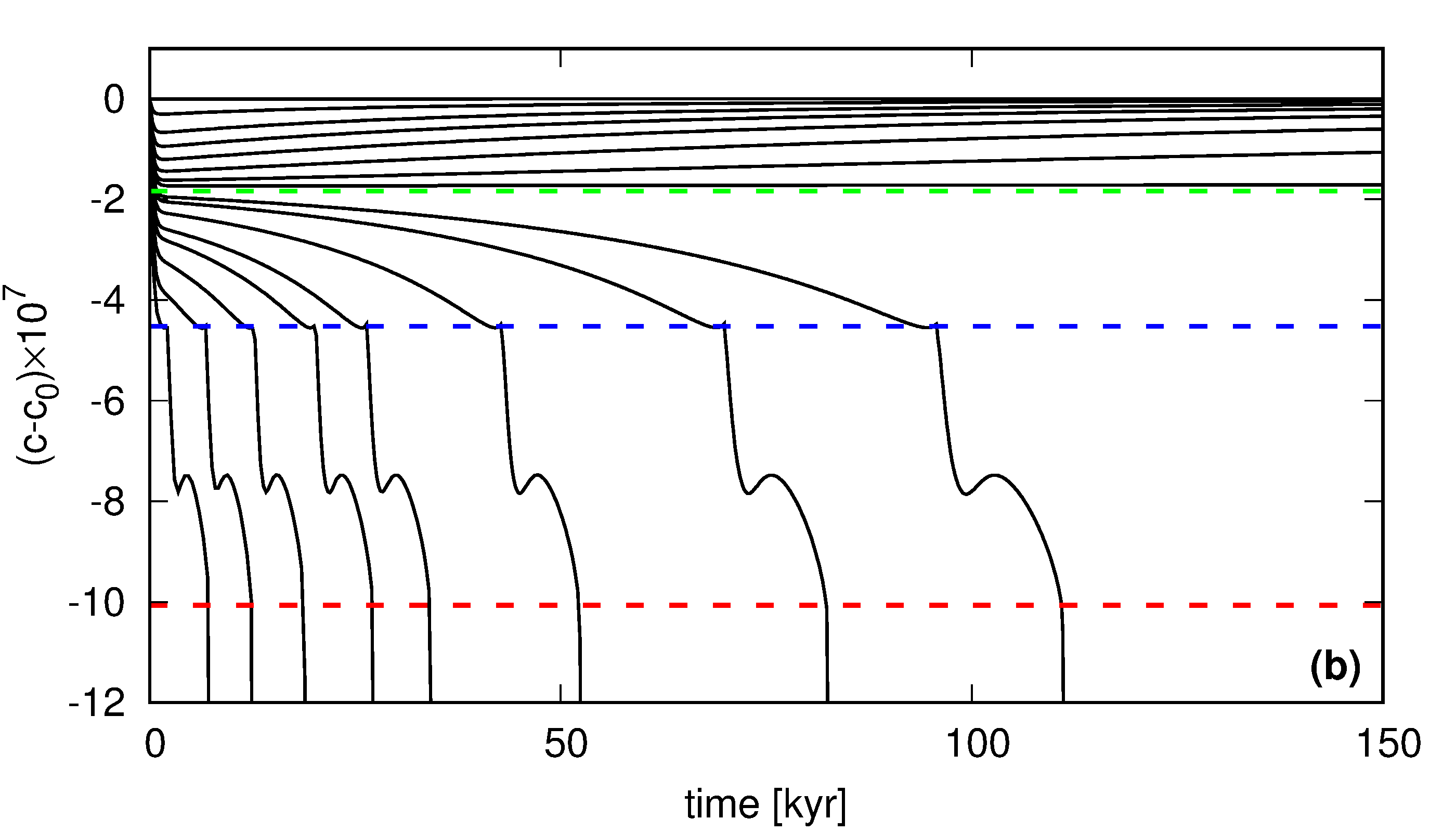}}
}
}
\caption{The evolution of the scale-free energy $\epsilon \equiv E K^2$ (panel~a) and the scale-free angular momentum $c \equiv C/K$ (panel~b) in time for different initial systems taken from the branch of periodic orbits presented in Fig.~\ref{fig:eq_ecc}b, for $e_1^{\idm{(eq)}}=0.008$. The migration parameters are $\tau_0=5~$kyr, $\alpha=1.0$, $\kappa=75.05$. The dashed lines correspond to the critical values of $\epsilon$ and $c$. The green line denotes the value of $\epsilon$ above which the systems tend towards the equilibrium. Below this {they evolve away from the resonance centre}. The blue line denotes the transition between the two modes of the resonance (the inner one and the outer one). For $\epsilon$ smaller than the level denoted by the red line the system is non-resonant. See the text for details. }
\label{fig:stability_ranges1}
\end{figure}

\begin{figure*}
\centerline{
\vbox{
\hbox{\includegraphics[width=0.85\textwidth]{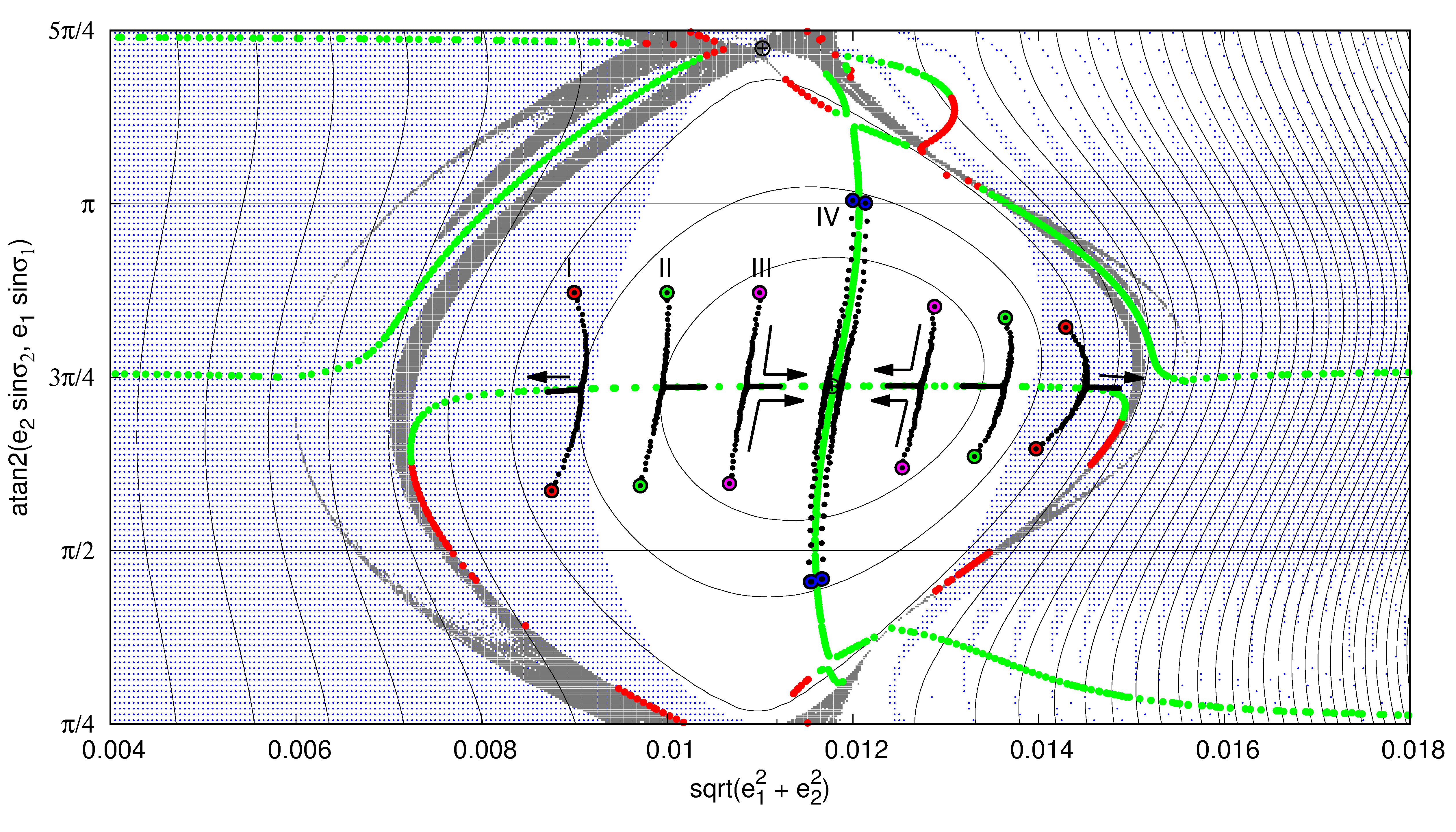}}
}
}
\caption{The energy diagram similar to Fig.~\ref{fig:energy_plots}c. Levels of constant values of the energy are shown with black solid curves. The green and the red curves denote stable and unstable periodic orbits, respectively. Grey dots indicate chaotic systems ($SN \geq 100$) without the dissipative terms. Blue dots denote initial systems that evolve away from the resonance when the migration terms are added. The parameters of migration are $\tau_0=10\,$kyr, $\alpha=1$ and $\kappa=75.05$. Evolution of four example systems that undergo the migration is illustrated with black points. The initial positions of each system are marked with big circle symbols, red, green, magenta and blue as well as labelled with I, II, III and~IV. The arrows show the direction of the evolution.
}
\label{fig:critical_energies}
\end{figure*}

As illustrated with the help of the example in Section~2, a system which enters the resonance not necessarily stays in it. Green area in Fig.~\ref{fig:eq_ecc}b shows the {stability zone} (the region of stable capture). It means that when starting from a given point from the green area of the branch of periodic orbits (corresponding to given equilibrium eccentricities) and with the appropriate values of the migration parameters $\alpha$ and $\kappa$, the system evolves towards the equilibrium. As seen in Fig.~\ref{fig:eq_ecc}a there is a continuum of combinations of $(\alpha, \kappa)$ leading to the same values of the equilibrium eccentricities. The size of the {stability} region, though, depends on particular values of the migration parameters. For smaller values of $\alpha$ the region is narrower. In the example shown in Fig.~\ref{fig:eq_ecc}b the darker green corresponds to $\alpha=0.2$, while the lighter green denotes the stability region for $\alpha=1.8$. Naturally, if the system is located in the darker green area, it is located {inside the stability zone} for both values of $\alpha$.

The evolution of the period ratio and both eccentricities of three representative configurations is presented in Fig.~\ref{fig:conv_div_example}. The initial orbital parameters were chosen from the branch of periodic orbits which corresponds to the equilibrium value of $e_1 = 0.008$ (see Fig.~\ref{fig:eq_ecc}b). Black solid lines in each panel correspond to the evolution of the system in the equilibrium. Dashed curves are for the evolution of the system initially located in the {stability} region. Only epochs when $\phi_1=\pi$ are plotted, which means that a given parameter oscillates within the ranges given by the bottom and the upper dashed curves.
As we can see, the amplitudes of oscillations decrease in time, so the system tends towards the equilibrium. The third example (shown with black points for the epochs of $\phi_1=\pi$ and grey area for the evolution between those epochs) corresponds to the system which is located initially {outside the stability} region. We observe an increase in the amplitudes of the oscillations. After $\sim 95\,$kyr the system moves to the second regime of the resonance (in which both $\phi_1$ and $\Delta\varpi$ keep librating around $\pi$ -- not shown here) and at $t \sim 110\,$kyr the system leaves the resonance. Similar behaviour could be observed in the example simulated within the $N$-body model of motion (see Fig.~\ref{fig:migr_ex1}e). Nevertheless, in the $N$-body simulation the system initially started outside the resonance, and in the example shown in Fig.~\ref{fig:conv_div_example} it starts from the branch of periodic orbits, i.e., when the system is already in the resonance. The mechanism of leaving the resonance is, though, the same in both cases.

\subsection{The scale-free energy and angular momentum evolution}

As we already mentioned, the structure of the equilibria, periodic orbits and the energy levels at the representative plane depend on the value of the scale-free angular momentum $c$. In order to follow the evolution of a given system one can also use another quantity, the scale-free energy $\epsilon \equiv E\,K^2$, where $E$ is the averaged energy given by the Hamiltonian (Eqs.~\ref{eq:Hamiltonian} and~\ref{eq:disturbing_function}) computed for a given point of the phase space. A system in equilibrium has a maximal value of $\epsilon$, which we denote as $\epsilon_0$. If a system starts away from the equilibrium but within the {stability} region, $\epsilon$ will grow asymptotically up to $\epsilon_0$.

Figure~\ref{fig:stability_ranges1}a illustrates the evolution of $\epsilon$ for initial systems chosen from the branch of periodic orbits for the equilibrium $e_1 = 0.008$. The systems of initial $\epsilon$ close enough to $\epsilon_0$ (above the green dashed line) tend towards the equilibrium. Below the green line $\epsilon$ decreases {(i.e., the systems tend away from the resonance)}. At the level denoted by the blue dashed line the system moves from one mode of the resonance to the second one. After further decrease of $\epsilon$ the system leaves the resonance when $\epsilon$ is below the red dashed line.

Figure~\ref{fig:stability_ranges1}b presents the evolution of the scale-free angular momentum for the same set of initial configurations as discussed above. All the systems have the same initial value of $c = c_0$. Shortly after the beginning of a given simulation $c$ decreases. If it decreases below the green dashed line, $c$ will decrease further. If $c$ is above the green line after this initial decrease, the system tends towards the equilibrium {($c$ increases)}. Similarly to the left-hand panel, the blue and the red dashed lines denote, respectively, the transition between the two modes of the resonance and the value at which the system leaves the resonance.

\begin{figure*}
\centerline{
\vbox{
\hbox{\includegraphics[width=0.85\textwidth]{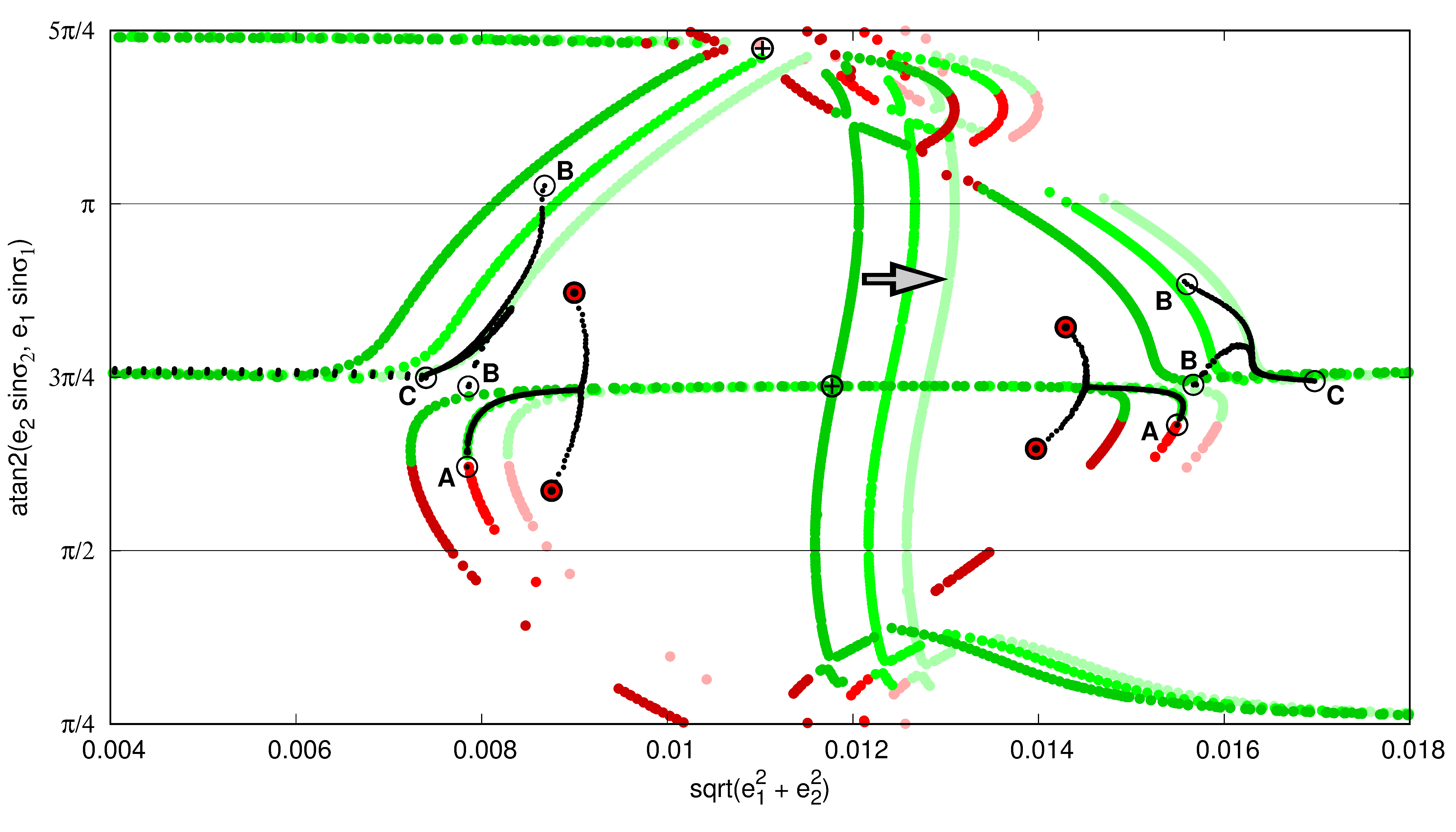}}
}
}
\caption{The evolution of the system which moves away from the resonance during the migration (system~I whose initial position at the diagram is shown with the red point in Fig.~\ref{fig:critical_energies}). For the reference the periodic orbits are shown for three values of $c$ which changes during the evolution. The darkest green/red curves denote the stable and unstable periodic orbits for the initial value of $c$. The medium green/red colour means the value of $c$ corresponding to the transition between the two modes of the resonance (see the text for detail, the system intersects the representative plane in points marked with a letter A -- just before the transition, and B -- just after the transition). The lightest green/red curves are computed for the moment of the evolution when the system leaves the resonance (the system intersects the representative plane in points marked with a letter C). A big grey arrow indicates the direction in which the structure of the periodic orbits moves during the evolution.
}
\label{fig:critical_energies2}
\end{figure*}

\subsection{The migration presented at the representative plane}

In this section we illustrate the evolution of the systems with respect to the branches of periodic orbits. Figures~\ref{fig:critical_energies} show the evolution of four systems of initial orbits chosen from the representative plane. The periodic orbits and chaotic regions are shown in the same manner as in Fig.~\ref{fig:energy_plots}c (with this difference that the chaotic orbits are marked only if $SN \geq 100$). Each system at the initial time intersects the plane in four points. Those points of intersection are denoted by circle symbols of different colours, red, green, magenta and blue, and labelled I, II, III and~IV, respectively.

The points of intersection corresponding to the starting epoch are found during the integration of the equations of motion without the migration terms for the initial values of the orbital parameters. Next, we integrate the equations of the system with migration (see the parameters of the migration listed in the caption) over the time interval of $100\,$kyr. We list the osculating values of the orbital elements with the output time-step of $100\,$yr. Next, for each epoch we take the orbital elements as a starting configuration for the integration of the equations of motion without the migration terms in order to find the points of intersection of the representative plane. This way we obtain tracks of the migrating system at the plane.

Lets first describe the evolution of the example~III (the magenta symbols). The system starts relatively close to the equilibrium. At first the system tends towards the horizontal branch of periodic orbits. After reaching the branch, the configuration moves towards the equilibrium. The evolution direction is shown with black arrows. Similar evolution one observes for the system~II whose initial position at the plane is shown with the green points. The configuration~IV (the blue symbols) also tends towards the equilibrium, however it starts very closely to the vertical branch of periodic orbits. We can then see very clearly that the evolution track is parallel to the vertical branch.

The system~I marked with the red symbols evolve differently. After reaching the horizontal branch it moves away from the equilibrium. That is an example of the evolution {away from the resonance centre}. We checked many initial systems taken from the representative plane and marked the ones that evolve {this way} with small blue dots. {Therefore, white zone centred around the equilibrium corresponds to the region of stable capture.} We can also find other {white areas at the plane} that are out of the resonance (in a higher eccentricity regime). They are localized around the branch of stable periodic orbits characterized by the $y$-axis value close to $\pi/4$. That means that $e_1 \approx e_2$ and $\Delta\varpi=0$. We conclude that it is possible to enter the resonance by moving along this branch. The fact that the white region is structured means that the entrance into the resonance depends on the phase in which the system reaches the separatrix. On the other hand, entering the resonance along the branch of periodic orbits with $e_1 \approx e_2$ and $\Delta\varpi=\pi$ (the $y$-axis value close $3\pi/4$) is not possible, when starting from the eccentricities higher than the resonant values.

\begin{figure*}
\centerline{
\vbox{
\hbox{
\includegraphics[width=0.49\textwidth]{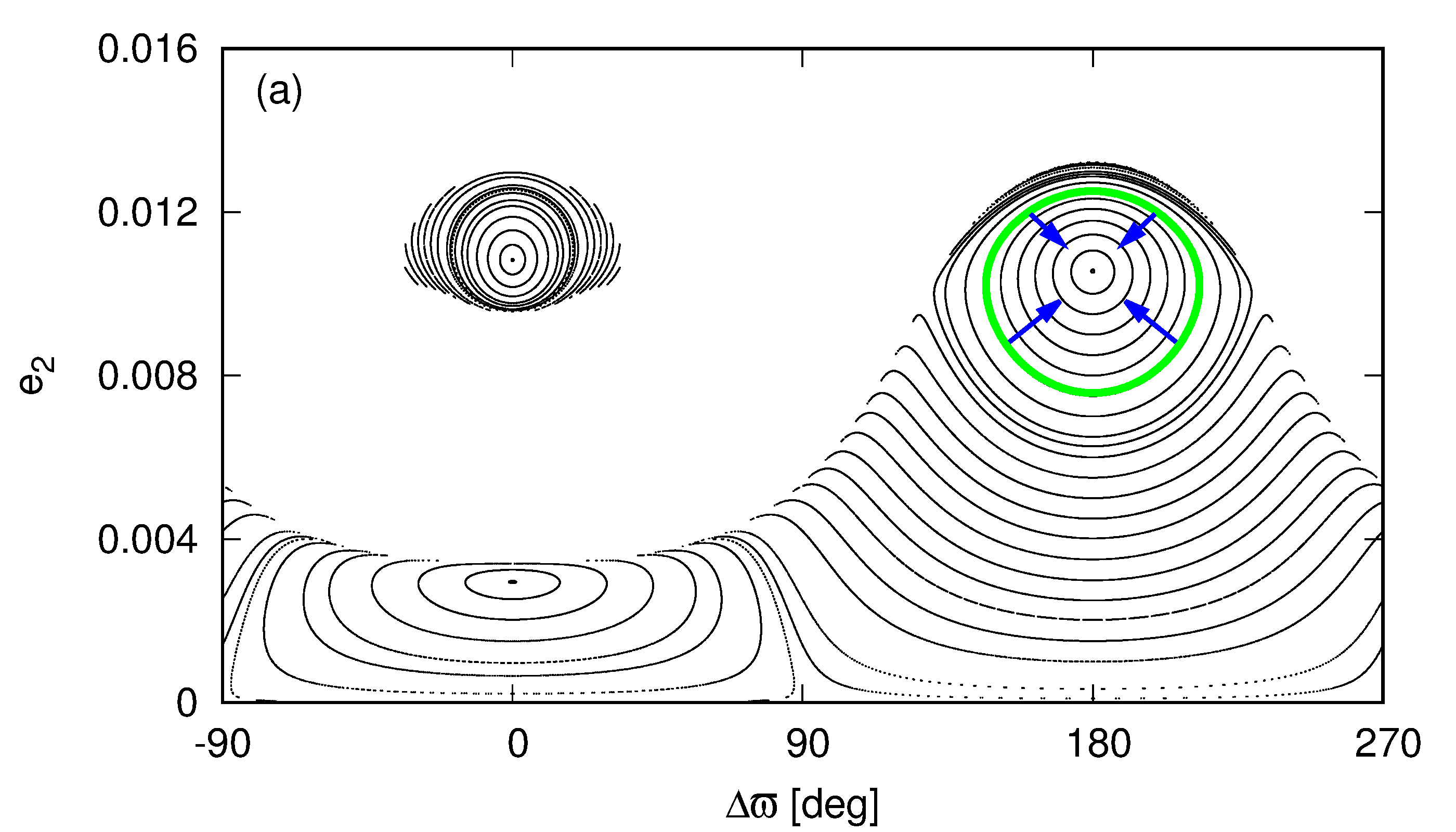}
\includegraphics[width=0.49\textwidth]{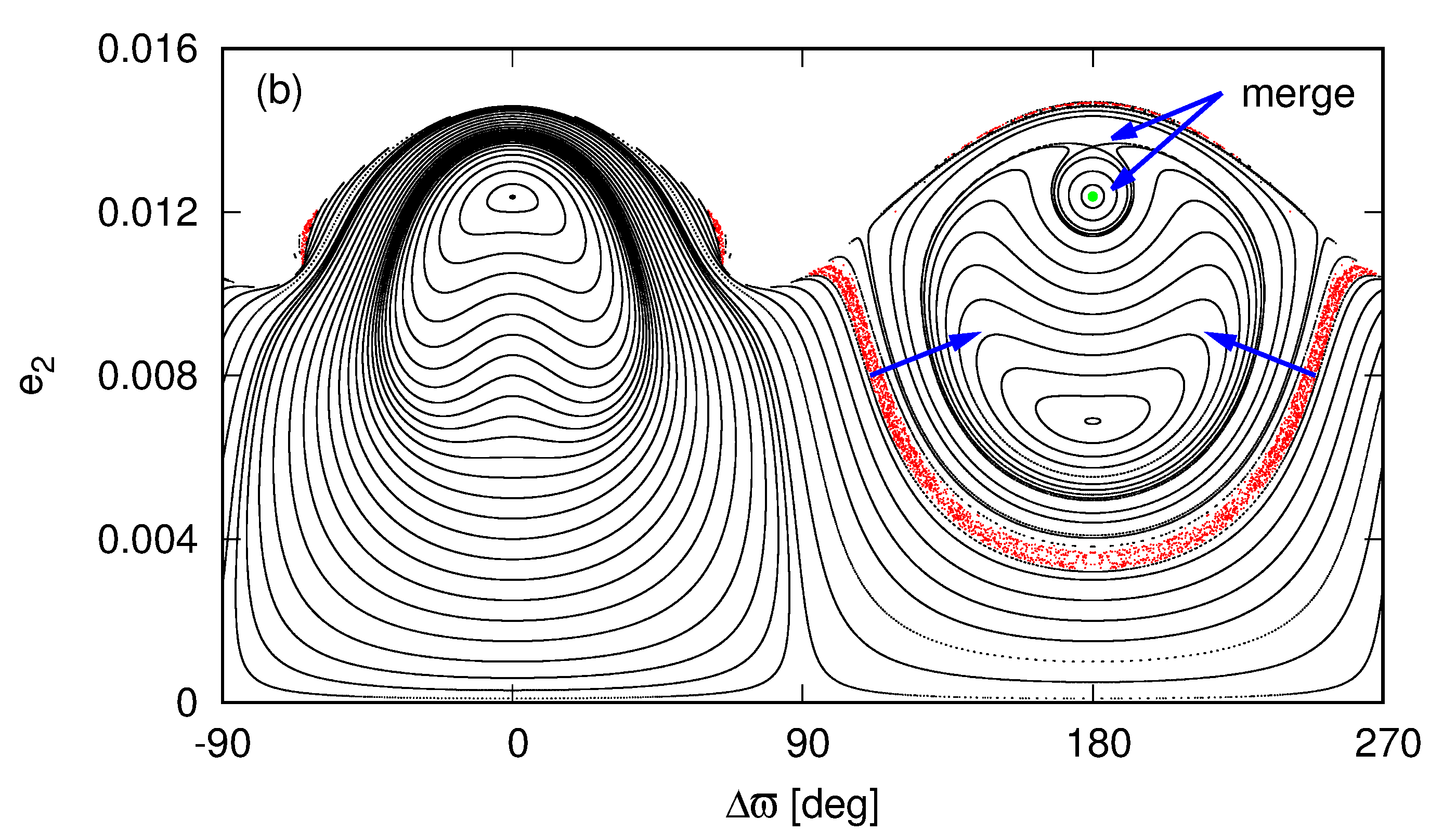}
}
\hbox{
\includegraphics[width=0.49\textwidth]{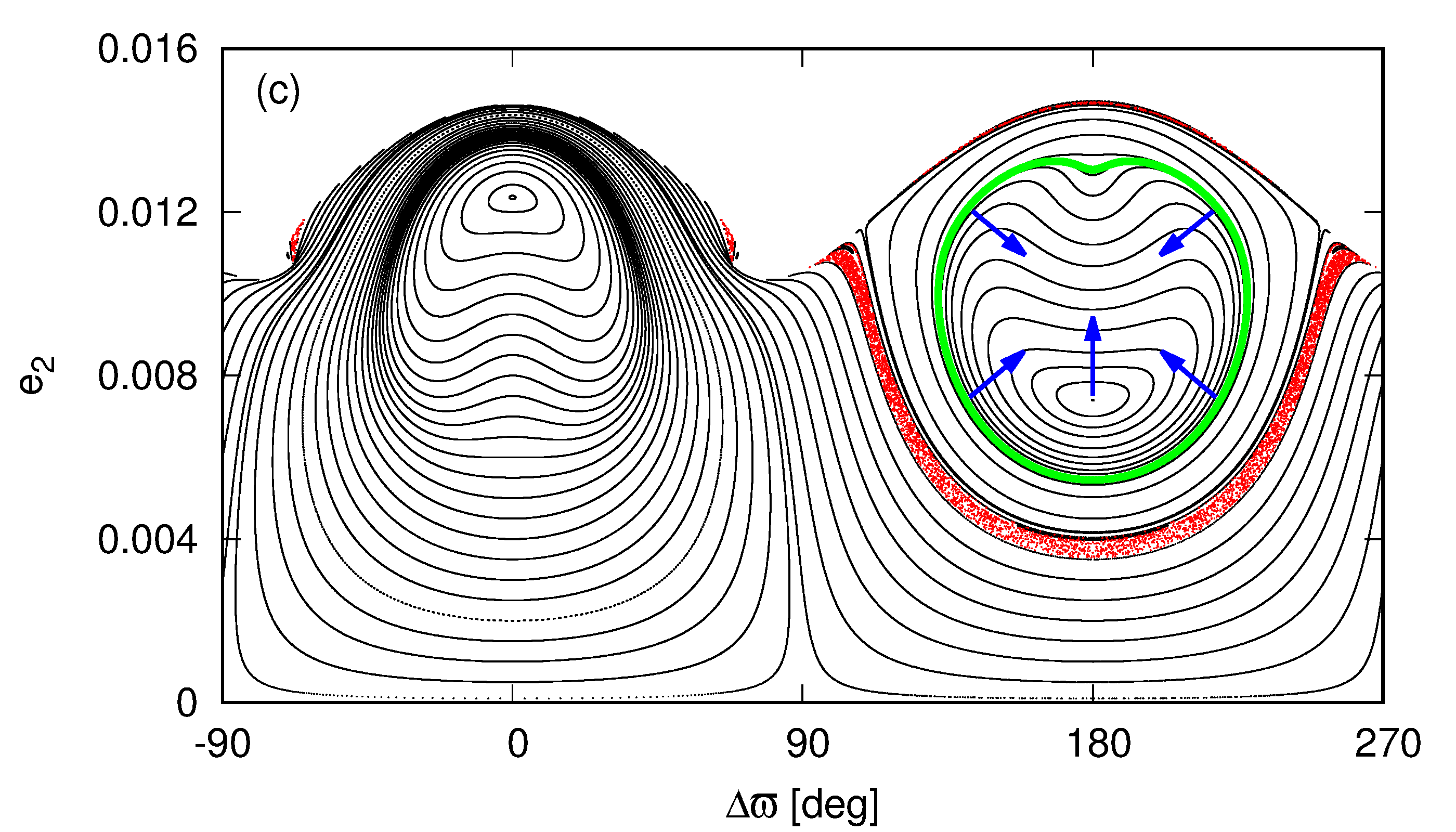}
\includegraphics[width=0.49\textwidth]{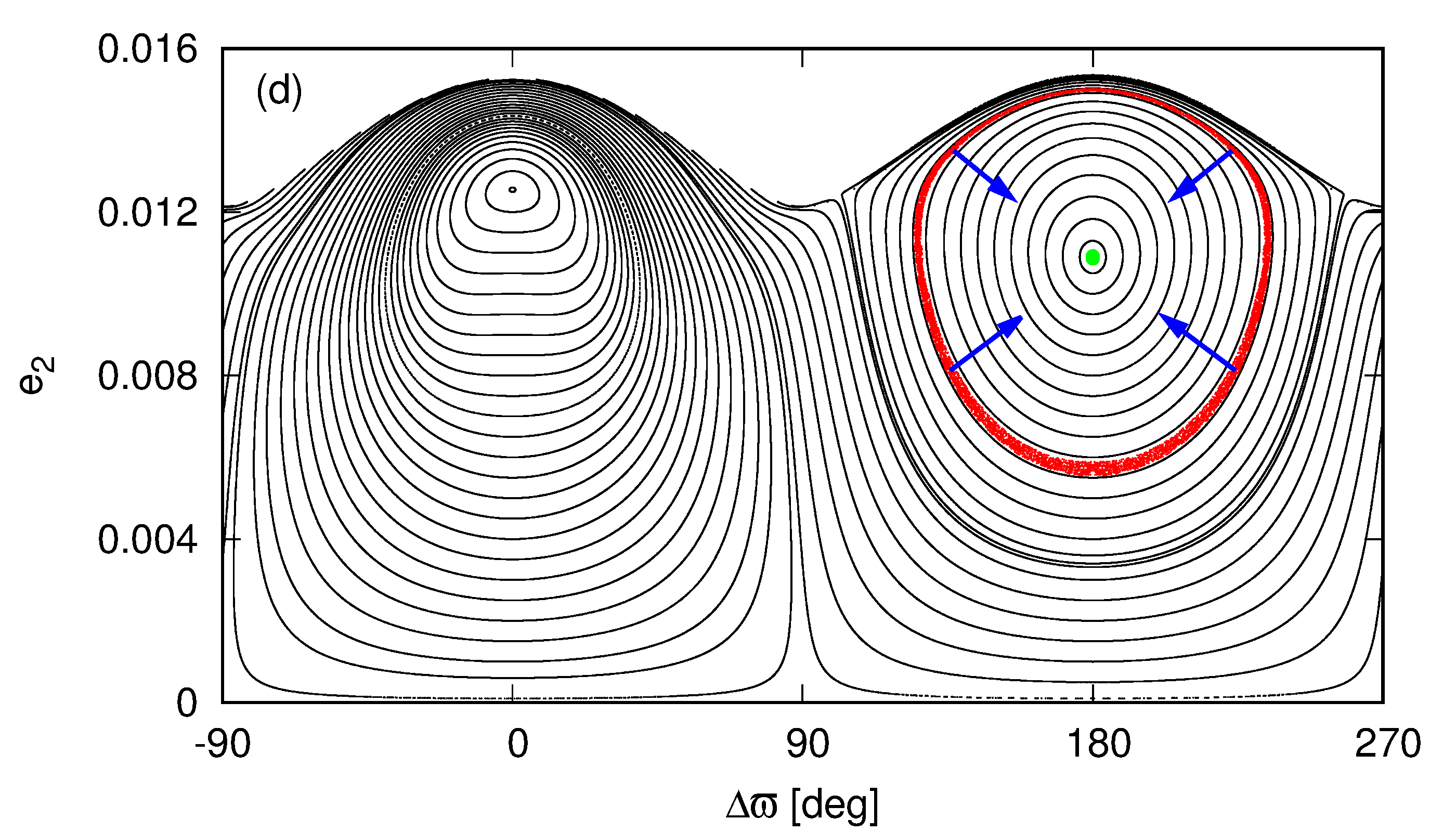}
}
}
}
\caption{Poincar{\'e} cross sections for the system whose evolution is illustrated in Fig.~\ref{fig:critical_energies2}. The sections are defined by $\phi_1=\pi$ and $\dot{\phi}_1 > 0$. See the text for details on the epochs choices.}
\label{fig:poincare}
\end{figure*}

We observe that after reaching the horizontal branch, the systems evolve along it (towards or away from the equilibrium). Nevertheless, an illustration of the further evolution of the system~I along the branches of periodic orbits is not easy. That stems from the fact that the scale-free angular momentum changes during the migration, and because of that the branches of periodic orbits change their position at the representative plane. The shift of the branches is shown in Fig.~\ref{fig:critical_energies2}. As $c$ decreases (see Fig.~\ref{fig:stability_ranges1}b), the equilibrium shifts towards higher values of the eccentricities. The position of the branches of periodic orbits in three representative epochs are shown with decreasing grades of green/red colours. The branches obtained for the initial epoch is marked with the darkest green/red colour. The branches for the epoch for which the system moves from the inner resonance to the outer resonance are denoted with lighter colours, while the epoch when the system leaves the resonance is illustrated with the lightest green/red colours.

The track of system~I is extended here to the whole time interval (in Fig.~\ref{fig:critical_energies} only a fragment of the evolution was shown). The system follows the horizontal branch until it reaches the bifurcation point (marked with letter A), which is a merging point of the stable and unstable periodic configurations. This is the moment of the transition between the two modes of the resonance. After the bifurcation the system moves close to the branch of periodic orbits which is the centre of the outer resonance (points B). Because the transition occurs very quickly, the system is not a periodic configuration just after that. Some amount of time is needed so the system moves closer to the branch of periodic orbits. When the system reaches the outer separatrix (points C), it leaves the resonance.

\subsection{The Poincar\'e cross sections}

The evolution of the system which is leaving the resonance can be also followed at the Poincar\'e cross sections. Figure~\ref{fig:poincare}a presents the section for the initial epoch (red points in Fig.~\ref{fig:critical_energies2}). The system is relatively far from the branch of periodic orbits. The phase trajectory for the nominal system is plotted with a {green} curve. Other curves are phase trajectories for the same energy. All of them fulfil the condition of $\phi_1 = \pi$ and $\dot{\phi}_1 > 0$. The fixed point towards which blue arrows point is the periodic orbits for the energy of the system. The arrows illustrate that the system will move towards the fixed point when the migration is added to the model. 

\begin{figure*}
\centerline{
\vbox{
\hbox{\includegraphics[width=0.9\textwidth]{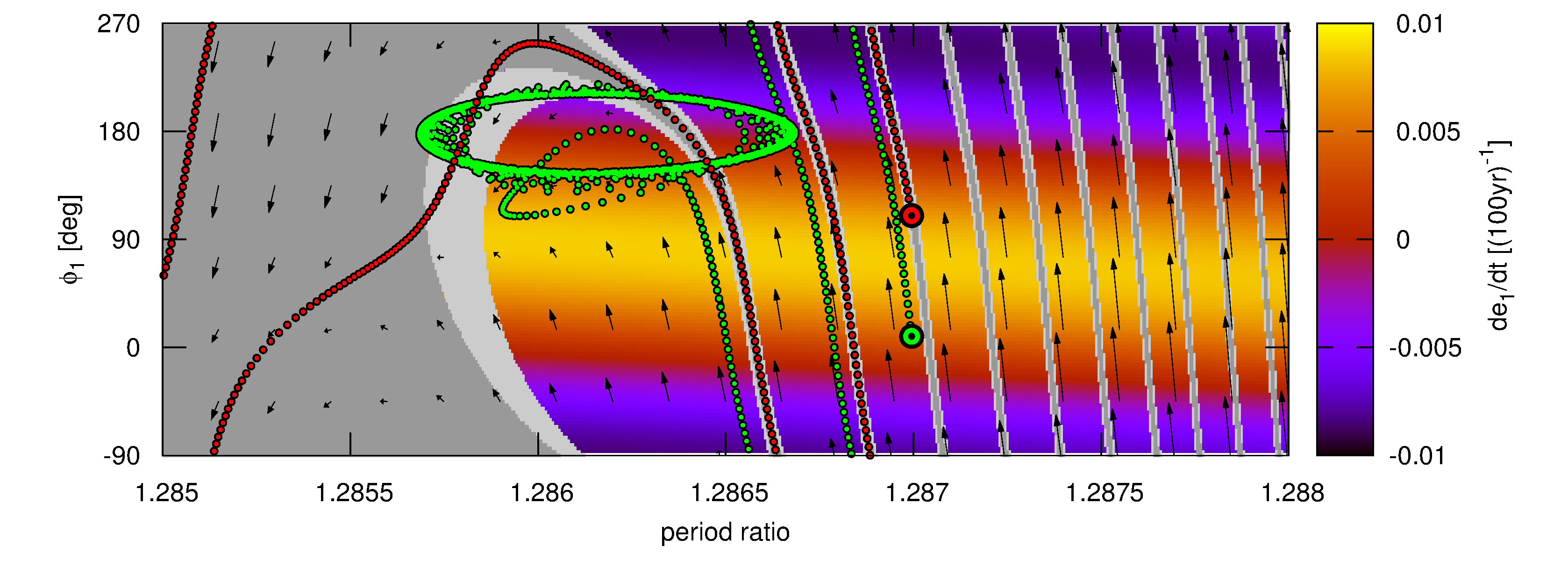}}
}
}
\caption{A palette of colours: the time variation of $e_1$ for different initial period ratios and $\phi_1$ under the assumption that the system starts from the branch of equilibria of 4:3~MMR. The arrows form a vector field, each of them points where a given system moves at the plane. The green and the red big symbols mark two representative initial positions at the diagram. The green and the red curves illustrate the evolution of those configurations, the green one is for the system that enters and stays in 9:7~MMR, the red curve presents the evolution of the system that passes through the resonance. The light grey colour denotes initial configurations that are captured in the resonance only temporarily, while the dark grey means that a given system passes through the resonance without being trapped at all. The planets masses $m_1 = m_2 = 6\,\mE$, and the parameters of migration are $\alpha=1$, $\kappa=75.05$, $e_1^{\idm{(eq)}}=0.008$, $\tau_0 = 1\,$kyr (i.e., initial $\tau_1=10\,$kyr, as initial $a_1=0.1\,\au$).
}
\label{fig:entering1}
\end{figure*}

Figure~\ref{fig:poincare}b shows the Poincar\'e section for the epoch shortly before the system reaches the bifurcation point (point A in Fig.~\ref{fig:critical_energies2}). The section is computed for the energy and the angular momentum equal to the values of the migrating system at this epoch. The nominal system is located in the fixed stable point of the section (a green point), which means that the system is a periodic configuration. There is an unstable fixed point close to the position of the system. During the evolution, those two points get closer one to another and shortly after the epoch for which we made the section, those two points merge. Red points correspond to chaotic motions. The arrows show the direction of the evolution of this chaotic region during the migration. The chaotic area is the border of the resonance. There is one more stable fixed point (below the two mentioned above) within the resonance.

The next panel~(c) in this figure corresponds to the epoch just after the bifurcation. The resonance region shrinks slightly. The two fixed points merged, and now there is only one periodic orbit for $\Delta\varpi = \pi$. The nominal system whose phase trajectory is shown with a green curve is not a periodic configuration. As we mentioned already, that stems from the fact that the bifurcation was very fast. The transition between panels~(b) and~(c) is the transition between the two modes of the resonance mentioned earlier in this work. 

During the further migration, the nominal system tends to the periodic configuration (as pointed by the arrows). The fixed point shifts towards higher values of the eccentricities. The last panel~(d) in Fig.~\ref{fig:poincare} shows the Poincar\'e cross section shortly before the system leaves the resonance. The nominal system is already in the fixed point of the section. The chaotic region gets narrower (see arrows in panel~d). After the chaotic area shrinks down and reaches the fixed point, the system leaves the resonance. In other words, for the energy of the nominal system there is no resonant region in the phase space.

\subsection{The entrance into the resonance}

The study presented in the previous section was devoted to the evolution of the system when it is already in the resonance. We studied on what conditions it can stay in the resonance and when it {evolves away from the equilibrium} and, as a consequence, leaves the resonance. As we showed, the final fate of the system depends on how close it is located to the equilibrium just after entering the resonance. One can easily guess that the initial distance from the equilibrium depends on the time-scale of migration. For slower migration the initial distance should be smaller. On the other hand, as the example shown in Fig.~\ref{fig:migr_ex1} may suggest, the dependence is not necessarily that simple. In this section we study the entrance into the resonance.

As shown in Fig.~\ref{fig:migr_ex2}, a system of the initial period ratio greater than $9/7$ approaches the branch of periodic orbits of 4:3~MMR before it reaches $P_2/P_1 \approx 9/7$. It is thus natural to consider the branch of equilibria of 4:3~MMR as a starting point of the migrating system which is modelled within the averaged equations of motion. A configuration that is in that equilibrium has $\Delta\varpi=\pi$ and $\phi_{4:3}^{(1)} = \pi$, while the resonant angles of 9:7~MMR rotate.

Figure~\ref{fig:entering1} illustrates the entrance of the system into 9:7~MMR when starting from the branch of equilibria of 4:3~MMR. The initial period ratio is being chosen from a relatively narrow range just before $9/7$, while $\phi_1$ is being chosen from the whole possible range of $(0, 2\pi)$. Each of the initial configurations represented by a point at the $(P_2/P_1, \phi_1)-$diagram is integrated in time over $2\,$kyr and the type of the system's behaviour is specified. The system can pass through the resonance (then such a configuration is marked with a dark grey point), it can enter the resonance temporarily (i.e., the system moves away from the centre of the resonance; such a configuration is marked with a light grey point) or the system can enter the resonance permanently (i.e., after entering the resonance the system moves towards the equilibrium of 9:7~MMR), in such a case no grey point is plotted at the diagram.

Configurations that are not destined for being permanently captured form stripes at the diagram in its right-hand part. The strips overlap with the vector filed shown with arrows. At a given point one can compute the variation of $P_2/P_1$ and $\phi_1$ in a given short unit of time. The arrows in the right-hand part of the diagram are almost vertical which means that the $\phi_1$ variation is much faster than the migration (expressed by the period ratio variation). Closer to the centre of the diagram $\phi_1$ evolves slower and thus a given system stays longer in a given part of the diagram. A palette of colours code $\dot{e}_1$ (black means the lowest values, while yellow denotes the highest values) and shows that the eccentricity is excited the most when $\phi_1 = \pi/2$, while is damped the most when $\phi_1 = -\pi/2$.

For the period ratio between $\sim 1.286$ and $\sim 1.2865$ and for $\phi_1 \approx \pi/2$ the eccentricities are accelerated the most, as $\dot{e}_1$ is the largest and the system stays in this region of the plane for a long time (the arrows are short). This region may be thus considered as the entry into the resonance. If a given system starts its evolution from a point which leads to that region of the diagram, the capture into 9:7~MMR is most likely.

Trajectories of two representative configurations are shown with green and red tracks. The green case corresponds to the system that happens to reach the entry into the resonance. The system is captured permanently. The red case corresponds to the system that passes through the resonance. It only differs from the green case by the initial value of $\phi_1$, while the initial period ratio is the same. 

\begin{figure}
\centerline{
\vbox{
\hbox{\includegraphics[width=0.48\textwidth]{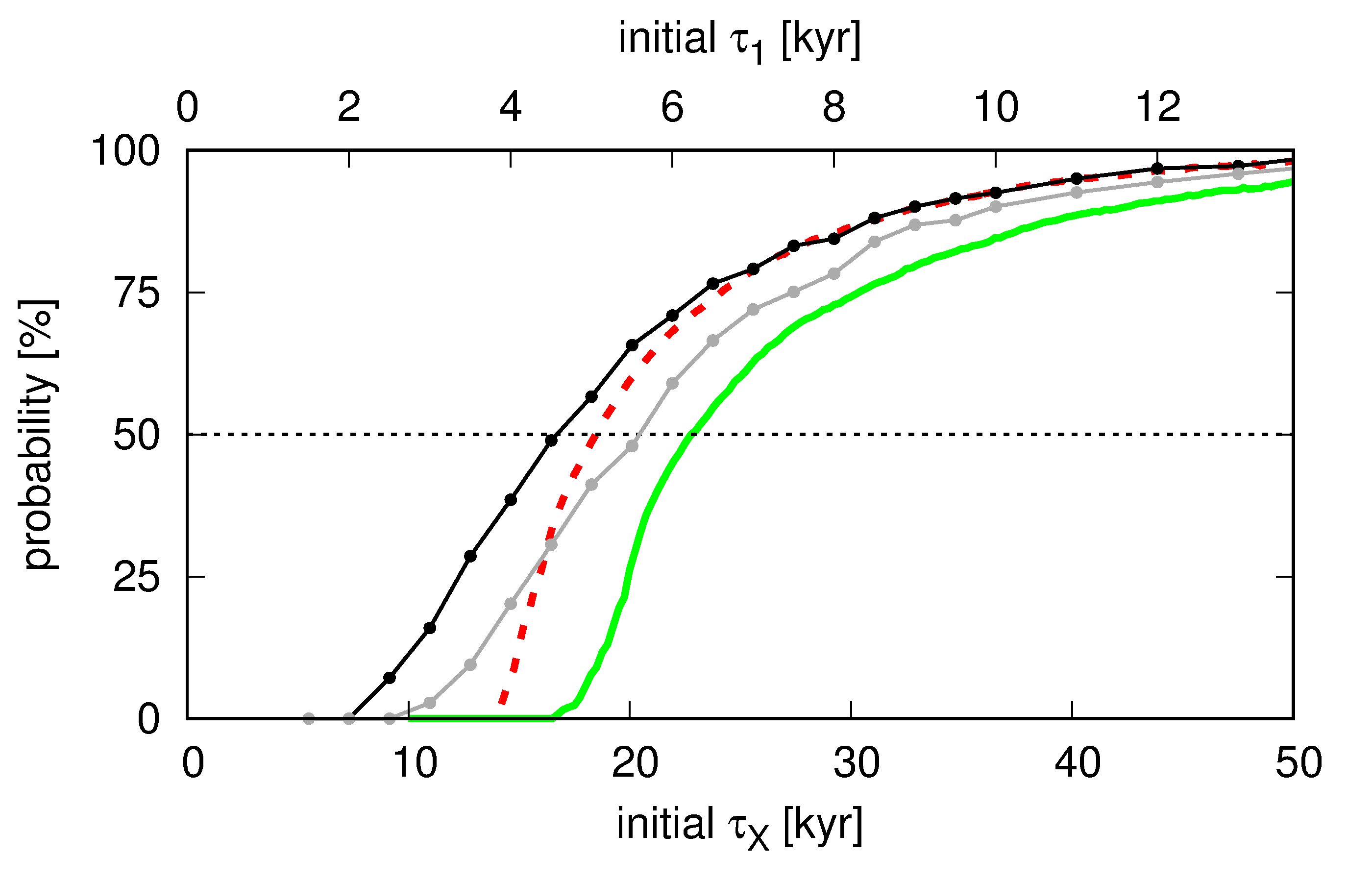}}
}
}
\caption{The capture probability as a function of the initial values of $\tau_1$ and $\tau_X$ (i.e., the values at $a_1=0.1\,\au$). The green solid curve illustrates the probability of the stable (permanent) capture, while the dashed red curve denotes the probability of any capture, stable or unstable. Both curves correspond to the results of the averaged approach. Black and grey points connected with lines show the probability of any capture obtained on the ground of the N-body simulations for two different {ranges of the initial period ratio: $(1.288, 1.29)$ and $(1.29, 1.292)$, respectively} (see the text for details). The masses of the planets $m_1 = m_2 = 6\,\mE$, and the migration parameters are $\alpha=1$, $\kappa=75.05$, $e_1^{\idm{(eq)}}=0.008$.
}
\label{fig:entering_prob_1}
\end{figure}

Because the initial $\phi_1$ and $P_2/P_1$ are arbitrary when the migration into the resonance is considered, one can say that the entrance into 9:7~MMR should be considered probabilistically. The probability of the stable (permanent) capture can be measured as the ratio between the area without the grey points and the total area for a given range of $P_2/P_1$ at the right-hand side of the diagram and for the whole range of $\phi_1$. For the case considered in Fig.~\ref{fig:entering1} that probability $\approx 84.6$~per cent. Similarly the probability of the unstable (temporary) capture can be measured as the ratio between the light grey area and the total area. The probability of any capture is simply a sum of those two probabilities. In the case considered the probability $\approx 92.8$~per~cent.

Naturally the probabilities depend on the time scale of migration, $\tau_0$, and the $\alpha$ and $\kappa$ values (or equivalently on $\alpha$ and $e_1^{\idm{(eq)}}$). Figure~\ref{fig:entering_prob_1} shows the capture probability as a function of $\tau_1$ (the top label) and $\tau_X \equiv -X/\dot{X}$, where $X \equiv P_2/P_1$ (the bottom label). The green curve corresponds to the stable capture probability, while the red dashed curve denotes the probability of any capture (stable or unstable). Predictably, the capture probability increases for slower migration. In the next section we try to verify the results shown in Fig.~\ref{fig:entering_prob_1} with the help of the N-body simulations.

\subsection{N-body verification of the capture probability}

In general, the capture probability depends not only on the migration parameters and planets' masses but also on the initial orbits. Moreover, the probability depends on the distribution of the initial orbits, both its character and ranges for particular orbital elements. In the previous analysis, we assumed that the systems initially start from the equilibrium of 4:3~MMR and that the initial period ratio is close to $9/7$. As the equilibria in the averaged model correspond to the periodic orbits in the full N-body model, a natural choice for the N-body verification would be that the systems start from the branch of periodic orbits of 4:3~MMR. The periodic configurations require $\Delta\varpi=\pi$ and particular combination of the mean anomalies. We mentioned already that $\Mmean_1=\pi, \Mmean_2=\pi$ as well as $\Mmean_1=\pi, \Mmean_2=0$ both lead to appropriate values of the resonant angles of 4:3~MMR, when $P_2/P_1<4/3$. We can choose the initial phase freely as far as we keep the eccentricities at the values corresponding to the periodic orbits. 

\begin{figure*}
\centerline{
\vbox{
\hbox{\includegraphics[width=0.45\textwidth]{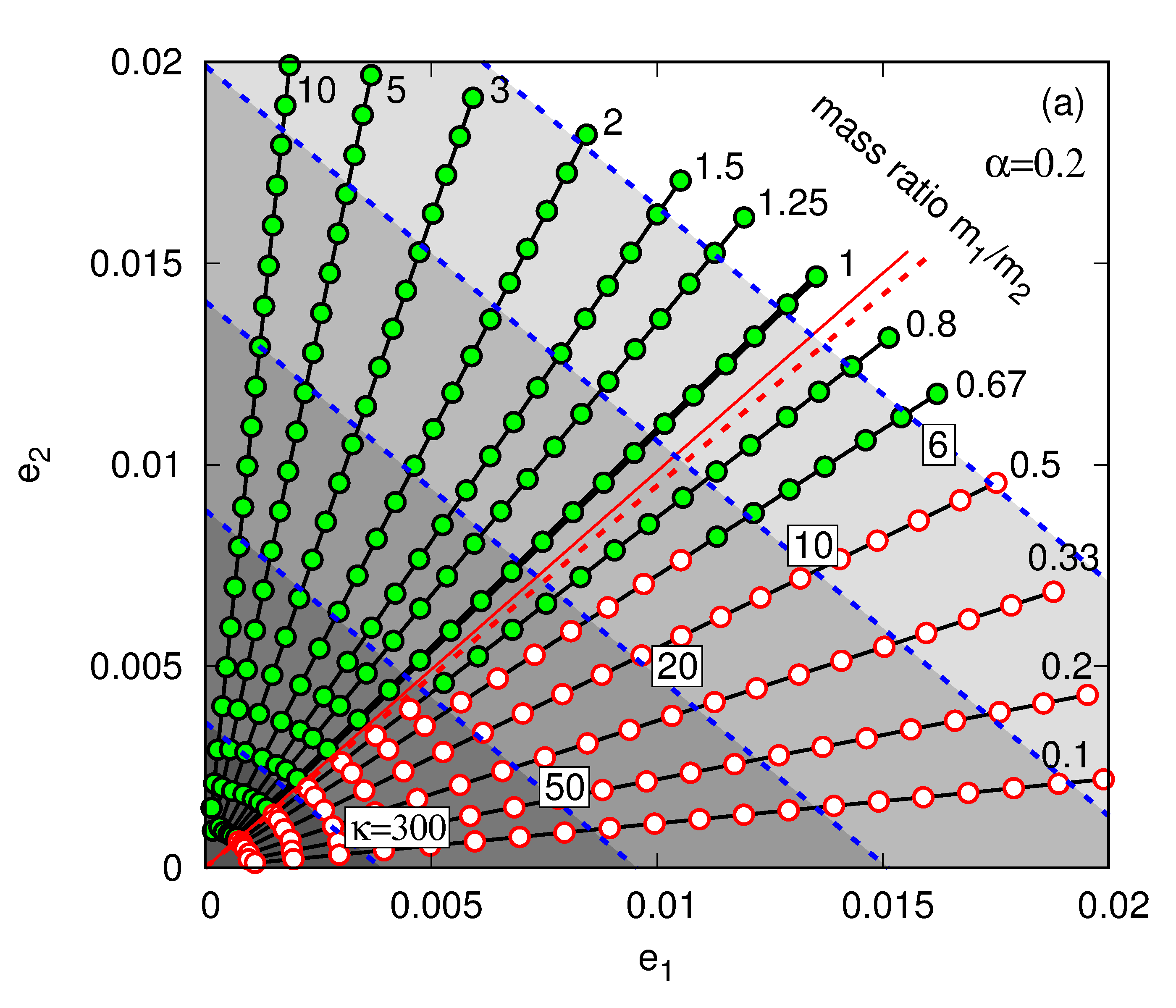}
\includegraphics[width=0.45\textwidth]{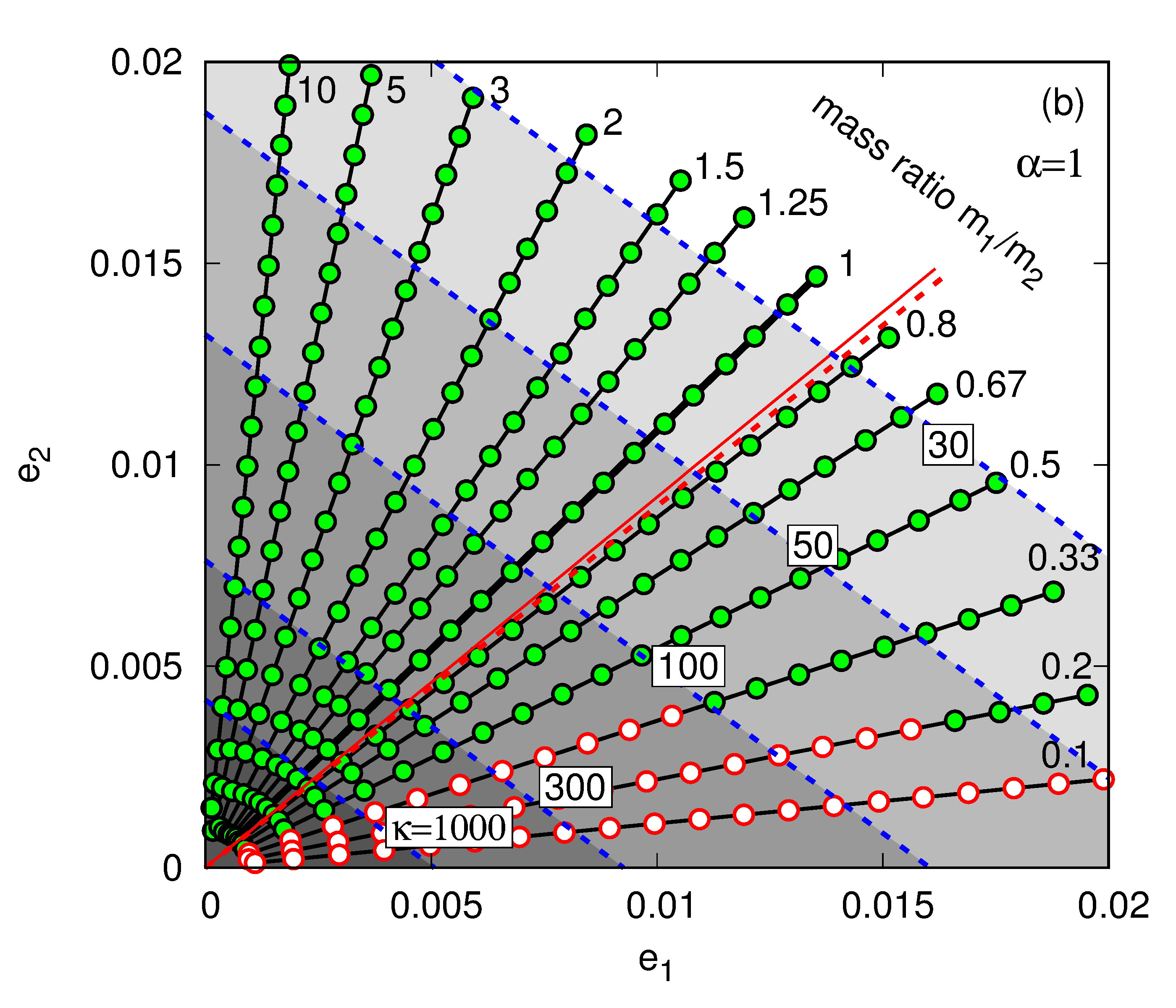}}
\hbox{\includegraphics[width=0.45\textwidth]{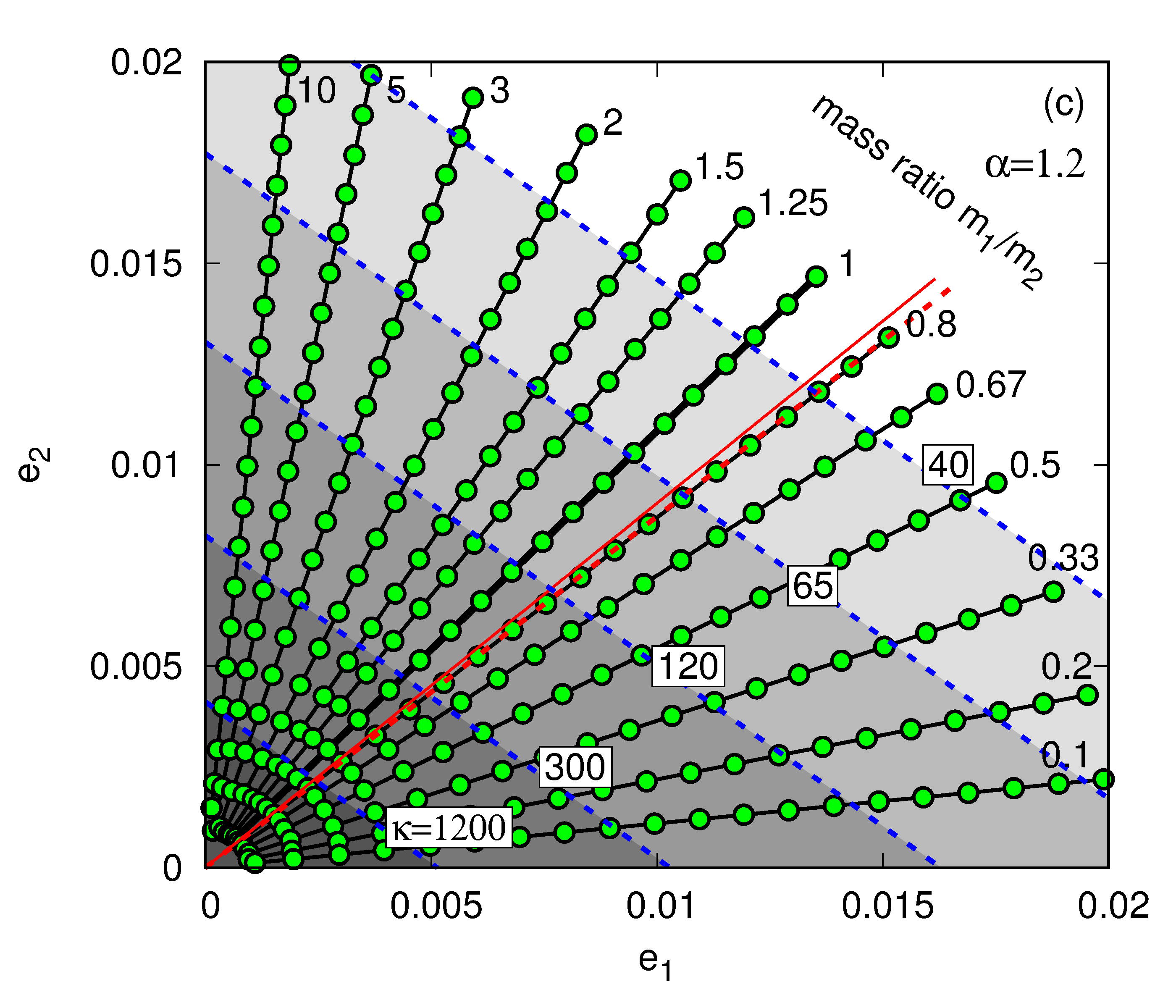}
\includegraphics[width=0.45\textwidth]{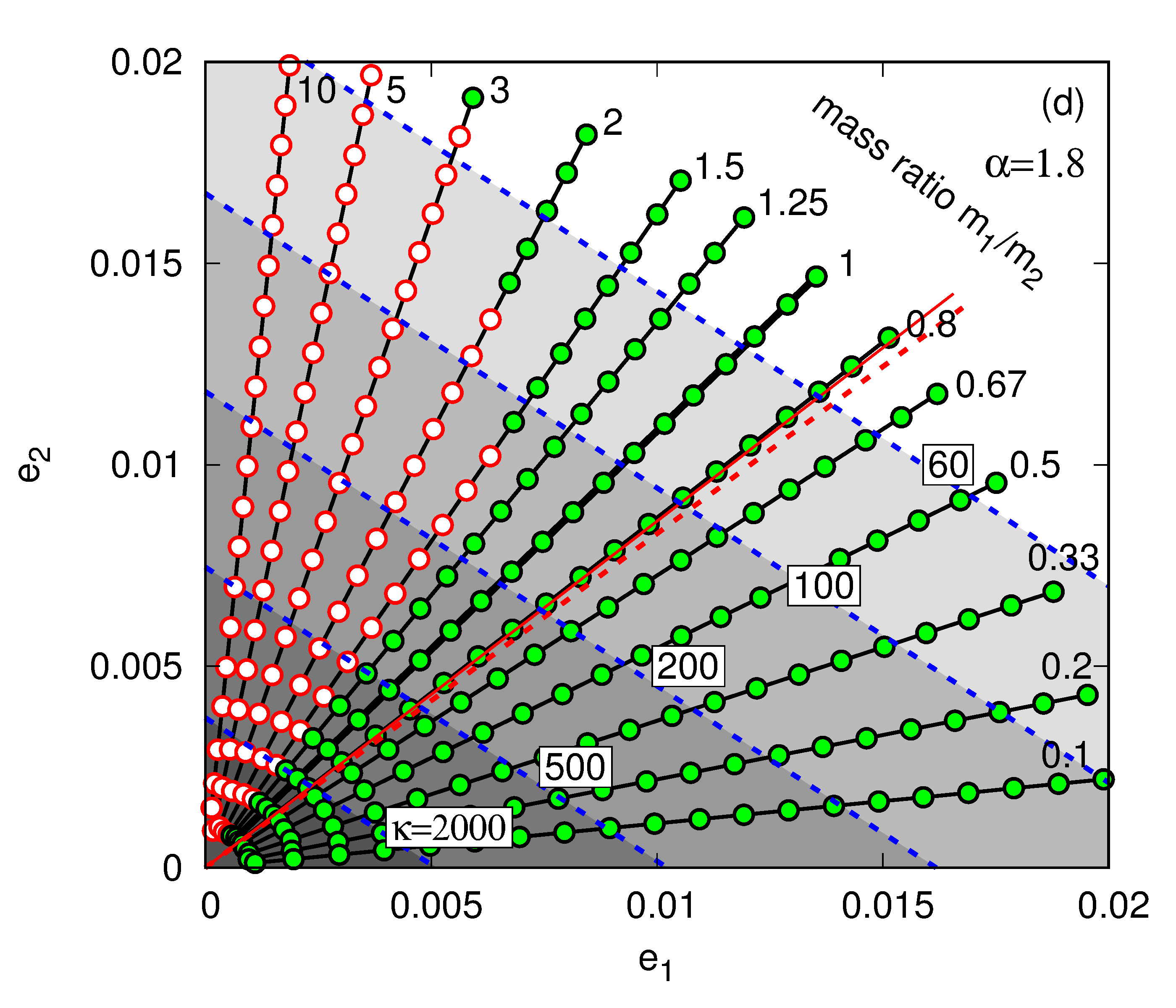}}
}
}
\caption{Black/green and red/white points along black solid lines denote the equilibria for $\Delta\varpi=\pi$ and $\phi_1=\pi$ for different mass ratios (labelled accordingly). For all the cases $m_1+m_2 = 12\,\mE$. Dashed blue lines denote levels of constant values of $\kappa$ (labelled accordingly) for which a given system reaches given equilibrium values of the eccentricities, i.e., a given point $(e_1, e_2)$ at the diagram. The black/green points indicate that the system remain resonant when the migration is added, while the red/white points mean that the system leaves the resonance. The dashed red line illustrates the results of \citep{Xu2016}, while the solid red line represents the results of \citep{Delisle2015}. For mass ratios higher than the one marked by the red line, the systems can remain resonant after adding the migration. For smaller mass ratios, the migration makes the systems leave the resonance. Subsequent panels, from (a) to (d),  present the results for $\alpha=0.2, 1.0, 1.2, 1.8$, respectively. 
}
\label{fig:eq_migr_stab}
\end{figure*}

Unfortunately, when starting from the exactly periodic system, the capture does not occur for $\tau_1 = 10\,$kyr (the averaged model gives almost certain capture for this value of $\tau_1$). It is necessary to increase $\tau_1$ by a factor of $3$ in order to make the capture possible. 
{Direct comparison of the N-body results with the results presented in Fig.~\ref{fig:entering1} cannot be done, as a given value of $\phi_1$ can be achieved for different combinations of $(\Mmean_1, \Mmean_2)$. Moreover, a choice of $(e_1, e_2)$ other than the one for a periodic orbit would be arbitrary. Instead, we performed simulations in which the orbital elements are being chosen randomly from given ranges, i.e., $e_1, e_2 \in [0, 0.002]$, $\Mmean_1, \Mmean_2 \in [0, 2\pi]$, $P_2/P_1 \in [1.288, 1.29]$ -- for the first set of simulations and $P_2/P_1 \in [1.29, 1.292]$ -- for the second set. Initial $a_1 = 0.1\,\au$, $\Delta\varpi = \pi$ for both sets. For each given $\tau_0$ we integrate $3000$ initial systems over time $t=\tau_0$ ($P_2/P_1$ approaches 9/7 in $t \sim 0.1\,\tau_0$ when starting from the initial values given above) and note final $P_2/P_1$. If the period ratio is below $1.275$ at the end of the simulation we count the system as non-resonant. Otherwise, the system is considered as captured into resonance. Therefore, we compute the probability of any capture (stable or unstable). The results are presented in Fig.~\ref{fig:entering_prob_1}. The black points connected with lines correspond to the first range of the period ratio, while the grey symbols indicate the capture probability obtained for the second range of $P_2/P_1$.}

{The probabilities obtained on the ground of the N-body model agree with the results of the averaged model (black and grey curves should be compared with the dashed red curve), although the differences for fast migration ($\tau_1 \sim 4\,$kyr) are noticeable. The systems that start farther from 9:7~MMR enter the resonance less likely. The explanation of this fact is the following. As it was shown in Fig.~\ref{fig:migr_ex2}, periodic orbits act as attractors in the phase space when the migration is included in the model. A system which starts farther from a given resonance approach the branch of periodic configurations closer when compared with a system whose initial $P_2/P_1$ is lower, i.e., closer to the nominal resonant value. As we mentioned already, the closer a given system is from the branch of periodic orbits, the less likely is the capture.}

\subsection{Dependence of the planets mass ratio}

Recently \cite{Xu2016} presented a study of the migration induced formation of second order MMRs. Although they were focused on 3:1, 5:3 and 7:5~MMRs, their results (especially the ones for 5:3 and 7:5) should be qualitatively transferable into 9:7~MMR. They used an analytic model of the system (similar to the averaged model used in this work, but of a lower order in the eccentricities) and found that a system initially in the equilibrium (from the branch of equilibria which exists for $\Delta\varpi=\pi$ and $\phi_1=\pi$) of the second order resonance not necessarily stays in the resonance when the migration is added. They show that when the time-scales of circularisation for both planets are comparable $\tau_{\idm{e,1}} \approx \tau_{\idm{e,2}}$, the systems can remain resonant only for $m_1 \gtrsim m_2$. Similar criterion was found by \cite{Delisle2015}. They found that the equilibrium is stable when $\tau_{\idm{e,1}}/\tau_{\idm{e,2}} > (e_1^{\idm{(eq)}}/e_2^{\idm{(eq)}})^2$. Because $e_1^{\idm{(eq)}}/e_2^{\idm{(eq)}} \approx m_2/m_1$ for 9:7~MMR in the low eccentricity limit, both criteria are equivalent. Below we try to verify those stability limits.

Figure~\ref{fig:eq_migr_stab} illustrates the dependence of the stability of equilibria when the migration is added to the model on the mass ratio and on $\alpha$. For each point $(e_1, e_2)$ from a given branch of equilibria (black/green and red/white points) which was found for a given value of $m_1/m_2$ and $\alpha$ we find such $\kappa$ for which those $e_1$ and $e_2$ are the equilibrium eccentricities. Next, we integrate the equations of motion and verify if the system evolves towards or away from the equilibrium. If the system moves towards the centre of the resonance, we mark the point as black/green, otherwise the point is red/white.

Although our parametrisation is different from the ones used in \citep{Delisle2015,Xu2016}, we can easily find that $q_{\idm{e}} \equiv \tau_{\idm{e,2}}/\tau_{\idm{e,1}} = (7/9)^{\alpha}$. For a given value of the ratio $q_{\idm{e}}$ one can find the critical value of $q \equiv m_1/m_2$ below which the system cannot stay in the resonance permanently. For all the $\alpha$ values considered here this critical value is between $\sim 0.8$ and $\sim 0.9$. The dashed and solid red lines plotted in each panel of Fig.~\ref{fig:eq_migr_stab} corresponds to the branch of equilibria for this critical value of the mass ratio computed on the ground of the criteria from \citep{Delisle2015,Xu2016}.

The results of \cite{Delisle2015,Xu2016} are confirmed in our study for $\alpha=0.2$ (Fig.~\ref{fig:eq_migr_stab}a), although we can see that for $e \gtrsim 0.01$ the equilibria are stable against the migration even if for lower eccentricities they were unstable. For $\alpha=1$ (Fig.~\ref{fig:eq_migr_stab}b) the border which stems from \citep{Delisle2015,Xu2016} does not overlap with our results. For slightly higher $\alpha=1.2$  (Fig.~\ref{fig:eq_migr_stab}c) the equilibria are stable within the whole range of the mass ratio. For $\alpha=1.8$ (Fig.~\ref{fig:eq_migr_stab}d) the situation reverses when compared to panels~(a) and~(b). The equilibria are unstable against the migration for high mass ratio, while for low $m_1/m_2$ they are stable. This last case shows that the permanent capture of the planets with $m_1/m_2$ as small as $0.1$ is possible for high values of $\alpha$.

\begin{center}
\begin{minipage}[l]{\linewidth}
\begin{center}
\includegraphics[width=0.92\textwidth]{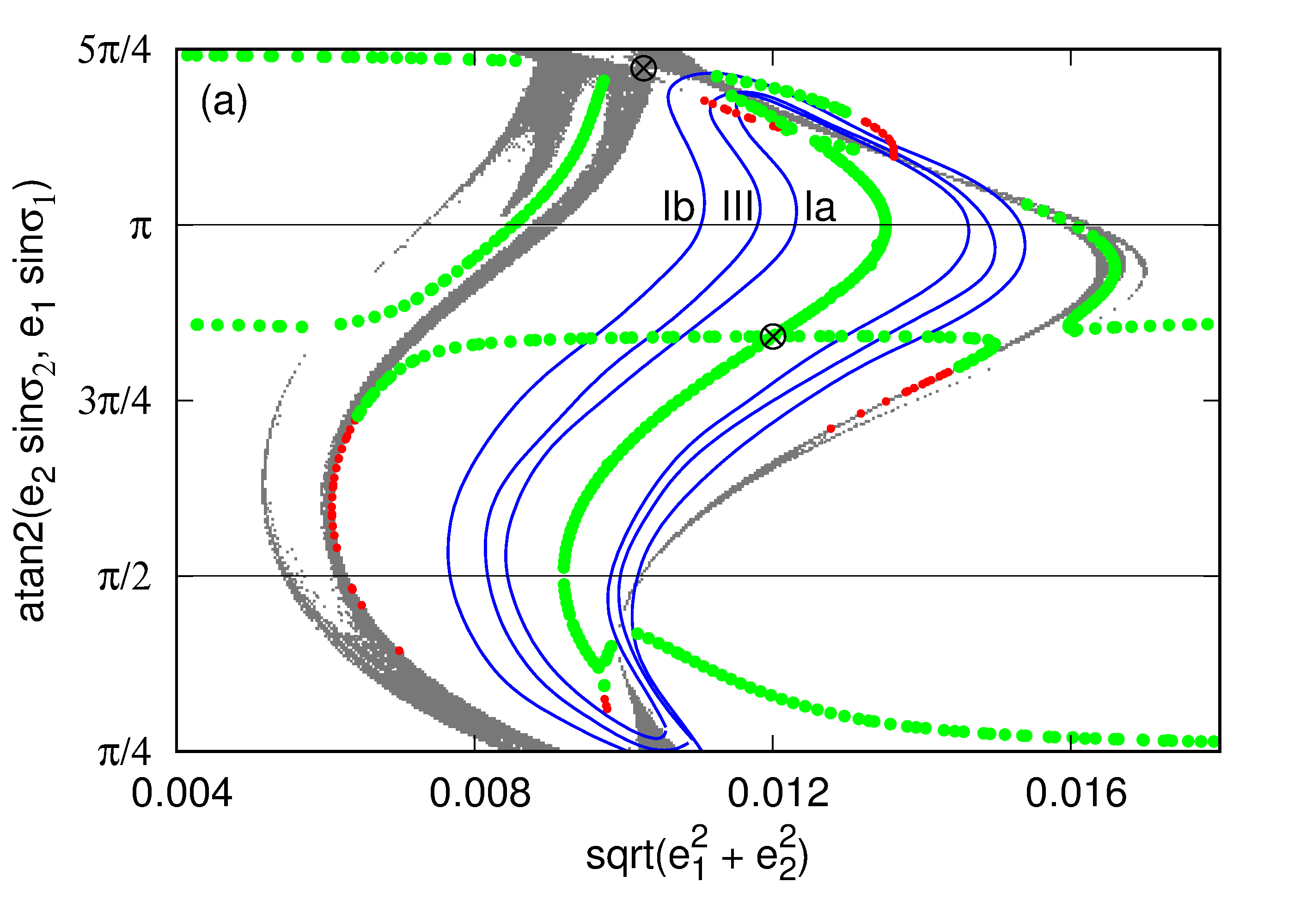} 
\includegraphics[width=0.92\textwidth]{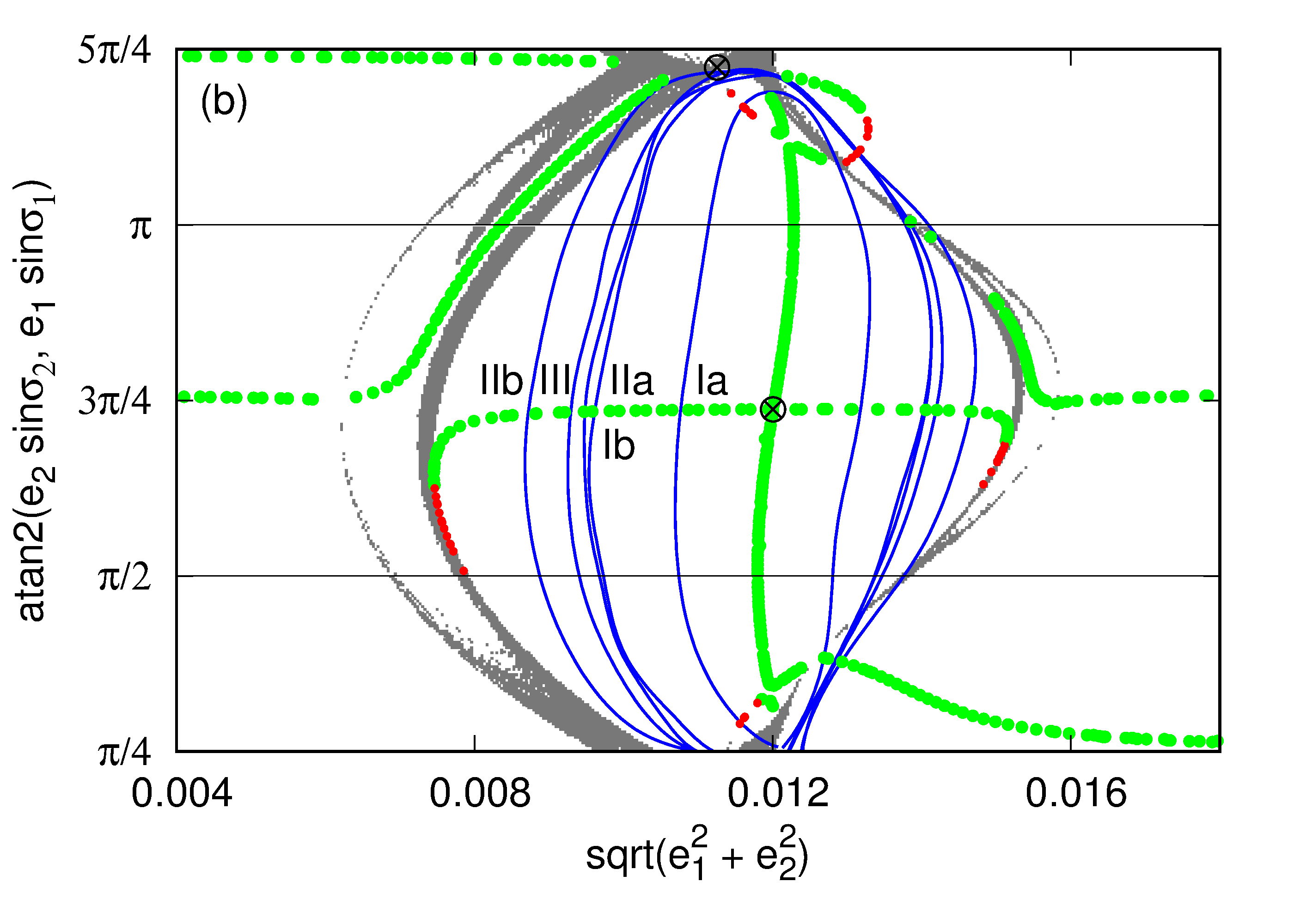} 
\includegraphics[width=0.92\textwidth]{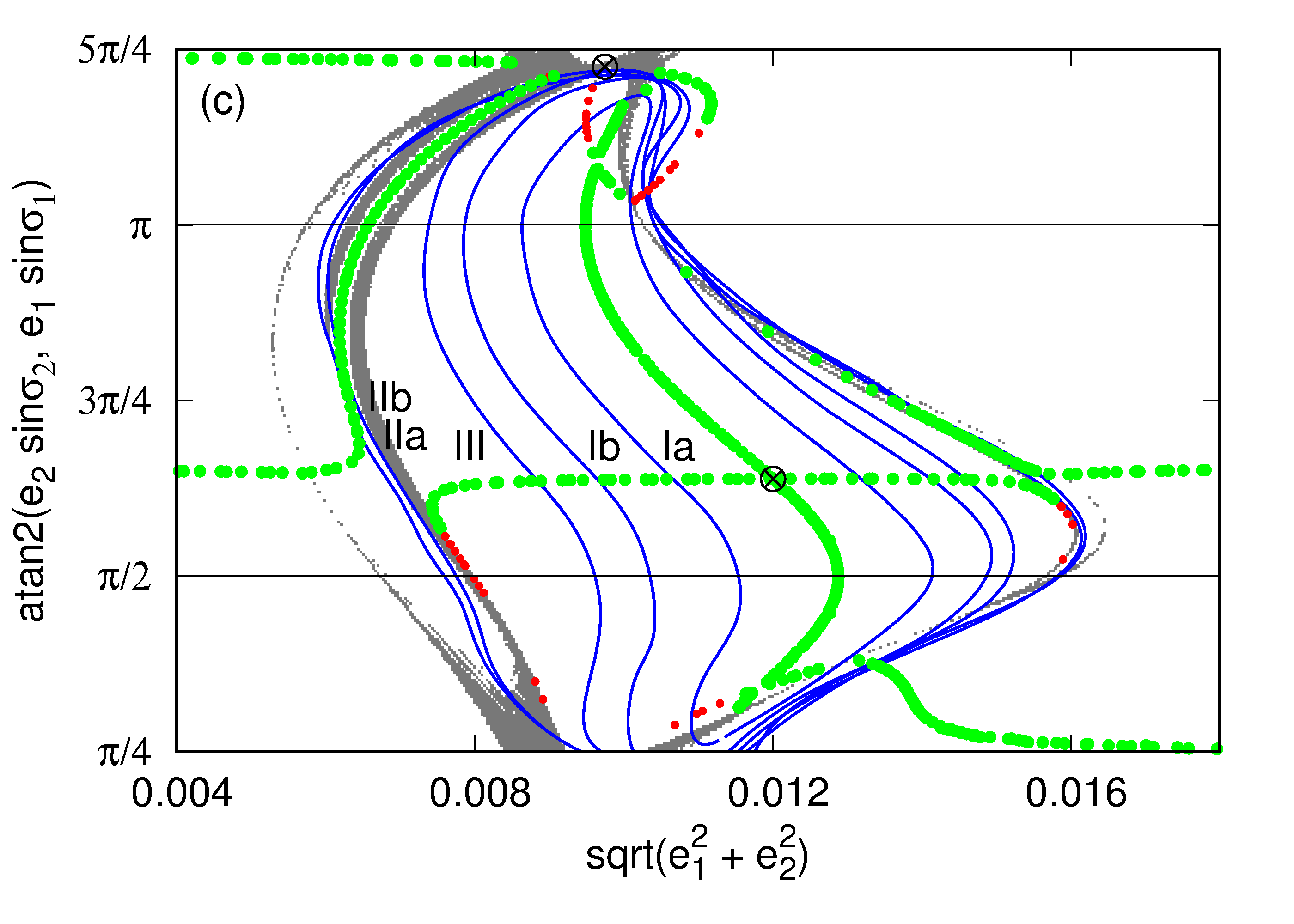}
\end{center}
\captionof{figure}{The energy diagrams similar to Fig.~\ref{fig:critical_energies} obtained for $m_1/m_2 = 0.5, 1, 2$ (panels~a, b and~c, respectively). For all three cases $m_1+m_2 = 12\,\mE$. Green and red points denote stable and unstable periodic orbits. Cross/circle symbols denote the equilibria. Grey dots indicate chaotic evolution. Blue curves mark borders of the stability zone, each curve is for different model of the migration (denoted with Ia, Ib, IIa, IIb and III, see the text for explanation). For all the cases $\alpha=1$.}
\label{fig:diff_models}
\end{minipage}
\end{center}

\subsection{Dependence on the migration model}

The differences discussed above stem from different models of the migration used in this work and in the cited papers. Here we use the force given by Eq.~\ref{eq:migration_force} together with the assumption of $\tau_a = \tau_a(r)$. That leads to Eq.~\ref{eq:migration_terms} for the $a$ and $e$ migration induced evolution (in the mean sense). \cite{Delisle2015} and \cite{Xu2016} use slightly different formulae:
\begin{equation}
\dot{a}^{\idm{(migr)}} = -\frac{a}{\tau_a} \left(1 + \gamma\,\kappa\,e^2\right), \quad
\dot{e}^{\idm{(migr)}} = -\frac{e\,\kappa}{\tau_a}.
\label{eq:dot_a_simpler}
\end{equation}

The force given by Eq.~\ref{eq:migration_force} and $\tau_a = \mbox{const}$ (or $\tau_a = \tau_a(a)$) leads to similar expressions, apart from the numeric factor $\gamma$. Instead of $2$, the averaging over the mean anomaly gives $5/8$. In other words, if the force which mimics the migration is a linear combination of the $\vec{v}$ and $\vec{v}_c$ vectors \citep[e.g.,][]{Papaloizou2000,Beauge2006,Moore2013,Voyatzis2016}, then $\gamma=5/8$. When $\tau_a$ is a function of $r$, the averaging gives $\gamma = \frac{5}{8}(1 - \frac{4}{5}\alpha)$. As recently shown \citep{Goldreich2014,Deck2015} a numeric value of this factor may be important for the stability of the equilibrium. In the literature $\gamma = 0$ is used as well \citep[e.g.,][]{Lee2002,Kley2004,Libert2009,Bodman2014}. Below we study the stability of the equilibrium for different mass ratios and different models of the migration.

Models~Ia and~Ib are characterized by $\gamma=0$ (Eq.~\ref{eq:dot_a_simpler}), where $\tau_a = \mbox{const}$ for model~Ia and $\tau_a = \tau_a(a)$ for model~Ib. The factor $\gamma=5/8$ for models~IIa and IIb (IIa means that $\tau_a = \mbox{const}$, while in model~IIb $\tau_a = \tau_a(a)$). The last model studied (III) has $\dot{a}$ and $\dot{e}$ given by Eq.~\ref{eq:migration_terms}.

Figure~\ref{fig:diff_models} shows the differences in the stability zone around the equilibrium  resulting from the five models. The blue curves in Fig.~\ref{fig:diff_models} denote the borders between the dotted-blue and white regions inside the resonance as presented in Fig.~\ref{fig:critical_energies}. For the mass ratio $m_1/m_2 = 0.5$ (panel~a) the stability zone disappears within models~IIa,b. Models~Ia,b and model~III give similar results, all of them indicate that the equilibrium is stable and the zone spans significant part of the resonant region. When $\alpha=1$, model~III has $\gamma=1/8$ which is only slightly more than $0$ (as for models~Ia,b). For equal masses (panel~b) all models give the stable equilibrium, although the sizes of the stability zone differs from one model to another. Still the zone spans significant part of the resonance. In the last case (panel~c;for $m_1/m_2=2$), models~IIa,b leads to the stability zone as wide as the whole resonance. The other three models with smaller values of $\gamma$ gives narrower stability zones, although again for all the models the equilibrium is stable when the migration is considered.

\begin{figure}
\centerline{
\vbox{
\hbox{\includegraphics[width=0.43\textwidth]{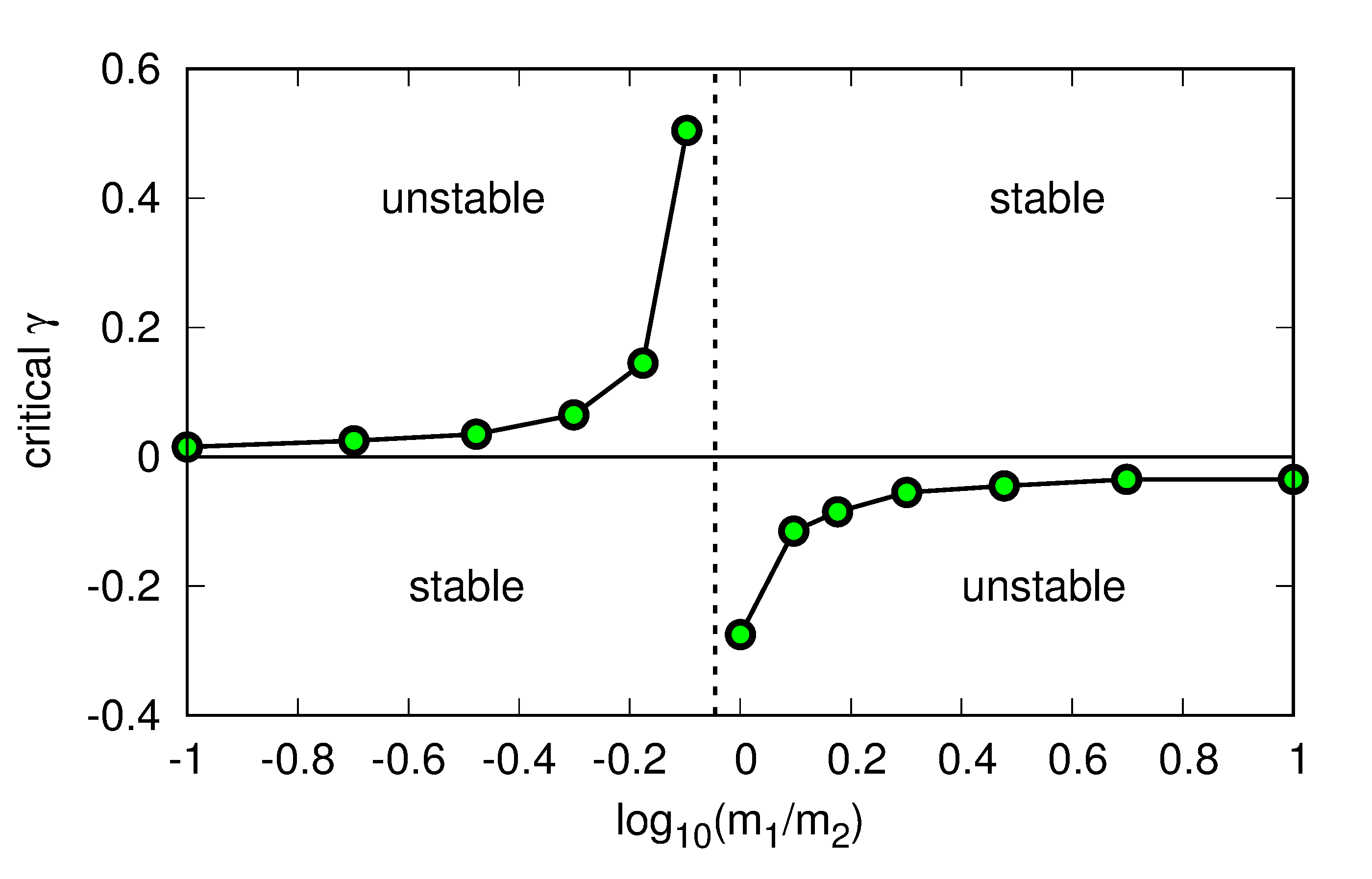}}
}
}
\caption{
The critical value of $\gamma$ as a function of the mass ratio.
}
\label{fig:eq_stab}
\end{figure}

Figure~\ref{fig:eq_migr_stab} shows that for small values of $\alpha=0.2$ the equilibrium is stable if $m_1/m_2 \gtrsim 1$, while for higher $\alpha=1.8$ the situation reverses. For an intermediate value of $\alpha=1.2$ the equilibrium is stable in the whole range of the mass ratio. Because $\gamma = \gamma(\alpha)$ in the model with $\tau_a = \tau_a(r)$, this transition can be understood as the dependence of the stability on $\gamma$. Figure~\ref{fig:eq_stab} illustrates the dependence. For the mass ratio $\gtrsim 0.9$ the equilibrium is stable for $\gamma$ greater the the critical value, while for $m_1/m_2 \lesssim 0.9$ the stability requires $\gamma$ to be smaller than the critical value. Although the critical value depends on the mass ratio, it is close to $0$ almost in the whole range, apart from $m_1/m_2 \approx 0.9$. Negative values of $\gamma$, though, require an inverse profile of the surface density of the disc, when classic Type~I migration is considered \citep{Tanaka2002}, which can be achieved at the disc inner border. 
An interesting observation is that for $\gamma=0$ the equilibrium is stable in the whole range of the mass ratio.

\section{Conclusions}

We have studied the formation of 9:7~mean motion resonance between two migrating planets of a few Earth masses. We used a simple parametric model of the migration and circularisation incorporated into the N-body as well as the averaged equations of motion. We confirm and explain the probabilistic nature of the entrance into the resonance \citep[e.g.,][]{Quillen2006,Mustill2011}. We show that the capture can be permanent or temporary, which is in agreement with previous studies \citep[e.g.,][]{Delisle2015,Xu2016}. 

We demonstrate that the equilibrium for $\Delta\varpi=\pi$ and $\phi_1=\pi$ of the averaged conservative model (which corresponds to the periodic configuration within the N-body model of motion) plays a crucial role in the capture. The equilibrium is surrounded by the stability zone, the existence and the size of which depend on the mass ratio as well as on a particular model of migration. When $m_1<m_2$ and the disc surface density decreases with the distance from the star, the stability zone does not exist and the capture cannot be permanent. When the disc profile is reversed, the stability zone exists for $m_1<m_2$ and does not exist when $m_1>m_2$. For particular disc profiles the equilibrium may be stable in the whole range of the mass ratio. Generally, the stability zone spans only some part of the resonance, a region centred at the equilibrium. If the system, which just entered the resonance, is located inside the stability zone, it will move towards the equilibrium and the capture is permanent. Otherwise the system leaves the resonance despite the fact that the stability zone exists. 

We show that 9:7~MMR consists of two modes that are separated one from another with the separatrix. We called them the inner mode and the outer mode. The inner mode is centred at the equilibrium of the averaged model, while the outer mode is centred at a branch of periodic configurations. We illustrate that a system in resonance evolves along branches of periodic orbits of the averaged system. Shortly after entering the resonance the system achieves the branch which goes through the equilibrium (the one with $e_1/e_2 \approx \mbox{const}$, i.e., the horizontal branch at the polar $\mathcal{S}$-plane) and evolves towards or away from the equilibrium. In the latter case, when reaching the separatrix the system moves from the inner mode of the resonance to the outer one. The system evolves further along another branch of periodic orbits, until it passes through the outer separatrix and leaves the resonance. We illustrate the process of leaving the resonance at the representative plane as well as at the Poincar\'e cross sections.

We study the probability of permanent and temporary capture into the resonance and show that the system needs to enter the resonance in particular phase (a particular value of the resonant angle $\phi_1$). Therefore, even small changes in the initial orbits may lead to different final configurations. The phase-sensitivity of the process of entering the resonance has its consequence in the probabilistic nature of the capture. We illustrate that the system which starts the migration with $P_2/P_1 > 9/7$ tends towards the branch of periodic orbits of 4:3~MMR. The further from 9:7~MMR the system starts, the closer to the branch of 4:3~MMR it is when $P_2/P_1$ approaches $9/7$. We show that the proximity to the branch of 4:3~MMR periodic orbits makes the capture into 9:7~MMR more difficult. Therefore, the capture probability depends strongly on the initial orbits.

The results presented in this paper illustrate the complexity of the process of the migration induced formation of 9:7~MMR (and possibly other second order resonances). Further studies based on simplified parametric as well as hydrodynamical models are needed in order to understand the formation of such compact resonant configurations.

\section*{acknowledgements}

{CM thanks Alexander Mustill for valuable comments that helped to improve the manuscript.}
This work was supported by Polish National Science Centre MAESTRO grant DEC-2012/06/A/ST9/00276.

\bibliographystyle{mn2e}
\bibliography{ms}
\label{lastpage}
\end{document}